\documentclass[12pt]{article}
\usepackage{amsmath, amssymb, amsthm, verbatim, tikz, natbib, booktabs, bbm, xcolor,float}
\usetikzlibrary{arrows.meta, matrix}
\usetikzlibrary{decorations.markings}
\tikzset{
	vtx/.style={draw, rounded corners=2.5pt, minimum width=12.5mm,
		minimum height=7mm, inner sep=1pt, font=\footnotesize},
	vtxph/.style={draw=gray, dashed, rounded corners=2.5pt, minimum width=12.5mm,
		minimum height=7mm, inner sep=1pt, font=\footnotesize, text=gray},
	intra/.style={-{Stealth[length=2.2mm]}, semithick},
	inter/.style={-{Stealth[length=2.2mm]}, semithick, blue!65!black},
	miss/.style={-{Stealth[length=2.2mm]}, semithick, dashed, gray}
}
\usepackage{enumerate, algorithm, algpseudocode, comment, csquotes, dsfont}
\usepackage{float}
\usepackage{tikz}
\usepackage{rotating}
\usepackage[title]{appendix}
\usepackage{xr-hyper}
\usepackage[colorlinks=true, urlcolor=blue!80!black, citecolor=blue!50!black, linkcolor=blue!50!black]{hyperref}
\AtBeginDocument{
\makeatletter
\newcount\MT@frozen \MT@frozen=0
\providecommand{\MTfreeze}[2]{%
  \@ifundefined{r@#1}{\global\advance\MT@frozen by 1\relax
    \global\@namedef{r@#1}{#2}}{}}

\MTfreeze{S1}{{S1}{2}{Construction of the market-level allocation instances}{section.1}{}}
\MTfreeze{appendix:ipf}{{S2}{6}{A worked example of the dataset construction}{section.2}{}}
\MTfreeze{appendix:placebo}{{S3}{9}{Why individual households cannot be tracked}{section.3}{}}
\MTfreeze{S-proof-DPWC2}{{S4}{10}{Proof of Theorem \ref {DP-Rationalizability-WC-2}}{section.4}{}}
\MTfreeze{fig-device}{{S1}{14}{Two forbidden cases}{figure.1}{}}
\MTfreeze{RF1}{15}
\MTfreeze{fig-full}{{S2}{15}{The complete mixed graph of the running example ($G^{\ast }$: the path $v_{1}$---$v_{2}$---$v_{3}$, $k=2$). Bottom layer: the switches. Middle layer: the counter switches with $j=1$. Top layer: the counter switches with $j=2$, and, above them, the relay}{figure.2}{}}
\MTfreeze{S-mip}{{S5}{19}{A mixed-integer formulation of Condition \ref {cond2}}{section.5}{}}
\MTfreeze{S-relations}{{S6}{22}{Relations between the properties}{section.6}{}}
\MTfreeze{prop-nonnested}{{S.1}{22}{}{proposition.1}{}}
\MTfreeze{prop-joint}{{S.2}{23}{}{proposition.2}{}}
\MTfreeze{S-examples}{{S7}{24}{Examples omitted from the main text}{section.7}{}}
\MTfreeze{example-sc}{{S.1}{24}{}{example.1}{}}
\MTfreeze{example-wc-not-pi}{{S.2}{25}{}{example.2}{}}
\MTfreeze{example-joint}{{S.3}{26}{}{example.3}{}}
\MTfreeze{ex-betweenness}{{S.4}{29}{Choosing admissible crossing points independently is not enough}{example.4}{}}
\MTfreeze{ex-unheld}{{S.5}{30}{The mixed graph is still not sufficient}{example.5}{}}

\AtEndDocument{\ifnum\MT@frozen>0
  \PackageWarningNoLine{main-refs}{Used FROZEN main-text numbers for
    \the\MT@frozen\space label(s): the live .aux was not found.^^J
    Numbers are correct as of the date in main-refs.tex.^^J
    If the main text changed, rerun make-main-refs.py}\fi}
\makeatother

{supp-refs.tex missing -- run make-refs.py}}}
\usepackage[letterpaper, margin=1.25in]{geometry}

\makeatletter
\g@addto@macro\UrlBreaks{\UrlOrds}
\makeatother

\tikzset{
	table/.style={
		matrix of nodes,
		row sep=-\pgflinewidth,
		column sep=-\pgflinewidth,
		nodes={rectangle, text width=3em, align=center},
		text depth=1.25ex,
		text height=2.5ex,
		nodes in empty cells
	}
}

\usepackage{etoolbox}

\theoremstyle{plain}
\newtheorem{theorem}{Theorem}
\newtheorem{proposition}{Proposition}
\newtheorem{lemma}{Lemma}
\newtheorem{corollary}{Corollary}

\theoremstyle{definition}
\newtheorem{definition}{Definition}
\newtheorem{condition}{Condition}

\newtheorem{example}{Example}
\AtEndEnvironment{example}{\hfill$\lozenge$}

\theoremstyle{plain}
\newtheorem*{decision}{Problem}
\newtheorem{observation}{Observation}
\newtheorem{claim}{Claim}
\newtheorem*{shortcircuit}{Short-circuit Lemma}

\newcommand{\new}[1]{#1}
\newcommand{\clr}[1]{#1}
\newcommand{\rev}[1]{#1}

\title{Revealed Preference Analysis of Shapley--Scarf Housing Markets}
\author{U. Salman\footnote{Department of Economics and Management, University of Padova} \and M. Lombardi\footnote{\textit{Corresponding author}: University of Liverpool and University of Napoli Federico II} \and F. Ciardiello\footnote{Department of Economics and Statistics, University of Salerno} \and R. Saulle$^*$}

\begin{document}
	\maketitle
	\begin{abstract}

			\noindent We ask what the reallocation of indivisible objects reveals about the market that produced it. A central authority assigns the objects, after which recipients exchange them among themselves. Preferences are never observed, and individuals of the same type have identical preferences. A reallocation is rationalizable as Pareto efficient and individually rational exactly when a directed graph built from the data is acyclic, and as weak core-stable exactly when a mixed graph admits an orientation acyclic within types and free of alternating cycles. Deciding the latter is NP-complete, and on the single-crossing domain, although the test becomes more restrictive, it turns out to be more useful in practice. Both properties are less demanding than the strict core, so their rejection eliminates a broader set of possibilities. We apply the tests to administrative data on social housing lettings in England.
		\bigskip
		
	\end{abstract}

	\section{Introduction}\label{Introduction}

	This paper studies the testable implications of Pareto efficiency and individual rationality, and of the weak core, when a central authority assigns indivisible objects and recipients then reallocate them among themselves. Following classical revealed preference analysis, we observe the assignment and the reallocation that follows it, but never preferences, and ask what the reallocation reveals about the market that produced it.

 Exchange of this kind is common and often officially allowed. Social
 housing is assigned to families, public-sector posts to employees, and offices and equipment to staff; afterward tenants swap homes, employees trade posts, and longer chains form in which one individual gives up an object that a second takes. Social housing tenants in the United Kingdom swap through ``mutual exchange'' schemes with their landlords'
 consent.\footnote{See \url{https://england.shelter.org.uk/housing\_advice/council\_housing\_association/tenancy\_exchanges}}
 Public employees in Turkey may swap posts within an institution across
 locations, a practice known as
 Becayi\c{s}.\footnote{This procedure is rooted in Article 73 of Civil
  Servants Law No. 657, T\"urkiye. See
  \url{https://tr.wikipedia.org/wiki/Becayi\%C5\%9F}}

Whether such exchange serves the recipients well is not obvious, and the authority cannot see the answer directly. However, the authority has a decision to make. If the exchange has already produced a good outcome, imposing a formal mechanism costs money and political effort for little gain; if it has not, a mechanism could deliver real welfare. It would like to tell the two cases apart and cannot do so directly, because the exchanges are driven by preferences it does not observe.

 We do not try to recover those preferences. We ask instead what the exchanges rule out. The authority sees who held each object at the start and who holds it once the exchange is over, and nothing else. Its question is this:

 \begin{quote}
  \emph{Does there exist a profile of individual preferences such that the observed allocation is Pareto efficient and individually rational, or weak core-stable?}
 \end{quote}

 We say that the observed allocation is rationalizable under a given notion if the answer to the corresponding question is yes: that is, if there exists an admissible profile of individual preferences under which the allocation satisfies that notion. As it stands, the question has no content. For any reallocation at all, let
 every individual rank the object he ends up holding above all others: then nobody envies anybody and no group can improve on the outcome. Content must come from restricting the individual preferences. Afriat's theorem draws its power from repeated, independent observations of one consumer; here there is a single outcome, so the restriction has to come from elsewhere. Following \citet{echenique2013revealed}, we group individuals into types on the basis of observable characteristics and require individuals of the same type to have identical preferences. This is a strong assumption, and without it nothing is refutable. It is also what gives the exercise content: one individual's move reveals a comparison that binds every other member of his type, and what the data can confirm or contradict is whether these inherited comparisons fit together.

 \rev{The solution concept and the partition therefore enter the test
 together, and a rejection is a rejection of the two jointly. However, the
 discipline the framework supplies runs one way only: refining a partition
 can only increase rationalizable allocations, so rationalizability is weakly increasing in fineness, and the partition into singletons accepts everything. A verdict is reportable only alongside the partition that produced it, and the informative object is how far the types can be coarsened before the data reject. Section \ref{UKapplication} accordingly runs every test at three nested levels of granularity rather than one.}

 \rev{For one-sided markets, in which objects are reallocated among agents whose preferences are never observed, the question was opened by \citet{tai2022revealed}. He established that the exercise is possible in the Shapley--Scarf environment and carried it out in two settings: the strict core when utility is non-transferable, and the weak core when monetary transfers are available. We study the case left open by his analysis, which is also the case relevant for the markets that motivate us. A local authority does not price its dwellings and its tenants do not compensate one another, so transfers are unavailable, and the two properties a planner can act on there are Pareto efficiency together with individual rationality, and weak core-stability that would not allow any group of individuals to exchange their initial endowments and make all of their members strictly better off. This paper characterizes their testable implications, defines two tests and runs them on administrative data.}

 \rev{Both requirements are less stringent than the strict core, and therefore provide more informative insights. In \citet{tai2022revealed}'s framework, the set of strict core allocations is essentially unique. Consequently, rejecting strict core rationalizability only establishes that the observed reallocation does not correspond to that unique allocation. This outcome reflects the strength of the concept rather than a flaw, and it is the reason the strict core is appropriate when the planner's primary concern is immunity to costless regrouping. In practice, a housing authority is more likely to inquire only whether the outcome can be further improved or whether a subset of tenants might organize a trade and leave. Because the proposed properties are less demanding, their rejection eliminates a broader set of possibilities (Proposition \ref{prop-tai}).}

 \medskip
 We give three characterizations, and the paper is organized around them.

 \paragraph{Efficiency and individual rationality.}
 The first concerns Pareto efficiency together with individual rationality,
 which we abbreviate to PI. From the endowment, the final
 allocation and the type partition we build a directed graph on the objects, the \emph{aggregate allocation graph} (Algorithm \ref{alg:graph_construction}). If an individual ends up with an object different from his endowment, individual rationality says that he prefers the new one, and everyone of his type shares the same preferences; so one observed move gives a comparison that binds a whole type, and the graph collects these comparisons. Theorem \ref{PI} says that the allocation is PI-rationalizable if and only if this graph has no cycle. A cycle is a sequence of individuals, each of whom strictly prefers the object that the next one holds, and passing the objects around it makes everyone on it strictly better off, which is what Pareto efficiency forbids. The test is standard cycle detection on a graph with one vertex per object.

 \paragraph{The weak core.}
 The second characterization is the center of the paper. It concerns the
 weak core-stability which we abbreviate to WC: no group of individuals should be able to take back
 their own endowments, redistribute them among themselves, and all end up strictly better off. It is much harder to characterize than PI, because a rejection turns on comparisons the data never reveal: to rule out a group we need to know how a type ranks objects that none of its recorded moves touch. Theorem \ref{Th-WC} handles this with a second graph, a \emph{mixed graph} that keeps track of which type holds which object. Its directed arcs are the comparisons the data reveal; its undirected edges are the within-type comparisons the data leave open. The allocation is WC-rationalizable if and only if the undirected edges can be directed so that the comparisons of each type fit together into a ranking and no group can trade profitably. The latter requirement takes the form of a cycle condition in which revealed preferences and endowment transfers alternate (i.e., an alternating cycle).

 The condition has a drawback, and we state it plainly: it asks whether
 some orientation exists, so testing it means searching, whereas Theorem \ref{PI} only asks us to look. Section \ref{WCsection} establishes what the condition is worth. The search is far smaller than the one the definition calls for --- orientations of the open edges rather than preference profiles, about $10^{6}$ against about $10^{1187}$ in the largest market in our empirical illustration --- and unlike a profile, an orientation can be verified
 quickly. It is often unnecessary, since the arcs the data fix are present
 under every orientation, so a bad cycle among them rejects the instance at no cost (Proposition \ref{prop-fixedarc}); this settles $4{,}591$ of the $4{,}622$ rejections in our empirical illustration. More importantly, the existential clause in our characterizing condition cannot be removed: deciding WC-rationalizability is NP-complete, so unless
 $\mathrm{P}=\mathrm{NP}$ no polynomial-time property of the data
 characterizes the weak core-stability (Theorem \ref{DP-Rationalizability-WC-2},
 Corollary \ref{cor-nopoly}). The search belongs to the problem rather than to our statement of it.

 \paragraph{An application of the weak core test: single-crossing preferences.} The third characterization applies the weak core analysis to a restricted domain. When there is a natural order on households (say, by income or need) and on houses (by quality or size), it is natural to ask for preferences that are \emph{single-crossing}: if some (household-)type prefers the better object of a pair, then so does every higher type. On this domain, a single number summarizes how every type ranks a given pair, the point in the type order at which the ranking switches, and Theorem \ref{Th-SCWC} characterizes single-crossing WC-rationalizability through these crossing points. When the number of objects is the same across the instances an analyst faces, the test runs in polynomial time in the size of the market. In our empirical illustration, the number of objects is not the same across the instances, and yet the program still decides every instance in well under a second.\footnote{In our Supplementary Appendix, we write the characterization as a mixed-integer program, in which the crossing points are the integer variables and every requirement becomes a linear inequality.} 

 \paragraph{Application to data.}
 Finally, we run the three tests on the English social housing lettings
 register (CORE, 2021--22, accessed under a Special License agreement). We treat each local authority as a separate market, which leaves eighteen markets and $2{,}266$ individuals. The reallocation has to be reconstructed from two cross-sections, and the assumption that a type shares identical preferences is strong, so we do not read the results as a verdict on how English social housing is allocated. \rev{Three things emerge. The tests have content, accepting in some markets and rejecting in others, and the two verdicts differ in both directions across markets, which is Proposition \ref{prop-nonnested} appearing in the data. The verdicts respond to how the data are built and grouped, so with data of this kind the analyst should report a distribution of verdicts rather than one. And the theory earns its keep: Proposition \ref{prop-fixedarc} disposes of almost every rejection without search.}

 \paragraph{Organization of the paper.}
 Section \ref{RelatedLit} places the paper in the literature and Section
 \ref{Framework} sets out the environment and the two notions of
 rationalizability. Section \ref{Aggregate} builds the aggregate allocation
 graph and characterizes PI-rationalizability. Section \ref{WCsection} constructs the mixed graph, characterizes
 WC-rationalizability, establishes what the characterization adds beyond the definition, and shows that no polynomial-time test of the weak core can exist. Section \ref{SCsection} treats the single-crossing domain, Section \ref{UKapplication} runs the three tests on the English data, and Section \ref{Conclusion} concludes. Appendix \ref{Appendix} proves Theorems \ref{PI} and \ref{Th-WC}, and Appendix \ref{AppendixLemmas} proves Lemmas \ref{lem-cp-profile} and \ref{lem-weakcore-dag} and Theorem \ref{Th-SCWC}.

 The Supplementary Appendix contains the remaining material. Sections \ref{S1}--\ref{appendix:placebo} concern the data. Section \ref{S-proof-DPWC2} proves Theorem \ref{DP-Rationalizability-WC-2}, and Section \ref{S-mip} gives the mixed-integer formulation of the single-crossing test. Section \ref{S-relations} establishes that PI- and WC-rationalizability are logically independent and that their conjunction is weaker than joint rationalizability (Propositions \ref{prop-nonnested} and \ref{prop-joint}). Section \ref{S-examples} collects the five examples cited in the main text as Examples \ref{example-sc}--\ref{ex-unheld}.

	\section{Related Literature}\label{RelatedLit}

	Matching theory has developed mainly on its normative side: taking preferences as observed, it designs mechanisms that produce allocations with desirable properties. Top Trading Cycles \citep{shapley1974cores} and Deferred Acceptance \citep{gale1962college} now assign school seats, medical residencies, and social housing, and organize kidney exchange \citep{roth2004kidney} and refugee resettlement \citep{Teytelboym}. The positive side has received far less attention, and for object reallocation the basic questions remain open: what can be said about an outcome that has already been produced, and is it PI or WC-stable? We take the outcome as given and ask which preferences, if any, could have produced it. Five literatures bear on that exercise.

 \paragraph{Revealed preference and testable implications.}
 Our question is that of \citet{afriat1967construction}, transposed from consumption to reallocation.  \citet{bachmann2006testable} characterizes, in a finite exchange economy with individual-level data, when an observed allocation is rationalizable as Pareto efficient and individually rational --- the divisible-goods counterpart of Theorem \ref{PI}. Bachmann obtains the characterization by restricting the shape of each agent's preference: utility is strictly concave and strictly monotone. With a finite set of indivisible objects, there is no analogous restriction on preference shape: any linear ordering of $H$ is admissible. Consequently, every reallocation can be rationalized by the argument in Section~\ref{Framework}. \citet{chambers2025empirical} shows that an allocation is efficient for some preference consistent with the data if and only if it is efficient for the incomplete relation implied by the revealed comparisons. Our Theorem \ref{PI} has the same form.\footnote{Our sufficiency arguments in Theorem \ref{PI} extend an acyclic relation to a linear order using \citet{szpilrajn1930extension}; \citet{duggan1999general} gives the general form of such extension theorems.} \rev{\citet{de2023empirical} study the same data from a complementary perspective. They fix an assignment method, such as the jungle or market economy, and characterize the outcomes it can generate. We instead fix the observed outcome and ask which preferences could have generated it.}

 \paragraph{Shapley--Scarf housing markets.}
 Our environment is the housing market of \citet{shapley1974cores}, in which each agent owns one indivisible object. With strict preferences, the strict core is the singleton TTC outcome \citep{roth1977weak}, which is strategy-proof \citep{roth1982incentive} and the unique mechanism that is individually rational, Pareto efficient, and strategy-proof at once \citep{ma1994strategy}; \citet{sonmez1999strategy} links strategy-proofness to essentially single-valued cores and \citet{goel2026ttc} identify the domains on which TTC retains these properties. Adjacent problems include pseudo-market assignment \citep{hylland1979efficient}, the limits of randomized assignment \citep{zhou1990conjecture}, random serial dictatorship \citep{abdulkadirouglu1998random}, the mixed problem of existing tenants and newcomers that our empirical application resembles \citep{abdulkadirouglu1999house}, participation in centralized reassignment \citep{compte2008voluntary}, balanced exchange with multi-object holdings \citep{biro2022balanced}, exchange of school-specific characteristics alongside seats \citep{rodriguez2024school}, and public housing exchange itself \citep{yuan1996residence}. \citet{ekici2024pair} weakens Pareto efficiency to pair-efficiency and asks what a weaker requirement still rules out. We basically assume strict preferences throughout and so do not draw on the literature that relaxes them \citep{Alcalde,Jaramillo,Aziz,saban2013house,Plaxton}. However, that is the natural direction in which our graph conditions would have to be reworked. This literature takes preferences as given and asks which mechanism to run; we reverse the exercise.

 \paragraph{Revealed preference in matching markets.}
 \rev{The question begins with \citet{echenique2008matchings}, who asked which matchings can be stable when preferences are unobserved and gave the first testable implications of matching theory.} \citet{echenique2013revealed} introduced aggregate marriage data and gave graph-theoretic conditions under which they can be rationalized as stable; the type structure we adopt in Section~\ref{Framework} is theirs. \citet{demuynck2022revealed} simplify that characterization and extend it to non-rationalizable data, \citet{zhang2026revealed} treat two-sided markets with single-peaked preferences. \rev{The literature then moves by degrees towards markets in which fewer preferences are observed. \citet{hu2020revealed} take the intermediate step: their market is still two-sided, but only one side has preferences to recover, and they tie the revealed preference analysis to the lattice structure of the stable set. In a one-sided market, preferences are unobserved for all agents, and unlike in two-sided matching, there is no second side whose choices provide additional information about them. That step is taken by \citet{tai2022revealed}, the paper closest to ours and the one to which ours is most indebted. He adopts the same type structure, shows that rationalizability is testable in the Shapley--Scarf environment, and characterizes it for the strict core when utility is non-transferable and for the weak core when monetary transfers are available, the latter through a supporting price vector and an equivalent cyclic monotonicity condition. Our environment is the one without transfers, which is the relevant case for social housing and for public-sector posts, where no price is quoted, and no side payment is made. Neither of the properties we characterize follows from his analysis: Pareto efficiency with individual rationality is not treated there, and the weak core is treated only once transfers supply the structure that a price vector carries. Proposition \ref{prop-tai} locates our tests relative to his. Strict core rationalizability implies both PI- and WC-rationalizability, and the converse fails, so ours are the less demanding tests and their rejections rule out a broader set of possibilities, which provides more informative insights. They are not substitutes for one another either, since neither implies the other (Proposition \ref{prop-nonnested}).}

 \paragraph{Computational complexity of matching and allocation.}
 A substantial literature fixes preferences and asks how hard it is to \emph{compute} an allocation with a given property \citep{abraham2004pareto,fekete2003complexity,saban2015complexity,biro2021complexity,biro2010size,deng1994complexity}; \citet{manlove2013algorithmics} surveys the field. Theorem \ref{DP-Rationalizability-WC-2} asks the same question from the other side: we fix the allocation and ask how hard it is to decide whether \emph{some} preference profile makes it weak core-stable. This difference is what produces the gap between our two results. Checking Pareto efficiency at a known profile is easy \citep{abraham2004pareto}, and so is our test for PI-rationalizability, yet WC-rationalizability is NP-complete. The difficulty does not come from the weak core, which is easy to check once a profile is in hand; it comes from having to search over the profiles the analyst never sees. 

 \paragraph{Structural estimation of matching markets.}
 A parallel empirical literature recovers preferences in matching markets rather than testing them \citep{choo2006marries,galichon2022cupid,fox2018estimating,agarwal2015empirical}, surveyed by \citet{chiappori2016econometrics}. That literature and ours face the same issue and resolve it in opposite ways: structural estimation imposes a functional form and a distribution for the unobserved component and returns preference parameters, whereas we impose neither and return a yes-or-no verdict. The cost of our approach is that it is not quantitative. The gain is that a rejection is a rejection of the solution concept together with the type structure, and not of a parametric specification that the analyst chose.
	
	\section{The Framework} \label{Framework}
	
	Let \( I \) be a finite set of individuals and \( H \) a finite set of objects.  
	Object \( h \in H \) has \( q_{h} \in \mathbb{N}_{+} \) identical copies, with \( q = (q_{h})_{h \in H} \) and \(\sum_{h \in H} q_{h} = |I|\). Each individual \( i \in I \) has a complete, transitive, and antisymmetric preference relation \( \succ_{i} \) over \( H \); let \( \succ = (\succ_{i})_{i \in I} \) denote the preference profile.  
	We write \( h \succ_{i} h' \) if \( i \) strictly prefers \( h \) to \( h' \). \new{Because \( \succ_{i} \) is a linear order, the induced weak relation is simply \( h \succeq_{i} h' \) if and only if \( h \succ_{i} h' \) or \( h = h' \); we use it only where indifference between an object and itself has to be admitted.}
	
	Preferences \( \succ \) are unobserved.  
	Instead, we observe individual characteristics and partition \( I \) into a finite set of types \( T = \{t_{1}, \dots, t_{|T|}\} \) via a type function \( \tau : I \to T \).  
	For \( t \in T \), let \( I^{t} = \{ i \in I : \tau(i) = t \} \).  
	\new{We restrict attention to profiles in which individuals of the same type hold identical preferences: \( \tau(i) = \tau(j) \) implies \( \succ_{i} \, = \, \succ_{j} \). Such a profile assigns a single linear order over \( H \) to each type; we write \( \succ_{t} \) for the order common to the individuals of type \( t \), so that \( \succ_{\tau(i)} \, = \, \succ_{i} \) for every \( i \in I \), and we represent the profile by the type-indexed list \( \succ = (\succ_{t})_{t \in T} \).}  
	This restriction is necessary for testability. Without this assumption, any reallocation can be rationalized by assuming each individual ranks their reallocated object highest.
	
	An allocation is a mapping \( x : I \to H \) satisfying
	\[
	|\{i \in I : x(i) = h\}| = q_{h} \quad \forall h \in H.
	\]
	\new{The initial endowment is a mapping \( \omega : I \to H \) which is
		itself an allocation, so that}
	\[
	\new{|\{i \in I : \omega(i) = h\}| = q_{h} \quad \forall h \in H.}
	\]
	\new{The same objects are therefore present before and after the
		reallocation. We use this throughout: it is what makes a redistribution of
		a coalition's own endowments among its members feasible, and it is what
		allows the grand coalition to redistribute the endowments into any
		allocation whatsoever.}
	In this setting, we observe \( (x, \omega) \) and \( (I, H, q, \tau, T) \), but not \( \succ \).  
	We refer to \( \langle I, H, q, \tau, T, x, \omega \rangle \) as an \emph{aggregate allocation instance}.
	
	\new{We now record the two normative criteria. We characterize their empirical content via two rationalizability notions once $\succ$ is unobserved. The normative criteria are properties of an allocation \emph{given} a profile; the rationalizability questions ask whether \emph{some} admissible profile makes them hold.}

	\begin{definition}[Pareto efficiency (PE)]
		An allocation \( x' \) \emph{Pareto dominates} \( x \) at $\succ$   if
		\[
		x'(i) \succeq_{\tau(i)} x(i) \quad \forall i \in I,
		\]
		and \( x'(j) \succ_{\tau(j)} x(j) \) for some \( j \in I \).  
		An allocation is \emph{Pareto efficient} at $\succ$  if it is not Pareto dominated by any other allocation.
	\end{definition}
	\new{Pareto efficiency asks whether the grand coalition, acting on the whole endowment, has left anything on the table. Individual rationality asks the same question of each individual acting alone.}
	
	\begin{definition}[Individual rationality (IR)]
		An allocation \( x \) is \emph{individually rational} at $\succ$ if
		\[
		x(i) \succeq_{\tau(i)} \, \omega(i) \quad \forall i \in I.
		\]
	\end{definition}
	\new{Neither property can be checked directly, since $\succ$ is not observed. What the analyst can ask is whether the two hold jointly under some profile consistent with the type structure.}
	
	\begin{definition}[PI-rationalizability]
		An allocation \( x \) is \emph{PI-rationalizable} if there exists a
		profile \new{\( \succ=(\succ_{t})_{t\in T} \)} of linear orders over
		\( H \), one for each type, such that \( x \) is both PE and IR when every
		individual \( i \) is endowed with the order \new{\( \succ_{\tau (i)} \)}
		of his type.
	\end{definition}
	\new{All the other coalitions lie between the individual and the grand coalition, and the weak core asks the question of each of them.}
	
	\begin{definition}[weak core-stability]
		A nonempty set $S\subseteq I$ can \emph{strongly block} an allocation $x$ at $\succ$
		if there is an allocation $x'$ such that
		$|\{i\in S : x'(i)=h\}| = |\{i\in S : \omega(i)=h\}|$ for all $h\in H$,
		and every $i\in S$ strictly prefers $x'(i)$ to $x(i)$. An allocation $x$
		is in the \emph{weak core} at $\succ$ if no set of individuals can strongly block it.
	\end{definition}
	
	\begin{definition}[WC-rationalizability]
		An allocation $x$ is \emph{WC-rationalizable} if there exists a
		preference profile $\succ=(\succ_{t})_{t\in T}$, where $\succ_{t}$ is the common
		linear order of type $t$ over $H$, such that $x$ is in the weak core
		for $\succ$.
	\end{definition}
	
	\clr{Every allocation in the weak core is individually rational: an individual who preferred his endowment to his assignment would block on his own. So WC-rationalizability implies that the same profile makes $x$ individually rational.}
	
	\new{We close the section with an instance that will serve as a running example.}
	
	\begin{example}[\new{A running instance}]\label{ex-running}
		\new{Four individuals $i_{1},\dots,i_{4}$ hold two objects, $h_{1}$ and $h_{2}$, each available in two copies, so $q_{h_{1}}=q_{h_{2}}=2$. They form two types, $t_{1}=\{i_{1},i_{2}\}$ and $t_{2}=\{i_{3},i_{4}\}$.}
		
		\begin{center}
			\new{\begin{tabular}{cccc}
					\toprule
					Individual & Type & Endowment $\omega$ & Assignment $x$ \\
					\midrule
					$i_{1}$ & $t_{1}$ & $h_{1}$ & $h_{2}$ \\
					$i_{2}$ & $t_{1}$ & $h_{1}$ & $h_{1}$ \\
					$i_{3}$ & $t_{2}$ & $h_{2}$ & $h_{1}$ \\
					$i_{4}$ & $t_{2}$ & $h_{2}$ & $h_{2}$ \\
					\bottomrule
			\end{tabular}}
		\end{center}
		
		\new{Both individuals of type $t_{1}$ were endowed with $h_{1}$ and one of them moved to $h_{2}$; both individuals of type $t_{2}$ were endowed with $h_{2}$ and one of them moved to $h_{1}$. This table is the whole of the analyst's information: the orders $\succ_{t_{1}}$ and $\succ_{t_{2}}$ that produced the moves are not observed. We return to the instance at the end of Section \ref{Aggregate}.}
	\end{example}

	\section{The Aggregate Allocation Graph} \label{Aggregate}
	
	\clr{In this section, we characterize PI-rationalizability. The characterization rests on one inference, which we make first.}
	
	\clr{Suppose the observed allocation \(x\) is PI, and take an individual \(j\) who ends up with a different object than he started with: \(x(j)\neq \omega(j)\). Individual rationality requires that he prefer what he holds, \(x(j)\succ_{\tau(j)}\omega(j)\). The individuals of his type share his preference, so each of them ranks \(x(j)\) above \(\omega(j)\) as well. A single observed move thus reveals a comparison that binds an entire type.}
	
	\clr{The \emph{aggregate allocation graph} \(AG=(AV,AE)\) collects these comparisons. Its vertices are the objects. It has a directed edge from $h$ to $h'$ when two conditions hold in the data: some individual endowed with $h$ ends up with $h'$, and some individual of his type holds $h$ under $x$. Algorithm~\ref{alg:graph_construction} states the construction.}
	
	\clr{Both conditions matter, for different reasons. The first supplies the comparison. The second supplies someone for whom it currently binds: an individual who \emph{holds} $h$ and would rather have the $h'$ that someone else holds. This is the configuration Pareto efficiency turns on, and it is why the observed moves alone do not determine the graph.}
	
	\begin{algorithm}
		\caption{Construction of \( AG = (AV, AE) \)}
		\label{alg:graph_construction}
		\begin{algorithmic}[1]
			\State \textbf{Vertices:} For each \( h \in H \), create a vertex \( h \) and add it to \( AV \).
			\State \textbf{Edges:} For \( h, h' \in H \), if \new{there exist \( t \in T \) and} \(i,j \in I^{t} \) \new{with} \( \omega(j)=h \neq h'=x(j) \) and \( x(i)=h \), add edge \( (h, h') \) to \( AE \).
		\end{algorithmic}
	\end{algorithm}
	
	We write \( h \to h' \) if there exists \( (h, h') \in AE \). 
	
	\begin{definition}[Directed cycle and acyclicity]\label{def:acyclic}
		\new{Since every edge of \( AG \) joins two distinct objects, \( AG \) has no loops. A \emph{directed cycle} in \( AG \) is a sequence of pairwise distinct objects \( h_{0}, h_{1}, \dots, h_{k-1} \) with \( k \ge 2 \) such that}
		\[
		h_{0} \to h_{1} \to \cdots \to h_{k-1} \to h_{0},
		\]
		\new{each arrow being an edge of \( AE \). Its \emph{length} is \( k \); the shortest directed cycle, of length two, is a pair of opposite edges \( h \to h' \to h \) between two distinct objects. The graph \( AG \) is \emph{acyclic} if it contains no directed cycle.}
	\end{definition}

	\clr{Figure~\ref{allocationgraph} illustrates the construction. In panel (a), individual \(i_{1}\) starts with \(h_{1}\) and receives \(h_{2}\), so individual rationality gives \(h_{2}\succ_{\tau(i_{1})}h_{1}\). A second individual \(i_{2}\) of the same type holds \(h_{1}\) and inherits the comparison, so the graph has an edge from $h_{1}$ to $h_{2}$. Object \(h_{3}\) enters no comparison and stays isolated. In panel (b), the case of Example \ref{ex-running}, the same construction runs in both directions and closes a cycle.}
	
	\begin{figure}[htb]
		\caption{The aggregate allocation graph $AG = (AV, AE)$\new{: (a) a single revealed comparison, and (b) the instance of Example \ref{ex-running}}}\label{allocationgraph}
		\begin{center}
			\begin{tikzpicture}[xscale = .7, yscale = .7, >=stealth]
				\node[circle, draw] (a) at (0,0) {$h_{1}$};
				\node[circle, draw] (b) at (4,0) {$h_{2}$};
				\node[circle, draw] (c) at (2,-2) {$h_{3}$};
				\draw[->] (a) -- (b);
				\node at (2,-3.6) {(a)};
				\node[circle, draw] (d) at (9,0)  {$h_{1}$};
				\node[circle, draw] (e) at (13,0) {$h_{2}$};
				\draw[->, bend left=30] (d) to (e);
				\draw[->, bend left=30] (e) to (d);
				\node at (11,-3.6) {(b)};
			\end{tikzpicture}
		\end{center}
	\end{figure}
	
	\clr{We can now state the characterization. The condition is acyclicity of $AG$, and it can be checked in time polynomial in the size of the data.}
	
	\begin{theorem}\label{PI}
		An aggregate allocation \( x \) is PI-rationalizable if and only if \( AG = (AV, AE) \) is acyclic.
	\end{theorem}
	
	\begin{proof}
		See Appendix \ref{ProofPI}.
	\end{proof}
	
	\paragraph{Sketch of proof.} Necessity is the classical envy-graph argument. Every edge $(h,h')$ of $AG$ records a genuine strict preference: some individual endowed with $h$ moved to $h'$, and by type homogeneity the same-type individual who holds $h$ under $x$ inherits the comparison (Lemma \ref{edgepref}). A cycle in $AG$ would then let a group of individuals pass their assigned objects around the cycle and all strictly improve, contradicting Pareto efficiency; hence $AG$ is acyclic.
	
	Sufficiency is where the work lies. Each type has to rank all of $H$, including objects that no member of the type ever holds, and the rankings must make $x$ Pareto efficient and individually rational at the same time. They cannot therefore be built one type at a time. The construction gives every type a common order. Reversing the edges of $AG$ yields an acyclic relation, which Szpilrajn's theorem extends to a linear order $\succ_{AG}^{+}$ on $H$. Each type then ranks the objects it holds under $x$ above all others, with that top block ordered by the shared $\succ_{AG}^{+}$. Individual rationality is immediate, and the envy graph of the constructed profile sits inside $\succ_{AG}^{+}$, so it is acyclic and $x$ is efficient (Lemma \ref{lem-envy}).

	\clr{The running instance of Example \ref{ex-running} illustrates a rejection.
	Individual $i_{1}$ starts with $h_{1}$ and receives $h_{2}$ while $i_{2}$ of the
	same type holds $h_{1}$, so Algorithm \ref{alg:graph_construction} inserts
	$h_{1}\to h_{2}$; symmetrically $i_{3}$ and $i_{4}$ insert $h_{2}\to h_{1}$. The
	graph is panel (b) of Figure \ref{allocationgraph} and it has a cycle, so $x$ is
	not PI-rationalizable: under every profile consistent with the observed moves,
	$i_{2}$ and $i_{4}$ can swap $h_{1}$ and $h_{2}$ and both do strictly better.}

	\section{WC-rationalizability} \label{WCsection}
	
	Pareto efficiency and individual rationality look at the observed
	allocation as a whole. A different requirement is stability with respect
	to groups: no set of individuals should be able to break away, redistribute
	their own initial endowments among themselves, and all end up strictly
	better off. This is the \emph{weak core}, and it is the subject of this
	section, which is the main part of the paper.

	Section \ref{WCmixed} builds a \emph{mixed graph} that refines the aggregate
	allocation graph by recording which type holds which object, Section
	\ref{WCchar} states the characterization (Theorem \ref{Th-WC}), and Section
	\ref{WCbuys} establishes what the characterization
	adds beyond the definition and ends with a test that costs no search
	(Proposition \ref{prop-fixedarc}). Section \ref{WCNPC} shows that no condition
	as simple as Theorem \ref{PI}'s can exist for the weak core.

	\rev{We begin with an instance that Theorem \ref{PI} accepts and the weak core
	rejects. It is also the instance that the construction of Section \ref{WCmixed}
	is built to explain, and we return to it there.}

	\begin{example}\label{example-wc}
		Let $T=\{t_{1},t_{2}\}$ with
		\begin{equation*}
			I^{t_{1}}=\left\{ 1,2\right\} \text{ and } I^{t_{2}}=\left\{ 3,4\right\}
			\text{,}
		\end{equation*}
		and let there be four objects $h_{1},h_{2},h_{3},h_{4}$, each with a
		single copy ($q_{h_{j}}=1$ for $j=1,2,3,4$). Let the initial allocation
		$\omega$ and the observed allocation $x$ be
		\begin{equation*}
			\omega=\left(
			\begin{tabular}{cccc}
				$1$ & $2$ & $3$ & $4$ \\
				$h_{2}$ & $h_{4}$ & $h_{1}$ & $h_{3}$%
			\end{tabular}%
			\right) \text{ and  }x=\left(
			\begin{tabular}{cccc}
				$1$ & $2$ & $3$ & $4$ \\
				$h_{1}$ & $h_{2}$ & $h_{3}$ & $h_{4}$%
			\end{tabular}%
			\right) \text{.}
		\end{equation*}
		The aggregate allocation graph $AG=(AV,AE)$ of Section \ref{Aggregate}
		has exactly two edges, $h_{2}\rightarrow h_{1}$ (type $t_{1}$: individual
		1 moves from $h_{2}$ to $h_{1}$ while individual 2 holds $h_{2}$) and
		$h_{3}\rightarrow h_{4}$ (type $t_{2}$: individual 4 moves from $h_{3}$
		to $h_{4}$ while individual 3 holds $h_{3}$). The graph is acyclic, so by
		Theorem \ref{PI} the allocation $x$ is PI-rationalizable. A
		rationalizing profile appears below, with objects listed from best
		(top) to worst (bottom).
		\begin{equation*}
			\begin{tabular}{c|c}
				$\succ_{t_{1}}$ & $\succ_{t_{2}}$ \\ \hline
				$h_{1}$ & $h_{4}$ \\
				$h_{2}$ & $h_{3}$ \\
				$h_{3}$ & $h_{2}$ \\
				$h_{4}$ & $h_{1}$%
			\end{tabular}%
		\end{equation*}
	\end{example}
	
	The allocation $x$ of Example \ref{example-wc} is, however, not
	WC-rationalizable. \clr{Suppose that} $\succ$
	WC-rationalizes $x$. Because $x$ is individually rational, type $t_1$
	must strictly prefer $x(1)=h_1$ to $\omega(1)=h_2$, and type $t_2$ must
	strictly prefer $x(4)=h_4$ to $\omega(4)=h_3$. Since individual 2
	receives $h_2$ under $x$ and is of the same type as individual 1,
	individual 2 ranks $h_1$ above $h_2$. Similarly, individual 3 ranks
	$h_4$ above $h_3$. In summary, individual 2's endowment is
	$\omega(2)=h_4$ and he ranks $\omega(3)=h_1$ above $x(2)=h_2$, while
	individual 3's endowment is $\omega(3)=h_1$ and he ranks $\omega(2)=h_4$
	above $x(3)=h_3$. Therefore, individuals 2 and 3 strongly block $x$: by
	exchanging their endowments they both obtain a strictly better outcome, a
	contradiction.

	\subsection{\rev{The mixed graph}}\label{WCmixed}

	\clr{The blocking logic behind Example~\ref{example-wc}} is captured by a graph that refines the
	aggregate allocation graph by keeping track of \emph{which type holds
		which object}.We define $x_{t,h}=|\{i\in I^{t} : x(i)=h\}|$ that denotes
	the number of individuals of type $t$ assigned object $h$ under $x$.
	Given $x$ and $\omega$, the \emph{mixed graph}
	$\mathcal{G}=(\mathcal{V},\mathcal{E},\mathcal{A})$ has three building
	blocks: vertices, directed arcs, and undirected edges. To make the
	construction transparent, we introduce each building block formally and
	immediately illustrate it on the following running example.
	
	\begin{example}\label{example-wc-yes}
		There are two types, $t_{1}$ and $t_{2}$, and four objects: $h_{1}$ with
		two copies ($q_{h_{1}}=2$) and $h_{2},h_{3},h_{4}$ with one copy each.
		The five individuals, their types, their endowments, and their assigned
		objects are as follows.
		\begin{center}
			\begin{tabular}{ccccl}
				\toprule
				Individual & Type & Endowment $\omega(i)$ & Assignment $x(i)$ & \\
				\midrule
				$1$ & $t_{1}$ & $h_{1}$ & $h_{2}$ & mover \\
				$2$ & $t_{1}$ & $h_{1}$ & $h_{1}$ & stayer \\
				$5$ & $t_{1}$ & $h_{4}$ & $h_{3}$ & mover \\
				$3$ & $t_{2}$ & $h_{3}$ & $h_{4}$ & mover \\
				$4$ & $t_{2}$ & $h_{2}$ & $h_{1}$ & mover \\
				\bottomrule
			\end{tabular}
		\end{center}
	\end{example}
	
	\emph{Vertices.} The vertices are the occupied type--object pairs:
	\begin{equation*}
		\mathcal{V}=\left\{ (t,h)\in T\times H \;|\; x_{t,h}>0\right\} \text{.}
	\end{equation*}
	In Example \ref{example-wc-yes}, type $t_{1}$ obtains $h_{1}$ (individual
	2), $h_{2}$ (individual 1) and $h_{3}$ (individual 5), while type
	$t_{2}$ obtains $h_{1}$ (individual 4) and $h_{4}$ (individual 3). Hence
	$\mathcal{V}=\{(t_{1},h_{1}),(t_{1},h_{2}),(t_{1},h_{3}),
	(t_{2},h_{1}),(t_{2},h_{4})\}$.
	
	\emph{Intra-arcs.} The set of arcs is
	$\mathcal{A}=T(\mathcal{A}_{1})\cup \mathcal{A}_{2}$. The first family,
	\begin{equation*}
		\mathcal{A}_{1}=\left\{ \left( (t,h),(t,h')\right) \in \mathcal{V}^{2}
		\;|\; \omega(i)=h \text{ and } x(i)=h' \text{ for some } i\in I^{t},\;
		h\neq h'\right\} \text{,}
	\end{equation*}
	collects the revealed within-type comparisons: a voluntary move from $h$
	to $h'$ reveals, by individual rationality, that the type strictly
	prefers $h'$ to $h$. $T(\mathcal{A}_{1})$ is the transitive closure of
	$\mathcal{A}_{1}$ taken within each type, and we call its elements
	\emph{intra-arcs} (drawn in black). In Example \ref{example-wc-yes}
	there is exactly one intra-arc. Individual 1 gave up $h_{1}$ for
	$h_{2}$. This reveals that type $t_{1}$ prefers $h_{2}$ to $h_{1}$.
	Individual 2, of the same type, still holds $h_{1}$, so both endpoints
	$(t_{1},h_{1})$ and $(t_{1},h_{2})$ are vertices. This gives the black
	arc from $(t_{1},h_{1})$ to $(t_{1},h_{2})$. The other three moves also reveal
	preferences: individual 5 reveals $h_{3}$ over $h_{4}$ for $t_{1}$,
	individual 3 reveals $h_{4}$ over $h_{3}$ for $t_{2}$, and individual 4
	reveals $h_{1}$ over $h_{2}$ for $t_{2}$. None of them produces an
	intra-arc. In each case the object given up is not held by anyone of the
	mover's own type, so the pairs $(t_{1},h_{4})$, $(t_{2},h_{3})$ and
	$(t_{2},h_{2})$ are not vertices.
	
	\emph{Inter-arcs.} The second family of arcs,
	\begin{equation*}
		\mathcal{A}_{2}=\left\{ \left( (t,h),(t',h')\right) \in \mathcal{V}^{2}
		\;|\; t\neq t',\; \omega(i)=h \text{ and } x(i)=h' \text{ for some }
		i\in I^{t'}\right\} \text{,}
	\end{equation*}
	records the endowment transfers across types: some individual of type
	$t'$ initially held (a copy of) $h$, an object that type $t$ holds under
	$x$, and receives $h'$. These arcs are called \emph{inter-arcs} \new{(drawn in blue and dashed).} In Example \ref{example-wc-yes} there are five. Individual 1
	(type $t_{1}$) was endowed with $h_{1}$ which type $t_{2}$ holds and holds $h_2$: blue
	arc from $(t_{2},h_{1})$ to $(t_{1},h_{2})$. Individual 2
	(type $t_{1}$) was endowed with $h_{1}$ which type $t_{2}$ holds and holds $h_1$: blue
	arc from $(t_{2},h_{1})$ to $(t_{1},h_{1})$. Individual 3 (type $t_{2}$) was
	endowed with $h_{3}$ which type $t_{1}$ holds and holds $h_4$: blue arc from
	$(t_{1},h_{3})$ to $(t_{2},h_{4})$. Individual 4 (type $t_{2}$)
	was endowed with $h_{2}$ which type $t_{1}$ holds and holds $h_1$: blue arc from
	$(t_{1},h_{2})$ to $(t_{2},h_{1})$. Individual 5 (type $t_{1}$) was
	endowed with $h_{4}$ which type $t_{2}$ holds and holds $h_3$: blue arc from
	$(t_{2},h_{4})$ to $(t_{1},h_{3})$.
	
	\emph{Undirected edges.} Finally, the undirected edges connect same-type
	vertices for which the data reveal no comparison:
	\begin{equation*}
		\mathcal{E}=\left\{ \left\{ (t,h),(t,h')\right\} \subseteq \mathcal{V}
		\;\middle|\;
		\begin{array}{l}
			\left( (t,h),(t,h')\right) \notin T(\mathcal{A}_{1})\text{ and} \\
			\left( (t,h'),(t,h)\right) \notin T(\mathcal{A}_{1}),\; h\neq h'
		\end{array}
		\right\} \text{.}
	\end{equation*}
	\new{We draw these edges in red, dotted.} In Example
	\ref{example-wc-yes}, the only
	intra-arc compares $h_{1}$ with $h_{2}$ for type $t_{1}$, so $h_{3}$ is
	uncompared with both: red edges $\{(t_{1},h_{3}),(t_{1},h_{1})\}$ and
	$\{(t_{1},h_{3}),(t_{1},h_{2})\}$. For type $t_{2}$ there is no
	intra-arc at all, so its two held objects are uncompared: red edge
	$\{(t_{2},h_{1}),(t_{2},h_{4})\}$. Figure \ref{fig-wc-yes} displays the complete
	mixed graph of the example.
	
	\begin{figure}[htb]
		\begin{center}
			\begin{tikzpicture}[
				xscale=1.6,
				yscale=1.4,
				midarrow/.style={
					postaction={decorate},
					decoration={
						markings,
						mark=at position 0.5 with {\arrow[scale=1.2]{>}}
					}
				},
				bluearrow/.style={
					blue, dashed,
					postaction={decorate},
					decoration={
						markings,
						mark=at position 0.5 with {\arrow{>}}
					}
				}
				]
				\foreach \i in {1,2}
				\node at (0,-\i) {$t_\i$};
				\foreach \j in {1,2,3,4}
				\node at (\j,0) {$h_\j$};
				\foreach \x/\y in {1/1, 1/2, 1/3, 2/1, 2/4}
				\node[draw,circle,fill,inner sep=2pt] (\x\y) at (\y,-\x) {};
				
				\draw[midarrow] (11) -- (12);
                \draw[bluearrow] (21) to [bend left=20] (11);
				\draw[bluearrow] (21) to [bend right=20] (12);
				\draw[bluearrow] (12) to [bend right=20] (21);
				\draw[bluearrow] (24) to [bend right=20] (13);
				\draw[bluearrow] (13) to [bend right=20] (24);
				\draw[red, dotted, very thick] (13) to [bend right=35] (11);
				\draw[red, dotted, very thick] (13) -- (12);
				\draw[red, dotted, very thick] (21) -- (24);
			\end{tikzpicture}
		\end{center}
		\caption{The mixed graph of Example \ref{example-wc-yes}: intra-arc in
			black \new{(solid)}, inter-arcs in blue \new{(dashed)}, undirected edges in red
			\new{(dotted)}}\label{fig-wc-yes}
	\end{figure}
	
	\emph{Sub-mixed graphs and orientations.} For each type $t$, the
	\emph{type-$t$ sub-mixed graph} is the restriction of $\mathcal{G}$ to
	the vertices $\{(t,h)\in \mathcal{V}\}$ together with their intra-arcs
	and edges. An \emph{orientation} for $\mathcal{G}$ is a mapping $\gamma
	:\mathcal{E}\rightarrow \mathcal{V}^{2}$ that assigns to each undirected
	edge $\{v,v'\}$ one of its two directions, $(v,v')$ or $(v',v)$. If
	$\mathcal{E}$ is empty, $\gamma$ is the empty function. Orienting every
	edge turns the mixed graph into a digraph, denoted
	$\mathcal{G}_{\gamma}$, whose type-$t$ sub-digraph is denoted
	$\mathcal{G}_{\gamma}^{t}$. An orientation completes the revealed
	within-type comparisons into hypothesized ones: the analyst must decide,
	for every unrevealed same-type pair, which object the type prefers. In
	Example \ref{example-wc-yes}, an orientation must direct the three red
	edges of Figure \ref{fig-wc-yes}. We return to this choice after stating
	the characterization.
	
	\begin{definition}[Directed cycle in the oriented mixed graph]\label{def:cycle-mixed}
		\new{Fix an orientation \( \gamma \). The digraph \( \mathcal{G}_{\gamma} \) has vertex set \( \mathcal{V} \) and arc set \( \mathcal{A} \cup \gamma(\mathcal{E}) \), where \( \gamma(\mathcal{E}) = \{ \gamma(e) : e \in \mathcal{E} \} \) are the oriented edges. A \emph{directed cycle} in \( \mathcal{G}_{\gamma} \) is a sequence \( \langle v_{1}, v_{2}, \dots, v_{m} \rangle \) of pairwise distinct vertices of \( \mathcal{V} \), with \( m \ge 2 \), such that \( (v_{j}, v_{j+1}) \) is an arc of \( \mathcal{G}_{\gamma} \) for every \( j \), indices read modulo \( m \) so that \( (v_{m}, v_{1}) \) is also an arc. A sub-digraph of \( \mathcal{G}_{\gamma} \), in particular each type-\( t \) sub-digraph \( \mathcal{G}_{\gamma}^{t} \), is \emph{acyclic} if it contains no directed cycle.}
	\end{definition}

	\emph{Alternating cycles.} The last ingredient identifies, inside the
	oriented graph, the coalitions that could profitably trade.
	
	\begin{definition}[Alternating cycle]
		Let $\mathcal{G}=(\mathcal{V}, \mathcal{E},\mathcal{A})$ be the mixed
		graph associated with $x$ and $\omega$, and let $\gamma$ be an
		orientation for $\mathcal{G}$. \new{For this test, each oriented edge of $\mathcal{G}_{\gamma}$ counts as an intra-arc: like an intra-arc, it joins two vertices of the same type and encodes a (here hypothesized) within-type preference.} A \new{directed }cycle $\langle v_1,v_2,\dots,v_m
		\rangle$ in the digraph $\mathcal{G}_{\gamma}$ is \emph{alternating} if
		it contains neither two consecutive intra-arcs nor two consecutive
		inter-arcs.
	\end{definition}

	An alternating cycle traces a blocking coalition: each intra-arc
	certifies that a type strictly prefers the next object on the cycle, and
	each inter-arc hands over the endowment that makes the trade feasible.
	Example \ref{example-wc} illustrates this. Figure \ref{fig-wc-mixed}
	presents its aggregate allocation $X(x)$ (the matrix whose cell $(t,h)$
	counts the individuals of type $t$ assigned to object $h$) on the left
	and its mixed graph on the right. The edge set $\mathcal{E}$ is empty.
	
	\begin{figure}[htb]
		\begin{center}
			\begin{minipage}[b]{0.45\textwidth}
				\centering
				\begin{tikzpicture}[xscale=1,yscale=1]
					\foreach \i in {1,2}
					\node at (0,-\i) {$t_\i$} ;
					\foreach \j in {1,2,3,4}
					\node at (\j,0) {$h_\j$};
					\foreach \x/\y/\val in {1/1/1, 1/2/1, 1/3/0, 1/4/0, 2/1/0, 2/2/0, 2/3/1, 2/4/1}
					\node (\x\y) at (\y, -\x) {\val};
				\end{tikzpicture}
			\end{minipage}
			\hfill
			\begin{minipage}[b]{0.52\textwidth}
				\centering
				\begin{tikzpicture}[
					xscale=1,
					yscale=1,
					midarrow/.style={
						postaction={decorate},
						decoration={
							markings,
							mark=at position 0.5 with {\arrow[scale=1.2]{>}}
						}
					},
					bluearrow/.style={
						blue, dashed,
						postaction={decorate},
						decoration={
							markings,
							mark=at position 0.5 with {\arrow{>}}
						}
					}
					]
					\foreach \i in {1,2}
					\node at (0,-\i) {$t_\i$};
					\foreach \j in {1,2,3,4}
					\node at (\j,0) {$h_\j$};
					\foreach \x/\y/\val in {1/1/1, 1/2/1, 2/3/1, 2/4/1}
					\node[draw,circle,fill,inner sep=2pt] (\x\y) at (\y,-\x) {};
					
					\draw[midarrow] (12) -- (11);
					\draw[midarrow] (23) -- (24);
					
					\draw[bluearrow] (11) to [bend right=25] (23);
					\draw[bluearrow] (24) to [bend right=25] (12);
				\end{tikzpicture}
			\end{minipage}
		\end{center}
		\caption{Aggregate allocation $X(x)$ and mixed graph $\mathcal{G}=(\mathcal{V},\mathcal{E},\mathcal{A})$ of Example \ref{example-wc}}\label{fig-wc-mixed}
	\end{figure}
	
	The blocking by individuals 2 and 3 is represented in the mixed graph by
	the cycle $\langle (t_1,h_2), (t_1,h_1), (t_2,h_3), (t_2,h_4),
	(t_1,h_2)\rangle$. The intra-arc from $(t_1,h_2)$ to $(t_1,h_1)$
	certifies that type $t_1$ strictly prefers $h_1$ to $h_2$; the inter-arc
	from $(t_1,h_1)$ to $(t_2,h_3)$ records that individual 3, of type
	$t_2$, initially held $h_1$ and eventually received $h_3$; the intra-arc from $(t_2,h_3)$ to
	$(t_2,h_4)$ certifies that type $t_2$ strictly prefers $h_4$ to $h_3$;
	and the inter-arc from $(t_2,h_4)$ to $(t_1,h_2)$ records that
	individual 2, of type $t_1$, initially held $h_4$ and had $h_2$ in the final allocation. Black and blue arcs
	alternate throughout the cycle: it traces a coalition whose members can
	cyclically redistribute their endowments so that every participant
	strictly improves.

	\subsection{\rev{The characterization}}\label{WCchar}

	WC-rationalizability turns out to be exactly the existence of a
	well-behaved orientation.
	
	\begin{condition}\label{cond1}
		An allocation $x$ satisfies Condition \ref{cond1} provided that for its
		associated mixed graph $\mathcal{G}=(\mathcal{V},\mathcal{E},\mathcal{A})$
		there exists an orientation $\gamma:\mathcal{E}\rightarrow\mathcal{V}^{2}$
		such that:
		\begin{itemize}
			\item[i)] for all $t\in T$, the type-$t$ sub-digraph
			$\mathcal{G}_{\gamma}^{t}$ is acyclic;
			\item[ii)] $\mathcal{G}_{\gamma}$ does not contain any alternating
			cycles.
		\end{itemize}
	\end{condition}
	
	Part i) requires that the completed within-type comparisons form a
	well-defined ranking for every type. Part ii) rules out every coalition
	that could cyclically redistribute endowments with all members strictly
	improving. In Example \ref{example-wc}, the edge set $\mathcal{E}$ is
	empty, so the only candidate orientation is the empty one, and the
	alternating cycle displayed above shows that Condition \ref{cond1}
	fails --- in line with the fact that $x$ is not WC-rationalizable.
	
	\begin{theorem}
		\label{Th-WC}
		An allocation $x$ is WC-rationalizable if and only if $x$ satisfies
		Condition \ref{cond1}.
	\end{theorem}
	
	\begin{proof}
		See Appendix \ref{appendix B}.
	\end{proof}

	\paragraph{Sketch of proof.} Necessity is direct: given a rationalizing profile,
	orient each open edge the way that profile ranks its endpoints. Part i) then holds
	because a linear order has no cycles, and an alternating cycle would exhibit a
	coalition that redistributes its endowments with every member strictly improving,
	contradicting the weak core. Sufficiency is where the work lies, because an
	orientation ranks only the objects a type \emph{holds}, while a profile must rank
	all of $H$. Part i) makes each type's arcs acyclic, so Szpilrajn's theorem extends
	them to a linear order. The objects a type does not hold are then inserted below
	the held ones in a common order, which cannot create a block because no member of
	the type can be made to want them. The absence of alternating cycles is finally
	translated into the absence of a strongly blocking coalition through Lemma
	\ref{lem-weakcore-dag}, which recasts the weak core as acyclicity of a digraph on
	individuals rather than a condition on $2^{|I|}$ coalitions.

\bigskip
    
	Returning to Example \ref{example-wc-yes}, an orientation must direct the three
	red edges of Figure \ref{fig-wc-yes}, and the choices are not independent.
	Direct $\{(t_{2},h_{1}),(t_{2},h_{4})\}$ so that $t_{2}$ prefers $h_{4}$ to
	$h_{1}$. This forces $\{(t_{1},h_{3}),(t_{1},h_{2})\}$: the opposite direction
	would close the alternating cycle $(t_{1},h_{3})\rightarrow(t_{1},h_{2})
	\rightarrow(t_{2},h_{1})\rightarrow(t_{2},h_{4})\rightarrow(t_{1},h_{3})$,
	which is the blocking coalition in which individuals $5$ and $4$ exchange their
	endowments $h_{4}$ and $h_{2}$ and both improve. So $t_{1}$ prefers $h_{3}$ to
	$h_{2}$, and this in turn forces $\{(t_{1},h_{1}),(t_{1},h_{3})\}$, since the
	opposite direction would close a directed cycle inside the type-$t_{1}$
	sub-digraph and no transitive preference could rank the three objects. The
	resulting digraph, in Figure \ref{fig-wc-yes-oriented}, satisfies both parts of
	Condition \ref{cond1}: the sub-digraphs order $h_{3}\succ h_{2}\succ h_{1}$ and
	$h_{4}\succ h_{1}$, and every remaining cycle runs through two consecutive
	inter-arcs and so is not alternating. By Theorem \ref{Th-WC} the allocation is
	WC-rationalizable, one rationalizing profile being
	$h_{3}\succ_{t_{1}}h_{2}\succ_{t_{1}}h_{1}\succ_{t_{1}}h_{4}$ and
	$h_{4}\succ_{t_{2}}h_{1}\succ_{t_{2}}h_{2}\succ_{t_{2}}h_{3}$.

	\begin{figure}[htb]
		\begin{center}
			\begin{tikzpicture}[
				xscale=1.6,
				yscale=1.4,
				midarrow/.style={
					postaction={decorate},
					decoration={
						markings,
						mark=at position 0.5 with {\arrow[scale=1.2]{>}}
					}
				},
				bluearrow/.style={
					blue, dashed,
					postaction={decorate},
					decoration={
						markings,
						mark=at position 0.5 with {\arrow{>}}
					}
				},
				]
				\foreach \i in {1,2}
				\node at (0,-\i) {$t_\i$};
				\foreach \j in {1,2,3,4}
				\node at (\j,0) {$h_\j$};
				\foreach \x/\y in {1/1, 1/2, 1/3, 2/1, 2/4}
				\node[draw,circle,fill,inner sep=2pt] (\x\y) at (\y,-\x) {};
				
				\draw[midarrow] (11) -- (12);
                \draw[bluearrow] (21) to [bend left=20] (11);
				\draw[bluearrow] (21) to [bend right=20] (12);
				\draw[bluearrow] (12) to [bend right=20] (21);
				\draw[bluearrow] (24) to [bend right=20] (13);
				\draw[bluearrow] (13) to [bend right=20] (24);
				\draw[midarrow] (11) to [bend left=35] (13);
				\draw[midarrow] (12) -- (13);
				\draw[midarrow] (21) -- (24);
			\end{tikzpicture}
		\end{center}
		\caption{The digraph $\mathcal{G}_{\gamma}$ of Example
			\ref{example-wc-yes}: the three red edges are now directed by the
			orientation $\gamma$ constructed in Steps 1--3 and they became black.}\label{fig-wc-yes-oriented}
	\end{figure}

	\rev{Before discussing the implications of the characterization, we compare WC-rationalizability with the other requirements. Two of the three
	answers are in Section \ref{S-relations} of the Supplementary Appendix. PI- and
	WC-rationalizability are logically independent, so neither test stands in for the
	other (Proposition \ref{prop-nonnested}): Example \ref{example-wc} is PI-rationalizable and not
	WC-rationalizable, and Example \ref{example-wc-not-pi} runs the other way. The reason is that the two
	notions discipline different trades --- Pareto efficiency restricts exchanges of
	\emph{assigned} objects, the weak core exchanges of \emph{endowments} --- and that
	weak core allocations may fail Pareto efficiency is known since
	\citet{roth1977weak}. Passing both of our tests is also weaker than it looks:
	nothing requires the profile that makes $x$ efficient to be the profile that makes
	$x$ weak core-stable, and an instance can be PI-rationalizable and
	WC-rationalizable without being \emph{jointly} rationalizable by one profile
	(Proposition \ref{prop-joint}). Running both tests and finding that each accepts therefore does
	not establish that the reallocation is consistent with a market that is at once
	efficient and weak core-stable. In Section \ref{UKapplication} the fraction of
	jointly rationalizable instances is bounded above by the smaller of the two
	reported frequencies and may lie strictly below it. We do not characterize joint
	rationalizability, and regard it as the natural next question.}

	\new{\paragraph{Relation to the strict core}
		\citet{tai2022revealed} tests the same data against the \emph{strict}
		core. Our two properties are deliberately less demanding than his, and it is
		worth recording where they sit relative to it, since that ordering is what
		makes our rejections informative. Say that a nonempty set
		$S\subseteq I$ \emph{weakly blocks} an allocation $x$ at $\succ$ if there is an allocation
		$x^{\prime }$ such that $\left\vert \left\{ i\in S:x^{\prime }(i)=h\right\}
		\right\vert =\left\vert \left\{ i\in S:\omega(i)=h\right\} \right\vert $ for all
		$h\in H$, $x^{\prime }(i)\succeq_{\tau (i)}x(i)$ for all $i\in S$, and
		$x^{\prime }(j)\succ_{\tau (j)}x(j)$ for at least one $j\in S$. An allocation is
		in the \emph{strict core} at $\succ$ if no nonempty set of individuals weakly blocks it,
		and it is \emph{SC-rationalizable} if there is a profile $\succ$ for
		which that allocation is a strict core allocation at $\succ$.}
	
	\begin{proposition}\label{prop-tai}
		\new{
			\begin{itemize}
				\item[i)] If $x$ is SC-rationalizable, then $x$ is both PI-rationalizable and
				WC-rationalizable, and a single profile witnesses all three properties.
				\item[ii)] The converse fails: there is an aggregate allocation instance that
				is both PI-rationalizable and WC-rationalizable but not SC-rationalizable.
		\end{itemize}}
	\end{proposition}
	
	\begin{proof}
		\new{Part i). Let $\succ$ be a profile for which $x$ lies in the
			strict core at $\succ$. For individual rationality, suppose $\omega(i)\succ_{\tau
				(i)}x(i)$ for some $i\in I$; then $S=\left\{ i\right\} $ with $x^{\prime
			}(i)=\omega(i)$ weakly blocks $x$, a contradiction. For Pareto efficiency,
			suppose $x^{\prime }$ Pareto dominates $x$ and take $S=I$. Since $x^{\prime }$
			and $\omega$ are both allocations, $\left\vert \left\{ i\in I:x^{\prime
			}(i)=h\right\} \right\vert =q_{h}=\left\vert \left\{ i\in I:\omega(i)=h\right\}
			\right\vert $ for every $h\in H$, so the reallocation is feasible for $S$;
			together with $x^{\prime }(i)\succeq_{\tau (i)}x(i)$ for all $i$ and
			$x^{\prime }(j)\succ_{\tau (j)}x(j)$ for some $j$, this makes $I$ a weakly
			blocking coalition, a contradiction. For the weak core, every strong block is a
			weak block, so no coalition strongly blocks $x$. Hence the single profile
			$\succ$ makes $x$ Pareto efficient, individually rational and weak core-stable.
			Part ii) is Example \ref{example-sc} of the Supplementary Appendix.}
	\end{proof}
	
	\new{Part ii) is Example \ref{example-sc} of the Supplementary Appendix, and the reason the
		converse fails is worth naming. A coalition weakly blocks when one member gains
		and the others merely do not lose, and a member loses nothing when the trade
		hands him the object he was going to hold anyway. The strict core rules out
		deviations of this kind. The weak core does not, since it requires every member
		to gain strictly. That is the right requirement when the planner wants the
		outcome to survive regrouping that costs its participants nothing. The weak core
		answers a different question, namely whether some group would have reason to
		organize a trade, and it is that question we take up here.}

	\subsection{\texorpdfstring{\new{What the characterization adds}}{What the characterization adds}}\label{WCbuys}

		Both the definition of WC-rationalizability and Theorem \ref{Th-WC} begin
		with ``there exists'', so it is fair to ask what the theorem has bought.
		We stress three things.

	\emph{First, the search becomes much smaller.} The definition searches
	over profiles, of which there are $\left( \left\vert H\right\vert
	!\right) ^{\left\vert T\right\vert }$. Condition \ref{cond1} searches over
	orientations, of which there are $2^{\left\vert \mathcal{E}\right\vert }$.
	The reduction is large. The search contains only
	comparisons between objects that a type actually \emph{holds} under $x$,
	and among those, only the ones that the type's own recorded moves leave
	open. Everything the data reveal is already fixed as an arc, and it is
	never searched over. Three markets from our empirical application show the size of
	the difference.
	
	\begin{center}
		\begin{tabular}{l rr rr}
			\toprule
			Market & $\left\vert H\right\vert$ & $\left\vert T\right\vert$ &
			profiles & orientations \\
			\midrule
			Leeds & $15$ & $98$ & $10^{1187}$ & $10^{6}$ \\
			Birmingham & $14$ & $89$ & $10^{974}$ & $10^{5}$ \\
			Lambeth & $6$ & $6$ & $10^{17}$ & $10^{2}$ \\
			\bottomrule
		\end{tabular}
	\end{center}
	
	\noindent Across all the instances we test in our empirical illustration, $\left\vert \mathcal{E}
	\right\vert$ has median $7$ and never exceeds $53$. Both counts are
	exponential in the worst case, and Theorem \ref{DP-Rationalizability-WC-2}
	says that they must be. But what is being searched over has changed. It is
	no longer the preferences of every type over every object. It is the
	handful of within-type comparisons that the data leave open.
	
	\emph{Second, the condition supplies a certificate.} Suppose someone hands
	us a profile $\succ$ and claims that it rationalizes $x$. To check the
	claim from the definition we would have to rule out every blocking
	coalition, and there are $2^{\left\vert I\right\vert }$ of them. A profile
	is therefore not, by itself, evidence that can be checked quickly. An
	orientation is. Condition \ref{cond1} can be checked easily whenever an orientation is given. 

	\emph{Third, the condition separates the two ways rationalization can
		fail.} The definition returns one verdict and says nothing about its
	source. Condition \ref{cond1} distinguishes a type that cannot rank its
	own objects consistently, which is part i), from a group that can trade
	profitably, which is part ii). The next proposition uses this distinction and turns it into a simple test whenever the data
	are informative enough. 
	
	\begin{proposition}
		\label{prop-fixedarc}
		Let $\mathcal{G}_{\mathcal{A}}=(\mathcal{V},\mathcal{A})$ denote the
		\emph{fixed sub-digraph} of $\mathcal{G}$, obtained by deleting the
		undirected edges.
		\begin{itemize}
			\item[i)] If some type-$t$ restriction of $\mathcal{G}_{\mathcal{A}}$
			contains a cycle, or if $\mathcal{G}_{\mathcal{A}}$ contains an
			alternating cycle, then $x$ is not WC-rationalizable. Both conditions
			are decidable in time polynomial in the size of $\mathcal{G}$.
			\item[ii)] If moreover $\mathcal{E}=\varnothing $, the converse holds as
			well. Hence on instances with $\mathcal{E}=\varnothing $,
			WC-rationalizability is decidable in polynomial time.
		\end{itemize}
	\end{proposition}
	
	\begin{proof}
		i) Orienting an edge adds an arc and removes none, so
		$\mathcal{G}_{\mathcal{A}}$ is a subgraph of $\mathcal{G}_{\gamma }$ for
		every orientation $\gamma $, on the same vertex set and with the labels
		of its arcs unchanged. A cycle of $\mathcal{G}_{\mathcal{A}}$ within
		row $t$ is therefore a cycle of $\mathcal{G}_{\gamma }^{t}$, violating
		part i) of Condition \ref{cond1}; and an alternating cycle of
		$\mathcal{G}_{\mathcal{A}}$ is still an alternating cycle of
		$\mathcal{G}_{\gamma }$, violating part ii). No orientation satisfies
		Condition \ref{cond1}, and Theorem \ref{Th-WC} gives the claim. For
		decidability, row cycles are found by cycle detection in each row. For
		alternating cycles, split every vertex $v$ into two copies,
		$v^{\text{intra}}$ and $v^{\text{inter}}$, and reroute each intra-arc
		$(u,v)$ to $(u^{\text{inter}},v^{\text{intra}})$ and each inter-arc
		$(u,v)$ to $(u^{\text{intra}},v^{\text{inter}})$. A cycle of the split
		digraph is precisely an alternating cycle of $\mathcal{G}_{\mathcal{A}}$.
		
		ii) If $\mathcal{E}=\varnothing $ the empty function is the only
		orientation and $\mathcal{G}_{\gamma }=\mathcal{G}_{\mathcal{A}}$, so
		Condition \ref{cond1} holds if and only if neither cycle occurs.
	\end{proof}
	
	Part i) is a one-sided test. It can reject but it can never accept, and it
	costs no search. It is nevertheless what settles most rejections in
	practice: in our empirical application it certifies $4{,}591$ of the $4{,}622$
	rejected instances. Part ii) says when the same computation decides the
	question outright. The instances it covers are those in which the recorded
	moves of every type already order all of the objects that the type holds,
	so that nothing is left for the analyst to complete.

	Two lessons follow. Even though Theorem \ref{Th-WC} asks for a search, in
	practice the search is rarely the thing that does the work, because the
	arcs the data fix are usually enough on their own. And the difficulty is
	confined to a part of the problem we can point to, namely the comparisons
	the data leave open. This is the first indication that the search in
	Condition \ref{cond1} is well targeted, and the next subsection shows that
	it cannot be avoided in general.

	\subsection{\texorpdfstring{\new{Why no simpler test exists}}{Why no simpler test exists}}\label{WCNPC}

	The test of Theorem \ref{Th-WC} is not of the form Theorem \ref{PI} takes. It
	contains an existential clause, and the number of candidate orientations grows
	exponentially. One might suspect that this is an artifact of how we state the
	result, and that a directly checkable condition is waiting to be found. It is
	not. Deciding WC-rationalizability is NP-complete, so unless
	$\mathrm{P}=\mathrm{NP}$ no property of the data checkable in polynomial time
	characterizes it (Corollary \ref{cor-nopoly}).\footnote{For an introduction to
		NP-completeness see \citet*{gareyjohnson1979}. In short: P collects the
		decision problems solvable in polynomial time; NP those whose yes instances
		can be \emph{verified} in polynomial time; a problem is NP-complete if it lies
		in NP and every problem in NP reduces to it in polynomial time.} We prove it by
	reduction from the maximum independent set problem in Section \ref{S-proof-DPWC2} of the
	Supplementary Appendix. The result is not one of the paper's three
	characterizations and we claim no contribution to the theory of computation.
	Its purpose is to justify Theorem \ref{Th-WC}: the search belongs to the
	problem, so any correct characterization of the weak core must contain
	something of the same kind.

	\begin{decision}[DP-Rationalizability-WC]
		\textit{Instance}: An aggregate allocation instance
		$\langle I, H, q, \tau, T, x, \omega \rangle$.
		
		\smallskip
		\noindent\textit{Question}: Is $x$ WC-rationalizable, i.e.\ does there
		exist a preference profile $\succ=(\succ_{t})_{t\in T}$, where each $\succ_{t}$ is a
		linear order over $H$, such that $x$ is a weak core-stable allocation at $\succ$?
	\end{decision}
	
	\begin{theorem}\label{DP-Rationalizability-WC-2}
		The decision problem DP-Rationalizability-WC is NP-complete.
	\end{theorem}
	
	\begin{proof}
		See Section \ref{S-proof-DPWC2} of the Supplementary Appendix.
	\end{proof}
	
	The following consequence is the precise sense in which the weak core
	admits no test of the kind Theorem \ref{PI} provides for PI-rationalizability.
	
	\begin{corollary}\label{cor-nopoly}
		Unless $\mathrm{P}=\mathrm{NP}$, there is no property $\Phi$ of aggregate
		allocation instances such that (i) whether $\Phi$ holds at
		$\langle I,H,q,\tau ,T,x,\omega \rangle$ can be decided in time polynomial
		in the size of the instance, and (ii) $x$ is WC-rationalizable if and only
		if $\Phi$ holds.
	\end{corollary}
	
	\begin{proof}
		Such a $\Phi$ would decide DP-Rationalizability-WC in polynomial time,
		which by Theorem \ref{DP-Rationalizability-WC-2} is impossible unless
		$\mathrm{P}=\mathrm{NP}$.
	\end{proof}
	
	Corollary \ref{cor-nopoly} does not say that Condition \ref{cond1} is the only
	characterization of the weak core. Other equivalent conditions certainly exist.
	What it rules out is that any of them can be checked in polynomial time. We have two
	remarks. Membership in NP says that an orientation satisfying Condition
	\ref{cond1} is a certificate verifiable in polynomial time, so a rationalizing
	profile may be hard to find but is easy to check once exhibited. And
	NP-completeness concerns worst-case instances, not those an analyst meets: in
	our empirical application the tests decide every market within seconds (Section
	\ref{UKapplication}).

	\section{\rev{Single-crossing preferences}}\label{SCsection}
 In many of the markets we have in mind the objects are ranked by a common notion of
	quality: houses by size and condition, posts by desirability of location. When
	this is so it is natural to ask for preferences that are \emph{single-crossing},
	meaning that if some type prefers the better object of a pair then so does every
	higher type. This section characterizes WC-rationalizability by such a profile
	(Theorem~\ref{Th-SCWC}).

	The restriction is worth imposing for two reasons. It is a genuine economic
	assumption, so a rejection under it says something the general test does not, and
	Example \ref{ex-unheld} of the Supplementary Appendix shows that it can reject allocations the
	general test accepts. It also changes the shape of the problem: under
	Theorem~\ref{Th-WC} the analyst chooses a direction for each within-type
	comparison the data leave open, whereas here the analyst chooses one number per
	pair of objects, which settles how \emph{every} type ranks that pair at once. None
	of this undoes Theorem~\ref{DP-Rationalizability-WC-2}, which concerns the general
	problem. The restriction is motivated by economic considerations, but it also simplifies the problem enough to allow for a faster solution.

	Throughout this section we fix two linear orders: $\rhd _{H}$ over the
	objects and $\rhd _{T}$ over the types. The ranking $\rhd _{H}$ can be interpreted as a quality
	ranking: $h\rhd _{H}h^{\prime }$ means that $h$ is the higher-quality
	object. We rank types from the top: $\operatorname{rk}(t)$ is the
	position of type $t$ in $\rhd _{T}$, so $\operatorname{rk}(t)=1$ for the
	highest type and $\operatorname{rk}(t)=\left\vert T\right\vert$ for the
	lowest. We introduce three pieces of notation below. First, for a linear order
	$\succ_{t}$ over $H$, we write $h\succ _{t}h^{\prime }$ when type $t$ strictly
	prefers $h$ to $h^{\prime }$, that is, when $(h,h^{\prime })\in \succ_{t}$
	and $h\neq h^{\prime }$. Second, $H_{t}=\{h\in H:x_{t,h}>0\}$ is the set
	of objects that type $t$ holds under $x$. These are the vertices in row
	$t$ of the mixed graph. Third, an individual is a \emph{mover} if the
	object he receives under $x$ differs from the object of his endowment. The question of this section is whether
	$x$ is WC-rationalizable by a profile that is single-crossing with
	respect to the two fixed orders.
	
	\begin{definition}[Single-crossing profile]
		\label{def-sc-profile}
		A profile $\succ=(\succ_{t})_{t\in T}$ of linear orders over $H$ is
		\emph{single-crossing} with respect to $\rhd _{T}$ and $\rhd _{H}$ if
		for all objects $h,h^{\prime }\in H$ with $h\rhd _{H}h^{\prime }$ and
		all types $t,t^{\prime }\in T$ with $t^{\prime }\rhd _{T}t$,
		\begin{equation*}
			h\succ _{t}h^{\prime }\quad \Longrightarrow \quad
			h\succ _{t^{\prime }}h^{\prime }\text{.}
		\end{equation*}
	\end{definition}
	
	If some type prefers the higher-quality object of a pair, then
	so does every higher type. Moving down the type order, the preference
	between any two objects switches at most once  (from the
	higher-quality object to the lower-quality one) which is the
	single-crossing property.
	
	\begin{definition}[Single-crossing WC-rationalizability]
		\label{def-sc-wc}
		The allocation $x$ is \emph{single-crossing WC-rationalizable} with
		respect to $\rhd _{T}$ and $\rhd _{H}$ if there exists a single-crossing
		profile $\succ$ such that $x$ is in the weak core for $\succ$.
	\end{definition}
	
	\clr{Fix a pair of objects $h\rhd _{H}h^{\prime }$. By Definition \ref{def-sc-profile}, the types that rank $h$ above $h^{\prime }$ are the $k$ highest types, for some $k\in \{0,1,\ldots ,\left\vert T\right\vert \}$. So a single number $k$ settles how every type ranks this pair. We call it the \emph{crossing point} of the pair. Our problem reduces to choosing one crossing point per pair of objects so that the induced preferences place $x$ in the weak core.}
	
	\clr{The rest of this section proceeds as follows. We first define a \emph{crossing-point function}, which fixes a crossing point for every pair of objects, and show that it induces a single-crossing profile exactly when it satisfies a consistency requirement (Lemma~\ref{lem-cp-profile}). Two examples in the Supplementary Appendix then show why one cannot test single-crossing WC-rationalizability by choosing crossing points pair by pair: Example \ref{ex-betweenness} shows that the crossing points of different pairs must fit together, and Example \ref{ex-unheld} shows that Condition~\ref{cond1} on the mixed graph is not enough. We close with Condition~\ref{cond2} and the characterization result, Theorem~\ref{Th-SCWC}. A mixed-integer linear programming formulation can be found in our Supplementary Appendix to test Theorem \ref{Th-SCWC}}.
	
	\begin{definition}[Crossing-point function]
		\label{def-cpf}
		A \emph{crossing-point function} is a map $\kappa $ that assigns to
		every pair of objects $h,h^{\prime }\in H$ with $h\rhd _{H}h^{\prime }$
		an integer $\kappa (h,h^{\prime })\in \{0,1,\ldots ,\left\vert
		T\right\vert \}$. The \emph{induced family}
		$\succ^{\kappa }=(\succ_{t}^{\kappa })_{t\in T}$ of binary relations over $H$ is
		defined as follows: for all $h,h^{\prime }\in H$ with
		$h\rhd _{H}h^{\prime }$ and all $t\in T$,
		\begin{equation*}
			h\succ _{t}^{\kappa }h^{\prime }\ \text{ if }\
			\operatorname{rk}(t)\leq \kappa (h,h^{\prime })\text{,}
			\qquad \text{and}\qquad
			h^{\prime }\succ _{t}^{\kappa }h\ \text{ if }\
			\operatorname{rk}(t)> \kappa (h,h^{\prime })\text{.}
		\end{equation*}
	\end{definition}
	
	That is, on the pair $\{h,h^{\prime }\}$ the top $\kappa (h,h^{\prime
	})$ types rank the higher-quality object $h$ first, and the remaining
	types rank $h^{\prime }$ first. Both boundary values are allowed:
	$\kappa (h,h^{\prime })=0$ means that every type ranks $h^{\prime }$
	first, and $\kappa (h,h^{\prime })=\left\vert T\right\vert $ that every
	type ranks $h$ first. \clr{The next example shows how a crossing-point function produces a profile.} 
	
	\begin{example}[Reading a crossing-point function]
		Let $\rhd _{T}:t_{1}\rhd t_{2}\rhd t_{3}$ and
		$\rhd _{H}:h_{1}\rhd h_{2}\rhd h_{3}$. The
		crossing-point function $\kappa (h_{1},h_{2})=2$, $\kappa (h_{2},h_{3})=1$, $\kappa (h_{1},h_{3})=2$
		places a threshold in the agent order on each pair. On $\{h_{1},h_{2}\}$ the top two types $t_{1},t_{2}$ rank the higher-quality $h_{1}$ first, and $t_{3}$ ranks $h_{2}$ first.
		On $\{h_{2},h_{3}\}$ only $t_{1}$ ranks $h_{2}$ first. On
		$\{h_{1},h_{3}\}$ the top two rank $h_{1}$ first. Hence we have the following linear orders for types:
		\begin{equation*}
			\succ_{t_{1}}^{\kappa }:h_{1}\succ h_{2}\succ h_{3}\text{,}\qquad \succ_{t_{2}}^{\kappa }:h_{1}\succ h_{3}\succ h_{2}\text{,}\qquad \succ_{t_{3}}^{\kappa }:h_{3}\succ h_{2}\succ h_{1}\text{,}
		\end{equation*}
	\end{example}
	
	\clr{By construction each relation $\succ _{t}^{\kappa }$ ranks every pair of distinct objects in one direction, so it is complete and antisymmetric, but it need not be transitive, and then $\succ^{\kappa }$ is not a profile of linear orders. The consistency condition below rules this out, and Lemma \ref{lem-cp-profile} shows that it is exactly what is needed.}
	
	\begin{definition}[Consistency]
		\label{def-consistency}
		A crossing-point function $\kappa $ is \emph{consistent} if for all
		objects $h_{a},h_{b},h_{c}\in H$ with
		$h_{a}\rhd _{H}h_{b}\rhd _{H}h_{c}$,
		\begin{equation*}
			\min \{\kappa (h_{a},h_{b}),\kappa (h_{b},h_{c})\}\ \leq \ \kappa (h_{a},h_{c})\ \leq \ \max \{\kappa (h_{a},h_{b}),\kappa (h_{b},h_{c})\}\text{.}
		\end{equation*}
	\end{definition}
	
	\begin{lemma}
		\label{lem-cp-profile}Let $\kappa $ be a crossing-point function. The
		induced family $\succ^{\kappa }$ is a single-crossing profile with respect to $\rhd _{T}$ and $\rhd _{H}$ if and only if $\kappa $ is consistent. Conversely, every
		single-crossing profile with respect to $\rhd _{T}$ and $\rhd _{H}$
		arises from a unique consistent crossing-point function.
	\end{lemma}
	
	\begin{proof}
		See Appendix \ref{app-proof-cp}.
	\end{proof}
	
	\clr{The data already restrict the crossing points before we search for a rationalizing profile. A single individual is itself a coalition, so if $x$ is in the weak core for $\succ$, every mover's type must strictly prefer the object the mover receives to the object he was endowed with. Consider a mover of type $t$ with $\operatorname{rk}(t)=p$ whose move concerns the pair $h\rhd _{H}h^{\prime }$. If he receives the higher-quality $h$, then $h\succ _{t}h^{\prime }$, so type $t$ is in the top block of the pair and $\kappa (h,h^{\prime })\geq p$. If he receives the lower-quality $h^{\prime }$, then $h^{\prime }\succ _{t}h$, so type $t$ is outside the top block and $\kappa (h,h^{\prime })\leq p-1$. Collecting these bounds over all movers gives the following definition.}
	
	\begin{definition}[Admissible set]
		\label{def-admissible}
		For each pair of objects $h,h^{\prime }\in H$ with
		$h\rhd _{H}h^{\prime }$, let
		\begin{align*}
			\underline{k}(h,h^{\prime }) &=\max \left\{ \operatorname{rk}(\tau (i))
			\;|\; i\in I \text{ is a mover with } \omega (i)=h^{\prime } \text{ and }
			x(i)=h\right\} \text{,}\\
			\overline{k}(h,h^{\prime }) &=\min \left\{ \operatorname{rk}(\tau (i))-1
			\;|\; i\in I \text{ is a mover with } \omega (i)=h \text{ and }
			x(i)=h^{\prime }\right\} \text{,}
		\end{align*}
		with $\max \{\emptyset\} =0$ and
		$\min \{\emptyset\} =\left\vert T\right\vert $. The \emph{admissible set} of
		the pair is
		\begin{equation*}
			K(h,h^{\prime })=\left\{ k\in \{0,1,\ldots ,\left\vert T\right\vert \}
			\;|\; \underline{k}(h,h^{\prime })\leq k\leq
			\overline{k}(h,h^{\prime })\right\} \text{.}
		\end{equation*}
	\end{definition}
	
	\clr{The lower bound $\underline{k}(h,h^{\prime })$ comes from the movers who receive the higher-quality object of the pair, the upper bound $\overline{k}(h,h^{\prime })$ from those who receive the lower-quality one. If no mover concerns the pair, then $K(h,h^{\prime })=\{0,1,\ldots ,\left\vert T\right\vert \}$. Any single-crossing profile that places $x$ in the weak core must have $\kappa (h,h^{\prime })\in K(h,h^{\prime })$ for every pair. In particular, if some admissible set is empty --- that is, if $\underline{k}(h,h^{\prime })>\overline{k}(h,h^{\prime })$ --- then $x$ is not single-crossing WC-rationalizable.}
	
	\clr{One might expect the converse to be equally simple: for each pair of objects choose an admissible crossing point, orient the red edges of the mixed graph accordingly, and test Condition~\ref{cond1}. This fails, for two separate reasons, and Examples \ref{ex-betweenness} and \ref{ex-unheld} of the Supplementary Appendix give an instance of each. First, crossing points chosen pair by pair need not be consistent: each may lie in its own admissible set while no single single-crossing profile realizes them all. Second, even crossing points that are consistent and satisfy Condition~\ref{cond1} on the mixed graph can leave $x$ outside the weak core, because a blocking coalition may run through an object that no type holds and that the mixed graph therefore never represents.}
	
	\begin{definition}[Demand digraph]
		\label{def-demand}
		For an arbitrary profile $\succ$, the \emph{demand digraph} $D(\succ)$ has the set
		of individuals $I$ as its vertices and, for all individuals $i,j\in I$, an
		arc $i\rightarrow j$ whenever the type of $i$ strictly prefers the object
		of $j$'s endowment to the object of $i$'s own assignment:
		$\omega(j)\succ _{\tau (i)}x(i)$.
	\end{definition}
	
	Informally, an arc points from each individual to every individual whose
	endowment it would rather hold than its own assignment. A directed cycle
	is then a group of individuals who can pass their endowments around the
	cycle and all strictly improve at once, which is exactly a blocking
	coalition. Note that $D(\succ)$ lives on individuals, not on held objects: it
	sees every comparison in $\succ$, including a type's ranking of objects it
	does not hold---precisely the comparisons that Condition~\ref{cond1}
	misses in Example \ref{ex-unheld} of the Supplementary Appendix.


	\begin{lemma}\label{lem-weakcore-dag}
		For any profile $\succ$ (not necessarily single-crossing),
		the allocation $x$ lies in the
		weak core at $\succ$ if and only if the demand digraph $D(\succ)$ is acyclic (a
		self-loop counting as a cycle).
	\end{lemma}
	
	\begin{proof}
		See Appendix \ref{app-proof-dag}.
	\end{proof}
	
	The two requirements below are jointly necessary and sufficient. Both
	Condition~\ref{cond2} and Theorem \ref{Th-SCWC} are based on the fixed orders $\rhd _{T}$ and $\rhd _{H}$.
	
	\begin{condition}
		\label{cond2}There exists a consistent crossing-point function $\kappa $ such
		that the demand digraph $D(\succ^{\kappa })$ of the induced single-crossing
		profile $\succ^{\kappa }$ is acyclic.
	\end{condition}
	
	\begin{theorem}
		\label{Th-SCWC}The allocation $x$ is single-crossing WC-rationalizable with
		respect to $\rhd _{T}$ and $\rhd _{H}$ if and only if
		Condition~\ref{cond2} holds.
	\end{theorem}
	
	\begin{proof}
		See Appendix \ref{app-proof-SCWC}.
	\end{proof}
	
	Theorem~\ref{Th-SCWC} turns the search for a rationalizing profile into a finite
	feasibility problem. The candidate profiles are exactly the $\succ^{\kappa}$ for
	consistent $\kappa$, the admissible sets cut them down further, and there are at
	most $\prod|K(h,h^{\prime})|\leq(|T|+1)^{\binom{|H|}{2}}$ crossing-point functions
	to consider, each checkable in polynomial time. If an analyst faces a family of
	instances with the same number of objects throughout while the population grows,
	$\binom{|H|}{2}$ is a constant, the bound is polynomial in $|T|$, and the question
	is decided in polynomial time in the size of the market --- a reasonable
	description of many applications, where the objects are a fixed list of dwelling
	types or job grades. Our own application does not satisfy this condition, since
	the number of dwelling profiles ranges from $6$ to $15$ across markets, yet the
	test is immediate in practice: we test Condition~\ref{cond2} directly as a
	mixed-integer program (See our Supplementary Appendix) and every instance is decided in well under a second
	(Section~\ref{UKapplication}).


	\section{\texorpdfstring{\new{An application to social housing in England}}{An application to social housing in England}}\label{UKapplication}
	
	In this section, we conduct our revealed preference tests on a reallocation
	dataset that we construct from the continuous recording of lettings in social
	housing in England (CORE) for April 2021 to March 2022, accessed under a Special
	License agreement and restricted to general needs lettings by local
	authorities.\footnote{In the Supplementary Appendix, we explain how we construct
		our reallocation dataset in detail.} In the CORE data, each row records the
	characteristics of a tenant and a dwelling. There are no tenant or dwelling
	identifiers. Hence we cannot keep track of tenants' moves.\footnote{In our
		Supplementary Appendix, we show that individuals cannot be identified by using
		their characteristics with a placebo experiment.} For this reason, our dataset
	cannot fully identify inefficiencies in the English social housing market. Our
	goal is to illustrate our theoretical results with a real dataset, not to
	document inefficiencies in that market.
	
	Social housing in England is administered locally. Each local authority rents
	its own social houses and manages internal transfers. We therefore treat every
	local authority as a separate market. For each market, our tests need a
	population of individuals, each with an initial and a final dwelling. We do not
	have this, because there are no tenant or dwelling identifiers. To overcome
	this, we define an agent type and a dwelling profile for each row. An agent type
	combines seven recorded characteristics, and a dwelling profile combines four.
	We return to this choice later. We then split the year into two halves,
	April--September 2021 and October 2021--March 2022, and obtain two
	cross-sectional datasets. In the first half, a first-time entrant's dwelling
	profile is an initial endowment for its agent type. In the second half, the
	dwelling profile of an existing tenant who made a voluntary internal transfer is
	a final allocation. We now turn the two cross-sections into one reallocation
	dataset. We count each (agent type, dwelling profile) pair. We need $n$
	dwelling-profile copies reallocated to $n$ agents. To balance the two halves,
	and to reallocate the initial dwelling profiles to the same agent types, we use
	Iterative Proportional Fitting (IPF), rounding, and pairing within each
	type.\footnote{See Supplementary Appendix for an example about how the method
		works.} The pairing involves randomness. So each constructed dataset is one draw,
	and we generate one hundred per market. Of the $309$ authorities, $154$ have no
	general needs lettings of their own, and $137$ are too small once our
	restrictions are applied. This leaves $\mathbf{18}$ analyzable markets:
	$2{,}266$ individuals, between $50$ and $273$ per market, holding between $6$ and
	$15$ dwelling profiles.
	
	We must decide how finely to define agent types. We assume that individuals of
	the same agent type are identical and have the same preferences over dwelling
	profiles. This assumption is more credible when types are fine, that is, when
	each type has few members. However, fine types can be uninformative, because a
	type with few members reveals few comparisons between dwelling profiles. In the
	extreme, if every type has a single member, the data are trivially
	rationalizable. So we do not fix a single definition of types. We run every test
	at three nested levels.
	The coarsest uses age band and household composition ($6$--$35$ types per
	market). The middle adds economic status and vulnerability ($6$--$77$). The
	finest uses all seven characteristics ($6$--$99$). Within a market and draw, the
	individuals, the dwellings, and the two allocations are the same across the
	three levels. So any difference in the results comes from the type structure
	alone. The three levels are nested, not independent. Suppose a profile rationalizes an allocation for a coarse partition. Each fine type sits inside exactly one coarse type. We can give each fine type the preference of the coarse type that contains it. The individuals are the same, so their preferences do not change. Hence the allocation is still rationalized under the finer partition. Acceptance can therefore only rise as types get finer.
	
	\subsection{\texorpdfstring{\new{Results}}{Results}}
    
	Table~\ref{tab:main} collects the results. We read five things from it.

	\emph{First, the tests have content on data of this kind.} They do not
	accept everywhere and they do not reject everywhere. Under the coarsest
	partition, both properties are rejected in seventeen of the eighteen
	markets in every draw. Under the finest partition, three markets are
	accepted in all one hundred draws, and ten more are accepted in at least
	one draw. The characterizations are therefore not vacuous here, which is
	the first thing one would want to know about a test.
    \newpage

    \begin{table}[htb]
		\footnotesize
		\centering
		\caption{Rationalizability verdicts, market by market. Each cell reports how
			many of the $100$ pairing draws the test accepts, for the coarsest (C), middle
			(M), and finest (F) type partition. SC-WC is the single-crossing weak-core test,
			run on the weak-core-consistent instances with dwelling quality increasing in
			the number of bedrooms.}\label{tab:main}
		\begin{tabular}{l r rrr rrr rrr}
			\toprule
			& & \multicolumn{3}{c}{PI accepts} & \multicolumn{3}{c}{WC accepts} &
			\multicolumn{3}{c}{SC-WC accepts} \\
			\cmidrule(lr){3-5}\cmidrule(lr){6-8}\cmidrule(lr){9-11}
			Market & $N$ & C & M & F & C & M & F & C & M & F \\
			\midrule
			Leeds & 273 & 0 & 0 & 94 & 0 & 0 & 82 & 0 & 0 & 0 \\
			Sheffield & 247 & 0 & 0 & 0 & 0 & 0 & 0 & 0 & 0 & 0 \\
			Birmingham & 208 & 0 & 0 & 2 & 0 & 0 & 0 & 0 & 0 & 0 \\
			Stoke-on-Trent & 170 & 0 & 0 & 0 & 0 & 0 & 0 & 0 & 0 & 0 \\
			Hull & 150 & 0 & 0 & 0 & 0 & 0 & 0 & 0 & 0 & 0 \\
			Newcastle & 146 & 0 & 0 & 2 & 0 & 0 & 0 & 0 & 0 & 0 \\
			North Tyneside & 142 & 0 & 0 & 0 & 0 & 0 & 0 & 0 & 0 & 0 \\
			Gateshead & 127 & 0 & 48 & 100 & 0 & 0 & 100 & 0 & 0 & 0 \\
			Sandwell & 109 & 0 & 0 & 0 & 0 & 0 & 0 & 0 & 0 & 0 \\
			Rotherham & 107 & 0 & 0 & 17 & 0 & 0 & 11 & 0 & 0 & 0 \\
			Barnsley & 101 & 0 & 0 & 15 & 0 & 0 & 4 & 0 & 0 & 0 \\
			Doncaster & 96 & 0 & 0 & 17 & 0 & 0 & 18 & 0 & 0 & 0 \\
			Wigan & 89 & 0 & 0 & 100 & 0 & 0 & 100 & 0 & 0 & 0 \\
			Dudley & 75 & 0 & 9 & 31 & 0 & 4 & 23 & 0 & 0 & 0 \\
			Lambeth & 62 & 100 & 100 & 100 & 100 & 100 & 100 & 100 & 100 & 100 \\
			East Riding & 61 & 0 & 0 & 3 & 0 & 0 & 1 & 0 & 0 & 0 \\
			Derby & 53 & 0 & 56 & 56 & 0 & 64 & 64 & 0 & 0 & 0 \\
			Brent & 50 & 0 & 0 & 8 & 0 & 0 & 7 & 0 & 0 & 0 \\
			\bottomrule
		\end{tabular}
	\end{table}
    
\newpage

	\emph{Second, the two verdicts come apart, and in both directions.} This
	is Proposition~\ref{prop-nonnested} appearing in the data. Take Derby
	under the middle partition. The weak core test accepts sixty-four draws
	and the efficiency test accepts fifty-six, so at least eight of these
	datasets are consistent with stability but not with efficiency. Now take
	Leeds under the finest partition, where the efficiency test accepts
	ninety-four draws and the weak core test eighty-two: at least twelve
	datasets run the other way. Neither test's acceptances contain the
	other's. An analyst who ran only one of the two would reach conclusions
	the other test contradicts, so the two are not substitutes.

	\emph{Third, the answer depends on how the data are built and grouped.}
	Acceptance rises as types get finer, partly because the three levels are
	nested. It also depends on the pairing. Under the coarsest partition every
	market gives the same verdict in all one hundred draws. Under the finest
	partition only eight of eighteen do. For the other ten, a single
	constructed dataset could have reported either answer depending on the
	draw. We report this because it is a limitation of the exercise rather
	than of the tests: it says that with data of this kind the analyst should
	report a distribution of verdicts and not a single one.

	\emph{Fourth, the results of Section~\ref{WCsection} do real work here.}
	Proposition~\ref{prop-fixedarc} is the clearest case. The arcs fixed by
	the data already reject $4{,}591$ of the $4{,}622$ rejected instances, so
	in more than $99\%$ of the rejections we never examine a single
	orientation. The set $\mathcal{E}$ of open comparisons is also small in
	practice, with a median of $7$ and a maximum of $53$, which is why the
	remaining instances are settled quickly as well. The search that
	Theorem~\ref{Th-WC} calls for is thus a worst-case feature of the problem
	rather than a description of what the analyst usually does.

	\emph{Fifth, everything is fast.} Every instance is decided within
	seconds. The single-crossing test is the strongest illustration. We test it by formulating Condition~\ref{cond2} to a mixed-integer program. The test is run on the $778$
	instances that passed the weak core test, and it decides each of them in
	well under a second, even though the number of objects varies across
	markets and the polynomial-time guarantee of Section~\ref{SCsection} does
	not apply. This is the remark after Corollary~\ref{cor-nopoly} borne out:
	hardness rules out a general shortcut, not the instances an analyst
	actually meets.

	One substantive finding is worth recording separately. Only Lambeth passes
	the single-crossing test, and it passes in every draw and at every level of
	the type partition. It is also the only market that is consistent with both
	efficiency and stability throughout. Single-crossing is a demanding
	restriction on these data, which is what one would expect: it requires the
	recorded moves of every type to line up with one common quality ranking of
	dwellings, and the number of bedrooms is only a rough proxy for what makes
	a dwelling desirable.

	\section{Conclusion} \label{Conclusion}
	
	This paper asks what an observed reallocation of indivisible objects
	reveals about the market that produced it, when preferences are never
	observed and individuals are grouped into types that share them. We gave
	three characterizations.

	A reallocation is consistent with Pareto efficiency and individual
	rationality if and only if the aggregate allocation graph has no cycle
	(Theorem \ref{PI}). The condition is read directly off the data, and a
	cycle in the graph is exactly the multilateral trade that decentralized
	exchange is unable to arrange.

	A reallocation is consistent with the weak core if and only if the
	comparisons the data leave open can be directed so that each type's
	comparisons fit together and no group can trade profitably (Theorem
	\ref{Th-WC}). This condition asks for a search rather than an inspection,
	and we devoted Section \ref{WCsection} to what that search is worth. It is
	far smaller than the search in the definition, and it produces a
	certificate that can be checked quickly (Section \ref{WCbuys}). It is
	frequently unnecessary, since the arcs the data fix reject most instances
	on their own, and it disappears entirely when the data leave no comparison
	open (Proposition \ref{prop-fixedarc}). And it cannot be removed in
	general, because deciding WC-rationalizability is NP-complete (Theorem
	\ref{DP-Rationalizability-WC-2}), so any correct characterization would
	require something of the same kind.

	The two properties are logically independent, so neither test can stand in
	for the other (Proposition \ref{prop-nonnested}), and an instance can pass
	both without any single profile delivering both at once (Proposition
	S.2). Applying the weak core analysis to the single-crossing
	domain gives a third characterization, in terms of crossing points
	(Theorem \ref{Th-SCWC}).

	We took the three tests to the English social housing lettings register,
	treating each local authority as a separate market. Because the
	reallocation has to be reconstructed from two cross-sections, and because
	the assumption that a type shares one preference is strong, we do not read
	the results as a verdict on how English social housing is allocated. The
	exercise establishes three things instead. The tests have content, since
	they accept in some markets and reject in others, and the two verdicts
	differ in both directions across markets, which is Proposition \ref{prop-nonnested}
	appearing in the data. \rev{The verdicts also respond to how the data are
	built and grouped, so with data of this kind an analyst should report a
	distribution of verdicts rather than a single one.} And the results of
	Section \ref{WCsection} matter in practice, since Proposition
	\ref{prop-fixedarc} settles more than $99\%$ of the rejections with no
	search.

	Two questions remain open. Joint rationalizability is not characterized by
	the conjunction of our two conditions, and a characterization would have to
	couple the two graphs rather than inspect them one at a time. Second, our
	tests take one pairing of the two cross-sections as given. Asking whether
	the observed type-by-object marginals are rationalizable under \emph{some}
	pairing consistent with them is the right counterpart of our tests when
	even the pairing of endowments to assignments is unavailable, and we leave
	it for future work.

	\appendix

	\section{Proofs of Theorems \ref{PI} and \ref{Th-WC}}\label{Appendix}

	\subsection{Proof of Theorem \ref{PI}}\label{ProofPI}
\begin{proof}
\new{We first record a fact used in both directions. It says that Pareto
		efficiency is an object-level property here; the multiplicities $q_{h}$ make
		this something to be checked rather than assumed. Recall that the envy graph
		of $x$ at a profile $\succ$ has the objects as vertices, with an arc from $h$
		to $h^{\prime }$ whenever some individual $i$ with $x(i)=h$ satisfies
		$h^{\prime }\succ_{\tau (i)}h$.}
	
	\begin{lemma}\label{lem-envy}
		Let preferences be strict. An allocation $x$ is Pareto efficient if and only if
		its envy graph is acyclic.
	\end{lemma}
	
	\begin{proof}
		Suppose the envy graph contains a cycle $\langle h_{1},\dots ,h_{m}\rangle$.
		Its vertices are distinct, so for each $j$ we may pick an individual $i_{j}$
		with $x(i_{j})=h_{j}$ who strictly prefers $h_{j+1}$, and these individuals are
		distinct because their assignments are. Reassign each $i_{j}$ to the copy of
		$h_{j+1}$ vacated by $i_{j+1}$, indices modulo $m$, leaving everyone else in
		place. Every $q_{h}$ is respected, since the reassignment permutes $m$ copies
		among the $m$ individuals who held them, and each $i_{j}$ is strictly better
		off, so $x$ is Pareto dominated.
		
		Conversely, let $x'$ Pareto dominate $x$. Preferences are strict, so
		$D=\{i:x'(i)\neq x(i)\}$ is nonempty and every $i\in D$ strictly prefers
		$x'(i)$ to $x(i)$. Since $x$ and $x'$ assign the same multiset of objects, for
		each $i\in D$ we may pick $\sigma (i)\in D$ with $x(\sigma (i))=x'(i)$. As
		$\sigma$ maps the finite set $D$ into itself and never fixes a point, iterating
		it reaches a cycle $i_{1},\dots ,i_{m}$ with $x(i_{j+1})=x'(i_{j})$. The objects
		$x(i_{1}),\dots ,x(i_{m})$ are then distinct and each $i_{j}$ strictly prefers
		$x(i_{j+1})$ to $x(i_{j})$, which is a cycle in the envy graph.
	\end{proof}
	
		$(\Rightarrow)$ \clr{Let $\succ$ be a preference profile for which $x$ is PI. We show that $AG$ has no cycle.}
		
		\begin{lemma}\label{edgepref}
			Let $x$ be an IR allocation at $\succ$. If there exists $e=(h,h')\in AE$, then there exist $i,i'\in I$ such that $x(i')=h'\succ_{\tau(i)}h=x(i)$.
		\end{lemma} 
		
		\begin{proof}
			Let $e=(h,h')$ be constructed by Algorithm \ref{alg:graph_construction}. Hence, there exists some $i,i'\in I^{t}$ such that $\omega(i')=h\neq h'=x(i')$ and $x(i)=h$. Since $x$ is IR, \new{$h'\succ_{\tau(i')}h$}. As $i$ belongs to the same type $I^{t}$, it shares this ranking: $h'\succ_{\tau(i)}h$.
		\end{proof}
		
		\clr{Suppose for a contradiction that $AG$ has a cycle. Since $x$ is IR, Lemma \ref{edgepref} gives, for each edge $e=(h,h')\in AE$, an individual $i\in I$ with $x(i)=h$ and $h'\succ_{\tau(i)}h$.} \new{So every arc of $AG$ is an arc of the envy graph of $x$ at $\succ$, and a cycle in $AG$ is therefore a cycle in that envy graph. By Lemma \ref{lem-envy}, $x$ is not Pareto efficient, contradicting the assumption that $x$ is PI at $\succ$.}
		
		$(\Leftarrow)$\clr{Now assume that $AG$ has no cycle. We show that there exists a preference profile $\succ^{*}=(\succ^{*}_{t})_{t\in T}$ for which $x$ is PI.} 
		
		\clr{To do this, we construct the profile for each type. Let $H^{t}$ be the set of objects held by individuals of type $t$:}  
		
		\[
		H^{t} = \{ h \in H :\exists i\in I^{t}, x(i)= h \}.
		\]  
		
		\clr{To construct $\succ^{*}$ we define two binary relations. For every $t\in T$, let $M_{t}$ be the relation on $H$ given by:} 
		
		$$
		h M_{t} h' \text{ if and only if } \text{ there is an individual } i \in I^{t} \text{ such that } \begin{cases} h = x(i)\\ h' = \omega(i)\end{cases}
		$$
		
		\begin{lemma}\label{Pacyclic}
			If $AG$ is acyclic, then $M_{t}$ is acyclic.
		\end{lemma}
		\begin{proof} 
			Suppose for a contradiction that $M_{t}$ is cyclic. Then, there exists a sequence of distinct objects $h_{1},\dots,h_{m}$ in $H$ such that $h_{1}M_{t} h_{2} M_{t}\dots h_{m}$ and $h_{m} M_{t} h_{1}$. So, for $j\in \{1,\dots,m-1\}$,
			$$
			h_{j}M_{t}h_{j+1} \implies \exists i_{j}\in I^{t} \text{ such that } h_{j}=x(i_{j}) \text{ and } h_{j+1} =\omega(i_{j})
			$$
			and
			$$
			h_{m}M_{t}h_{1} \implies \exists i_{m}\in I^{t} \text{ such that } h_{m}=x(i_{m}) \text{ and } h_{1} =\omega(i_{m})
			$$ 
			Hence, $(h_{j+1},h_{j})\in AE$ for $j\in \{1,\dots,m-1\}$. Since $\exists i_{m}\in I^{t}$ such that $h_{m}=x(i_{m})$ and $h_{1}=\omega(i_{m})$, $(h_{1},h_{m})\in AE$ because $h_{1}=x(i_{1})$. Hence, $\langle h_{m},h_{m-1},\dots,h_{2},h_{1}\rangle$ is a cycle in $AG$, yielding a contradiction.
		\end{proof}
		
		\new{By Lemma \ref{Pacyclic}, $M_{t}$ is acyclic, so its transitive closure $T(M_{t})$ is irreflexive and transitive. Writing $\Delta = \{(h,h) : h \in H\}$, the reflexive closure $T(M_{t}) \cup \Delta$ is therefore reflexive, antisymmetric, and transitive, that is, a partial order on $H$. By Szpilrajn's extension theorem}\footnote{\new{See \cite{szpilrajn1930extension}; \citet{duggan1999general} gives the general form of such extension theorems.}}\new{ it extends to a linear order on $H$, and the asymmetric part of that linear order, denoted $M^{+}_{t}$, is complete, asymmetric, and transitive. Since $T(M_{t})$ is irreflexive, $M_{t} \subseteq T(M_{t}) \subseteq M^{+}_{t}$; in particular $h M_{t} h'$ implies $h M^{+}_{t} h'$ for all $h, h' \in H$.}

		\clr{Recall that $h'\rightarrow h$ means there is an edge from $h'$ to $h$ in $AG$, that is, $(h',h)\in AE$.} Let $\succ_{AG}$ be a binary relation defined for the objects $h,h' \in H$ in the following way: 
		
		$$
		h \succ_{AG} h' \text{ if and only if } h'\rightarrow h
		$$
		
		Since $AG$ is acyclic, $\succ_{AG}$ is acyclic as well.
		
		\new{Since $\succ_{AG}$ is acyclic, its transitive closure $T(\succ_{AG})$ is irreflexive and transitive, so the reflexive closure $T(\succ_{AG}) \cup \Delta$ is a partial order on $H$ (reflexive, antisymmetric, and transitive). By Szpilrajn's extension theorem, it extends to a linear order on $H$, whose asymmetric part $\succ_{AG}^{+}$ is complete, asymmetric, and transitive. Since $T(\succ_{AG})$ is irreflexive, $\succ_{AG} \subseteq T(\succ_{AG}) \subseteq \succ_{AG}^{+}$; in particular $h \succ_{AG} h'$ implies $h \succ_{AG}^{+} h'$ for all $h, h' \in H$.}

		Now we can construct the preference profile for every type $t\in T$ in the following way: Every object \( h \in H^{t} \) is strictly preferred to any object \( h' \notin H^{t} \), i.e.,  
		\[
		h \succ^{*}_{t} h' \quad \text{for all } h \in H^{t}, h' \notin H^{t}.
		\]  
		
		For every type $t\in T$ and objects $h,h' \in H^{t}$, $h \succ^{*}_{t} h'$ if and only if $h \succ_{AG}^{+} h'$. For every type $t\in T$ and objects $h,h' \in H\setminus H^{t}$, $h \succ^{*}_{t} h'$  if and only if $h M^{+}_{t} h'$.
		
		We need to show that $\succ^{*}_{t}$, for every $t\in T$, is complete, asymmetric and transitive. 
		
		\begin{enumerate}
			\item The relation $\succ^{*}_{t}$ is complete:
			\begin{itemize}
				\item If $h,h' \in H^{t}$, this follows from completeness of $\succ_{AG}^{+}$.
				\item If $h \in H^{t}$ and $h' \notin H^{t}$ then $h \succ^{*}_{t} h'$.
				\item If $h,h' \notin H^{t}$, this follows from completeness of $M^{+}_{t}$.
			\end{itemize}
			\item The relation $\succ^{*}_{t}$ is asymmetric:
			\begin{itemize}
				\item If $h,h' \in H^{t}$, this follows from asymmetry of $\succ_{AG}^{+}$.
				\item If $h \in H^{t}$ and $h' \notin H^{t}$ then $h \succ^{*}_{t} h'$ and not $h' \succ^{*}_{t} h$.
				\item If $h,h' \notin H^{t}$, this follows from asymmetry of $M^{+}_{t}$.
			\end{itemize}
			\item The relation $\succ^{*}_{t}$ is transitive. \new{Let $h \succ^{*}_{t} h' \succ^{*}_{t} h''$. We need to show that $h \succ^{*}_{t} h''$. Note first that, by construction, every object of $H^{t}$ is ranked above every object outside $H^{t}$, so $h'' \in H^{t}$ forces $h' \in H^{t}$, and $h' \in H^{t}$ forces $h \in H^{t}$. Then:}
			\begin{itemize}
				\item \new{if $h'' \in H^{t}$, then $h',h \in H^{t}$ as just noted, so $h \succ_{AG}^{+} h' \succ_{AG}^{+} h''$ and hence $h \succ_{AG}^{+} h''$ by transitivity of $\succ_{AG}^{+}$; therefore, by definition, $h \succ^{*}_{t} h''$.}
				\item \new{if $h'' \notin H^{t}$ and $h \in H^{t}$, then $h \succ^{*}_{t} h''$ directly, since every object of $H^{t}$ is ranked above every object outside it.}
				\item \new{if $h'' \notin H^{t}$ and $h \notin H^{t}$, then $h' \notin H^{t}$ as well, for otherwise $h' \in H^{t}$ would force $h \in H^{t}$. So $h,h',h'' \notin H^{t}$, and $h M^{+}_{t} h' M^{+}_{t} h''$ gives $h M^{+}_{t} h''$ by transitivity of $M^{+}_{t}$; hence $h \succ^{*}_{t} h''$.}
			\end{itemize}
		\end{enumerate}
		
		\clr{First we show that $x$ is IR under $\succ^{*}$. Take $i\in I^{t}$ with $h=x(i)$ and $h'=\omega(i)$; we must show $h\succ^{*}_{t}h'$.} If there exists $j\in I^{t}$ such that $x(j)=\omega(i)$ then $h'\rightarrow h$. Hence, $h\succ^{*}_{t} h'$. If there exist no $j\in I^{t}$ such that $x(j)=\omega(i)$, then $\omega(i)\notin H^{t}$. Hence, $h\succ^{*}_{t}h'$.
		
		Now, we need to show that $x$ is PE at $\succ^{*}$. Let $G^{*}=(V^{*},E^{*})$ be the envy graph of $x$ with the preference profile $\succ^{*}$ where the set of vertices $V^{*}$ is composed of objects and from each $v\in V^{*}$ there is an edge to $v'\in V^{*}$ if and only if there exist $i,i'\in I$ such that $x(i')=v'$, $x(i)=v$ and \new{$v'\succ^{*}_{\tau(i)} v$}. 	
		\begin{lemma}\label{AEE}
			Given an IR allocation $x$ at $\succ^{*}$, the aggregate allocation graph $AG=(AV,AE)$ and envy graph $G^{*}=(V^{*},E^{*})$ of $x$ with the preference profile $\succ^{*}$, we have the following: $AE \subseteq E^{*}$.
		\end{lemma}
		
		\begin{proof}
			Suppose for a contradiction that there exist $i,i'\in I$ such that $e=(x(i),x(i'))\in AE$ and $e\notin E^{*}$. Since $x$ is IR with $\succ^{*}$ and $e\in AE$, then by Lemma \ref{edgepref}, \new{$x(i')\succ^{*}_{\tau(i)} x(i)$}. Hence, $e \in E^{*}$, which is a contradiction.
		\end{proof}
		
		By Lemma \ref{AEE}, we know that either $AE=E^{*}$ or $AE \subset E^{*}$. \new{If $AE=E^{*}$, then the envy graph of $x$ at $\succ^{*}$ coincides with $AG$, which is acyclic by assumption; Lemma \ref{lem-envy} then gives that $x$ is PE with $\succ^{*}$.}
		
		Assume that $AE \subset E^{*}$. Suppose, for a contradiction, $x$ is not PE. \new{By Lemma \ref{lem-envy}} there exists a cycle $c=\langle e_1,\ldots, e_m \rangle$ in the envy graph $G^{*}=(V^{*},E^{*})$ such that for all $j = \{1,\ldots, m-1\}$, $e_{j}=(h_{j},h_{j+1})\in E^{*}$ and $e_{m}=(h_{m},h_{1})\in E^{*}$. If each edge in $c$ is also an edge in $AG$, then $c$ is also a cycle in $AG$, which is a contradiction. There must exist some $e_k$ such that $e_k \in E^{*}\setminus AE$.
		
		\new{We claim that every edge of $c$, whether or not it lies in $AE$, \clr{runs from a less-preferred to a more-preferred object under $\succ_{AG}^{+}$: if the edge is $(h_{k},h_{k+1})$, then $h_{k+1}\succ_{AG}^{+}h_{k}$.} Take first $e_{k}=(h_{k},h_{k+1})\in E^{*}\setminus AE$. By the definition of the envy graph, there is an individual $i$ with $x(i)=h_{k}$ and $h_{k+1}\succ^{*}_{t}h_{k}$ for $t=\tau (i)$. Since $i\in I^{t}$ is assigned $h_{k}$, we have $h_{k}\in H^{t}$; and since every object of $H^{t}$ is ranked by $\succ^{*}_{t}$ above every object outside $H^{t}$, the strict preference $h_{k+1}\succ^{*}_{t}h_{k}$ forces $h_{k+1}\in H^{t}$ as well. Both objects therefore lie in $H^{t}$, where $\succ^{*}_{t}$ agrees with $\succ_{AG}^{+}$ by construction, so $h_{k+1}\succ_{AG}^{+}h_{k}$. Take now $e_{k}=(h_{k},h_{k+1})\in AE$. Then $h_{k}\rightarrow h_{k+1}$ in $AG$, so $h_{k+1}\succ_{AG}h_{k}$ by the definition of $\succ_{AG}$, and $h_{k+1}\succ_{AG}^{+}h_{k}$ because $\succ_{AG}\subseteq \succ_{AG}^{+}$.}
		
		\new{Hence for all $j = \{1,\ldots, m-1\}$, $h_{j+1} \succ_{AG}^{+} h_{j}$, and $h_{1} \succ_{AG}^{+} h_{m}$. Chaining these along $c$ and using transitivity of $\succ_{AG}^{+}$ yields $h_{1}\succ_{AG}^{+}h_{1}$, contradicting its asymmetry.}
		
	\end{proof}

	\subsection{Proof of Theorem \ref{Th-WC}}\label{appendix B}
\begin{proof}
Throughout, $\mathcal{G}=\left( \mathcal{V},\mathcal{E},\mathcal{A}\right) $
	is the mixed graph associated with the allocation $x$, the endowment $\omega$
	and the aggregate allocation $X\left( x\right) $. Recall that $\omega$ is
	fixed, that $H_{t}=\left\{ h\in H\mid x_{t,h}>0\right\} $ denotes the set of
	objects held by type $t$ under $x$, and that $\mathcal{V}=\left\{ \left(
	t,h\right) \mid t\in T,\ h\in H_{t}\right\} $. Recall also that for all
	$\left\{ \left( t,h\right) ,\left( t,h^{\prime }\right) \right\} \in
	\mathcal{E}$ it holds that $h^{\prime }\neq h$.
	
	\new{We first record two immediate consequences of these definitions, both
		used repeatedly below.}
	
	\begin{observation}\label{obs-basic}
		\new{Let $t\in T$ and $i\in I^{t}$. Then:}
		\begin{itemize}
			\item[(i)] \new{if $x\left( i\right) \in H_{t}$, then $\left( t,x\left(
				i\right) \right) \in \mathcal{V}$;}
			\item[(ii)] \new{if $\omega\left( i\right),x\left( i\right) \in H_{t}$ and $\omega\left(
				i\right) \neq x\left( i\right) $, then $\left( \left( t,\omega\left( i\right)
				\right) ,\left( t,x\left( i\right) \right) \right) \in \mathcal{A}_{1}
				\subseteq T\left( \mathcal{A}_{1}\right) $.}
		\end{itemize}
	\end{observation}
	
	\begin{proof}
		\new{Part (i) holds because $i\in I^{t}$ is assigned $x\left( i\right) $, so
			$x_{t,x\left( i\right) }\geq 1>0$. For part (ii), $\left( t,\omega\left(
			i\right) \right) \in \mathcal{V}$ by hypothesis and $\left( t,x\left( i\right)
			\right) \in \mathcal{V}$ by part (i); since $i\in I^{t}$ with $\omega\left(
			i\right) \neq x\left( i\right) $, the pair meets the defining requirement of
			$\mathcal{A}_{1}$.}
	\end{proof}
	
	\bigskip
	
	\noindent
	$(\Rightarrow)$ Suppose that $x$ is WC-rationalized by $\succ$. Let $\gamma $
	be defined for all $\left\{ \left( t,h\right) ,\left( t,h^{\prime }\right)
	\right\} \in \mathcal{E}$ by
	\begin{equation}
		\gamma \left( \left\{ \left( t,h\right) ,\left( t,h^{\prime }\right) \right\}
		\right) =\left( \left( t,h\right) ,\left( t,h^{\prime }\right) \right)
		\iff \left( h^{\prime },h\right) \in \succ_{t}\text{.}
		\label{def-gamma}
	\end{equation}That is, the orientation is from $\left( t,h\right) $ to $\left(
	t,h^{\prime }\right) $ if and only if $h^{\prime }$ is strictly better than
	$h$ for $t$. \new{Since $\succ_{t}$ is complete and antisymmetric and
		$h\neq h^{\prime }$ for every edge of $\mathcal{E}$, the map $\gamma $ is well
		defined.}
	
	Let us first show part i) of Condition \ref{cond1}. Since $x$ is individually
	rational at $\succ$, any $i\in I^{t}$ with $\omega\left( i\right) \neq x\left(
	i\right) $ satisfies $\left( x\left( i\right) ,\omega\left( i\right) \right)
	\in \succ_{t}$. Hence, by the definition of $\mathcal{A}_{1}$,
	\begin{equation}
		\left( \left( t,h\right) ,\left( t,h^{\prime }\right) \right) \in \mathcal{A}
		_{1}\new{\ \Longrightarrow \ }\left( h^{\prime },h\right) \in \succ_{t}
		\text{.}
		\label{Nec-condition1-part_i}
	\end{equation} Since $\succ_{t}$ is
	transitive for each $t\in T$, (\ref{Nec-condition1-part_i}) extends to the
	within-type transitive closure: $\left( \left( t,h\right) ,\left( t,h^{\prime
	}\right) \right) \in T\left( \mathcal{A}_{1}\right) $ implies $\left(
	h^{\prime },h\right) \in \succ_{t}$. Together with (\ref{def-gamma}), every
	arc of $\mathcal{G}_{\gamma }^{t}$ is of the form $\left( \left( t,h\right)
	,\left( t,h^{\prime }\right) \right) $ with $\left( h^{\prime },h\right) \in
	\succ_{t}$. \new{As $\succ_{t}$ is a linear order, it follows that $\mathcal{G}_{\gamma }^{t}$ cannot contain a
		directed cycle.} It follows that the type-$t$ sub-digraph $\mathcal{G}_{\gamma
	}^{t}$ is acyclic, for each $t\in T$.
	
	Let us now show part ii) of Condition \ref{cond1}. Assume, to the contrary,
	that $\mathcal{G}_{\gamma }$ contains an alternating cycle $C$. \new{Since $C$
		contains neither two consecutive intra-arcs nor two consecutive inter-arcs,
		its arcs alternate between the two kinds; in particular $C$ has even length, and after relabelling we may write}
	\begin{equation*}
		\new{C=\left\langle \left( t_{1},h_{1}\right) ,\left( t_{1},h_{1}^{\prime
			}\right) ,\left( t_{2},h_{2}\right) ,\left( t_{2},h_{2}^{\prime }\right)
			,\ldots ,\left( t_{K},h_{K}\right) ,\left( t_{K},h_{K}^{\prime }\right)
			\right\rangle \text{,}}
	\end{equation*}\new{where, for each $l$ (indices modulo $K$), $\left( \left(
		t_{l},h_{l}\right) ,\left( t_{l},h_{l}^{\prime }\right) \right) $ is an
		intra-arc and $\left( \left( t_{l},h_{l}^{\prime }\right) ,\left(
		t_{l+1},h_{l+1}\right) \right) $ is an inter-arc.}
	
	\new{The intra-arc at $l$ gives $\left( h_{l}^{\prime },h_{l}\right) \in
		\succ_{t_{l}}$, by (\ref{def-gamma}) and (\ref{Nec-condition1-part_i}). The
		inter-arc at $l$ gives, by the definition of $\mathcal{A}_{2}$, an individual
		$i_{l}\in I^{t_{l+1}}$ with}
	\begin{equation*}
		\new{\omega\left( i_{l}\right) =h_{l}^{\prime }\qquad \text{and}\qquad x\left(
			i_{l}\right) =h_{l+1}\text{.}}
	\end{equation*}\new{The individuals $i_{1},\ldots ,i_{K}$ are pairwise
		distinct: if $i_{l}=i_{l^{\prime }}$ then $t_{l+1}=t_{l^{\prime }+1}$ and
		$h_{l+1}=h_{l^{\prime }+1}$, so the vertices $\left( t_{l+1},h_{l+1}\right) $
		and $\left( t_{l^{\prime }+1},h_{l^{\prime }+1}\right) $ of $C$ coincide, and
		hence $l=l^{\prime }$.}
	
	\new{Put $S=\left\{ i_{1},\ldots ,i_{K}\right\} $ and let $x^{\prime }$ assign
		$x^{\prime }\left( i_{l}\right) =\omega\left( i_{l+1}\right) $ for each $l$.
		This is a cyclic permutation of the endowments of $S$, so $x^{\prime }$ is
		feasible for $S$. Moreover, for each $l$,}
	\begin{equation*}
		\new{x^{\prime }\left( i_{l}\right) =\omega\left( i_{l+1}\right)
			=h_{l+1}^{\prime }\text{,}\qquad \left( h_{l+1}^{\prime },h_{l+1}\right) \in
			\succ_{t_{l+1}}\text{,}\qquad x\left( i_{l}\right) =h_{l+1}\text{,}}
	\end{equation*}\new{and $\tau \left( i_{l}\right) =t_{l+1}$, so every member of
		$S$ is strictly better off. Thus $S$ strongly blocks $x$ at $\succ$,
		contradicting the assumption that $x$ is in the weak core at $\succ$.} Hence,
	$\mathcal{G}_{\gamma }$ contains no alternating cycle, and so Condition
	\ref{cond1} is satisfied.
	
	\bigskip
	
	\noindent$(\Leftarrow)$ Assume that the mixed graph $\mathcal{G}=(\mathcal{V},
	\mathcal{E},\mathcal{A})$ associated with $x$, $\omega$ and $X(x)$ satisfies
	Condition \ref{cond1}. Then there exists an orientation $\gamma :\mathcal{E}
	\rightarrow \mathcal{V}^{2}$ such that parts i)--ii) of Condition \ref{cond1}
	are satisfied.
	
	\medskip
	
	\noindent \emph{Step 1: the within-type relations.} For each $t\in T$ let
	$\succ_{t}^{\gamma }\subseteq H_{t}\times H_{t}$ be defined, for all distinct
	$h,h^{\prime }\in H_{t}$, by
	\begin{equation}
		\left( h^{\prime },h\right) \in \succ_{t}^{\gamma }\quad \Longleftrightarrow
		\quad \left( \left( t,h\right) ,\left( t,h^{\prime }\right) \right) \in \gamma
		(\mathcal{E})\cup T(\mathcal{A}_{1})\text{.}
		\label{def-succ-gamma}
	\end{equation}
	
	\begin{claim}\label{claim-linear}
		\new{For every $t\in T$, $\succ_{t}^{\gamma }$ is a strict linear order on
			$H_{t}$.}
	\end{claim}
	
	\begin{proof}
		\new{Fix $t$ and let $h\neq h^{\prime }$ in $H_{t}$, so that $\left(
			t,h\right) $ and $\left( t,h^{\prime }\right) $ are distinct vertices of
			$\mathcal{V}$. By the definition of $\mathcal{E}$, either one of $\left(
			\left( t,h\right) ,\left( t,h^{\prime }\right) \right) $, $\left( \left(
			t,h^{\prime }\right) ,\left( t,h\right) \right) $ belongs to $T(\mathcal{A}
			_{1})$, or $\left\{ \left( t,h\right) ,\left( t,h^{\prime }\right) \right\}
			\in \mathcal{E}$, in which case $\gamma $ assigns it one of its two
			directions. In either case at least one of the two ordered pairs is an arc of
			$\mathcal{G}_{\gamma }^{t}$, so $\mathcal{G}_{\gamma }^{t}$ is a complete
			digraph on $\left\{ \left( t,h\right) \mid h\in H_{t}\right\} $ and
			$\succ_{t}^{\gamma }$ is complete.}
		
		\new{By part i) of Condition \ref{cond1}, $\mathcal{G}_{\gamma }^{t}$ is
			acyclic. In particular, it contains no cycle of length two, so at most one of
			the two ordered pairs is an arc: $\mathcal{G}_{\gamma }^{t}$ is a tournament
			and $\succ_{t}^{\gamma }$ is antisymmetric. Finally, a finite acyclic
			tournament is transitive. Indeed, if $\left( h^{\prime },h\right) \in
			\succ_{t}^{\gamma }$ and $\left( h^{\prime \prime },h^{\prime }\right) \in
			\succ_{t}^{\gamma }$ but $\left( h^{\prime \prime },h\right) \notin
			\succ_{t}^{\gamma }$, completeness gives $\left( h,h^{\prime \prime }\right)
			\in \succ_{t}^{\gamma }$ and $\mathcal{G}_{\gamma }^{t}$ contains the cycle
			$\left( t,h\right) \rightarrow \left( t,h^{\prime }\right) \rightarrow \left(
			t,h^{\prime \prime }\right) \rightarrow \left( t,h\right) $, a contradiction.
			Hence $\succ_{t}^{\gamma }$ is a strict linear order on $H_{t}$.}
	\end{proof}
	
	For each $t\in T$, fix any binary relation $\succ_{t}\subseteq H\times H$ such
	that:
	\begin{itemize}
		\item[1)] \new{the restriction of $\succ_{t}$ to $H_{t}$ equals
			$\succ_{t}^{\gamma }$, that is, $\succ_{t}\cap \left( H_{t}\times H_{t}\right)
			=\succ_{t}^{\gamma }$;}
		\item[2)] for all $h^{\prime }\in H_{t}$ and all $h\in H\backslash H_{t}$, it
		holds that $\left( h^{\prime },h\right) \in \succ_{t}$;
		\item[3)] \new{the restriction of $\succ_{t}$ to $H\backslash H_{t}$, that is,
			$\succ_{t}\cap \left( \left( H\backslash H_{t}\right) \times \left(
			H\backslash H_{t}\right) \right) $, is a strict linear order.}
	\end{itemize}
	By Claim \ref{claim-linear} and clauses 1)--3), $\succ_{t}$ is a strict linear
	order on $H$ which ranks every object held by $t$ above every object not held
	by $t$. Since the choice of $t$ was arbitrary, it follows that
	$\succ=(\succ_{t})_{t\in T}$ is a profile of strict linear orderings.
	
	\new{We record the property of this construction that does the work below.}
	
	\begin{observation}\label{obs-block}
		\new{Let $t\in T$, $h\in H_{t}$ and $h^{\prime }\in H$. If $\left( h^{\prime
			},h\right) \in \succ_{t}$, then $h^{\prime }\in H_{t}$, and therefore $\left(
			\left( t,h\right) ,\left( t,h^{\prime }\right) \right) \in \gamma (\mathcal{E}
			)\cup T(\mathcal{A}_{1})$ is an intra-arc of $\mathcal{G}_{\gamma }$.}
	\end{observation}
	
	\begin{proof}
		\new{If $h^{\prime }\in H\backslash H_{t}$, clause 2) gives $\left(
			h,h^{\prime }\right) \in \succ_{t}$, contradicting antisymmetry. So
			$h^{\prime }\in H_{t}$, and the conclusion follows from clause 1) and
			(\ref{def-succ-gamma}).}
	\end{proof}
	
	\medskip
	
	\noindent \emph{Step 2: reduction to a minimal blocking cycle.} We now show
	that $x$ is in the weak core at $\succ$. Suppose, to the contrary, that some
	nonempty set $S\subseteq I$ can strongly block $x$ at $\succ$. Then there
	exists an allocation $x^{\prime }$ such that, for all $h\in H$,
	\begin{equation*}
		\left\vert \left\{ i\in S:x^{\prime }(i)=h\right\} \right\vert =\left\vert
		\left\{ i\in S:\omega(i)=h\right\} \right\vert \text{,}
	\end{equation*}and $\left( x^{\prime }\left( i\right) ,x\left( i\right)
	\right) \in \succ_{\tau \left( i\right) }$ for all $i\in S$. Since $x^{\prime
	}$ reallocates only the initial objects of agents in $S$, it permutes the
	endowments of $S$; \new{as every permutation decomposes into cycles,} we may
	assume that $S=\left\{ 1,\ldots ,K\right\} $ with
	\begin{equation}
		x^{\prime }\left( i\right) =\omega\left( i+1\right) \qquad \text{and}\qquad
		\left( \omega\left( i+1\right) ,x\left( i\right) \right) \in \succ_{\tau
			\left( i\right) }\qquad \text{for all }i\in S\text{,}
		\label{eq-cycle}
	\end{equation}indices being taken modulo $K$, so that $K+1=1$.
	
	\new{Call a set $S$ with property (\ref{eq-cycle}) a \emph{blocking cycle}. We
		take $S$ to be a blocking cycle of \emph{minimum cardinality}, which is
		possible since $I$ is finite. The following lemma is the tool that makes
		minimality bite.}
	
	\begin{shortcircuit}\label{lem-shortcircuit}
		\new{Let $S=\left\{ 1,\ldots ,K\right\} $ be a blocking cycle and suppose that
			there are $m,n\in S$ with $m\neq n$, $\tau \left( m\right) =\tau \left(
			n\right) =:t$ and $\left( \omega\left( n+1\right) ,x\left( m\right) \right)
			\in \succ_{t}$. Then $S$ contains a blocking cycle of strictly smaller
			cardinality.}
	\end{shortcircuit}
	
	\begin{proof}
		\new{For $a,b\in S$ let $\left[ a,b\right] $ denote the cyclic segment
			$a,a+1,\ldots ,b$ (indices modulo $K$), and put $S^{\prime }=\left[ n+1,m
			\right] $. Define}
		\begin{equation*}
			\new{x^{\prime \prime }\left( m\right) =\omega\left( n+1\right) \text{,}\qquad
				x^{\prime \prime }\left( l\right) =\omega\left( l+1\right) \text{ for }l\in
				\left[ n+1,m-1\right] \text{.}}
		\end{equation*}\new{The objects distributed by $x^{\prime \prime }$ are
			$\omega\left( n+1\right) ,\omega\left( n+2\right) ,\ldots ,\omega\left(
			m\right) $, which are exactly the endowments of the members of $S^{\prime }$,
			so $x^{\prime \prime }$ is feasible for $S^{\prime }$. Every $l\in \left[ n+1,
			m-1\right] $ receives the same object as under $x^{\prime }$ and therefore
			strictly improves by (\ref{eq-cycle}), and $m$ strictly improves by
			hypothesis. Hence $S^{\prime }$ is a blocking cycle. Finally $\left\vert
			S^{\prime }\right\vert =\left( m-n\right) \bmod K\in \left\{ 1,\ldots
			,K-1\right\} $, because $m\neq n$ modulo $K$, so $\left\vert S^{\prime
			}\right\vert <K$.}
	\end{proof}
	
	\medskip
	
	\noindent \emph{Step 3: the case $K=1$.} \new{Suppose $K=1$, so that
		$S=\left\{ 1\right\} $ and, by (\ref{eq-cycle}), $\left( \omega\left( 1\right)
		,x\left( 1\right) \right) \in \succ_{t}$ with $t=\tau \left( 1\right) $. Note that $x\left( 1\right) \in H_{t}$; by Observation
		\ref{obs-block}, $\omega\left( 1\right) \in H_{t}$ and}
	\begin{equation*}
		\new{\left( \left( t,x\left( 1\right) \right) ,\left( t,\omega\left( 1\right)
			\right) \right) \in \gamma (\mathcal{E})\cup T(\mathcal{A}_{1})\text{.}}
	\end{equation*}\new{On the other hand $\omega\left( 1\right) \neq x\left(
		1\right) $, so Observation \ref{obs-basic}(ii) yields $\left( \left(
		t,\omega\left( 1\right) \right) ,\left( t,x\left( 1\right) \right) \right) \in
		T(\mathcal{A}_{1})$. The two arcs form a cycle of length two in
		$\mathcal{G}_{\gamma }^{t}$, contradicting part i) of Condition \ref{cond1}.
		Hence $K\geq 2$.}
	
	\medskip
	
	\noindent \emph{Step 4: consecutive members of $S$ have distinct types.}
	\new{Suppose $\tau \left( i\right) =\tau \left( i+1\right) =:t$ for some $i\in
		S$. By Observation \ref{obs-basic}(i), $x\left( i\right) ,x\left( i+1\right)
		\in H_{t}$, and by (\ref{eq-cycle}) together with Observation \ref{obs-block},
		$\omega\left( i+1\right) \in H_{t}$. We claim that}
	\begin{equation}
		\new{\left( x\left( i+1\right) ,\omega\left( i+1\right) \right) \in
			\succeq_{t}\text{,}}
		\label{eq-step4}
	\end{equation}\new{where $\succeq_{t}$ denotes the reflexive closure of
		$\succ_{t}$. Indeed, if $\omega\left( i+1\right) =x\left( i+1\right) $ then
		(\ref{eq-step4}) holds trivially; and if $\omega\left( i+1\right) \neq x\left(
		i+1\right) $ then, since $i+1\in I^{t}$ and $\omega\left( i+1\right) \in
		H_{t}$, Observation \ref{obs-basic}(ii) gives $\left( \left( t,\omega\left(
		i+1\right) \right) ,\left( t,x\left( i+1\right) \right) \right) \in T(
		\mathcal{A}_{1})$, whence $\left( x\left( i+1\right) ,\omega\left( i+1\right)
		\right) \in \succ_{t}$ by (\ref{def-succ-gamma}) and clause 1).}
	
	\new{Combining (\ref{eq-cycle}) for $i+1$, then (\ref{eq-step4}), then
		(\ref{eq-cycle}) for $i$, and using transitivity of $\succ_{t}$,}
	\begin{equation*}
		\new{\omega\left( i+2\right) \succ_{t}x\left( i+1\right) \succeq_{t}
			\omega\left( i+1\right) \succ_{t}x\left( i\right) \text{,}\qquad \text{so}
			\qquad \left( \omega\left( i+2\right) ,x\left( i\right) \right) \in
			\succ_{t}\text{.}}
	\end{equation*}\new{Applying the Short-circuit Lemma with $m=i$ and $n=i+1$,
		which is legitimate since $K\geq 2$ and $\tau \left( i\right) =\tau \left(
		i+1\right) =t$, produces a blocking cycle of cardinality $K-1$, contradicting
		the minimality of $S$. Hence $\tau \left( i\right) \neq \tau \left( i+1\right)
		$ for every $i\in S$.}
	
	\medskip
	
	\noindent \emph{Step 5: the associated vertices are pairwise distinct.}
	\new{For $i\in S$ write $t_{i}=\tau \left( i\right) $ and}
	\begin{equation*}
		\new{a_{i}=\left( t_{i},x\left( i\right) \right) \text{,}\qquad b_{i}=\left(
			t_{i},\omega\left( i+1\right) \right) \text{.}}
	\end{equation*}\new{By Observations \ref{obs-basic}(i) and \ref{obs-block},
		both are vertices of $\mathcal{V}$. We show that the $2K$ vertices
		$a_{1},b_{1},\ldots ,a_{K},b_{K}$ are pairwise distinct.}
	
	\new{First, $a_{i}\neq b_{i}$ for every $i$, since $\left( \omega\left(
		i+1\right) ,x\left( i\right) \right) \in \succ_{t_{i}}$ forces $\omega\left(
		i+1\right) \neq x\left( i\right) $. Now let $i\neq j$. Any coincidence among
		$a_{i},b_{i},a_{j},b_{j}$ forces $t_{i}=t_{j}=:t$, because the first
		coordinate of each vertex is the type. There are three cases, and each
		supplies the hypothesis of the Short-circuit Lemma, contradicting minimality:}
	\begin{itemize}
		\item \new{$a_{i}=a_{j}$, that is, $x\left( i\right) =x\left( j\right) $. Then
			$(\omega(j+1),x(i))=(\omega(j+1),x(j))\in \succ_{t}$; apply the Lemma with
			$m=i$ and $n=j$.}
		\item \new{$b_{i}=b_{j}$, that is, $\omega\left( i+1\right) =\omega\left(
			j+1\right) $. Then $(\omega(j+1),x(i))=(\omega(i+1),x(i))\in \succ_{t}$;
			apply the Lemma with $m=i$ and $n=j$.}
		\item \new{$a_{i}=b_{j}$, that is, $x\left( i\right) =\omega\left( j+1\right)
			$. Then $\omega\left( i+1\right) \succ_{t}x\left( i\right) =\omega\left(
			j+1\right) \succ_{t}x\left( j\right) $, so $\left( \omega\left( i+1\right)
			,x\left( j\right) \right) \in \succ_{t}$ by transitivity; apply the Lemma with
			$m=j$ and $n=i$.}
	\end{itemize}
	\new{Hence all $2K$ vertices are distinct.}
	
	\medskip
	
	\noindent \emph{Step 6: exhibiting the alternating cycle.} \new{Fix $i\in S$.
		By (\ref{eq-cycle}), $\left( \omega\left( i+1\right) ,x\left( i\right) \right)
		\in \succ_{t_{i}}$ with $x\left( i\right) \in H_{t_{i}}$, so Observation
		\ref{obs-block} gives the intra-arc}
	\begin{equation*}
		\new{a_{i}=\left( t_{i},x\left( i\right) \right) \longrightarrow \left(
			t_{i},\omega\left( i+1\right) \right) =b_{i}\text{.}}
	\end{equation*}\new{Next, $\omega\left( i+1\right) \in H_{t_{i}}$, the
		individual $i+1$ belongs to $I^{t_{i+1}}$ with $\omega\left( i+1\right) $ as
		endowment and $x\left( i+1\right) $ as assignment, and $t_{i+1}\neq t_{i}$ by
		Step 4. The defining requirement of $\mathcal{A}_{2}$ is therefore met, and we
		obtain the inter-arc}
	\begin{equation*}
		\new{b_{i}=\left( t_{i},\omega\left( i+1\right) \right) \longrightarrow \left(
			t_{i+1},x\left( i+1\right) \right) =a_{i+1}\text{.}}
	\end{equation*}\new{Concatenating over $i=1,\ldots ,K$ yields the closed walk}
	\begin{equation*}
		\new{\left\langle a_{1},b_{1},a_{2},b_{2},\ldots ,a_{K},b_{K}\right
			\rangle}
	\end{equation*}\new{in $\mathcal{G}_{\gamma }$, in which intra-arcs and
		inter-arcs strictly alternate. By Step 5 its $2K$ vertices are pairwise
		distinct, so it is a cycle, and it contains neither two consecutive intra-arcs
		nor two consecutive inter-arcs: it is an alternating cycle. This contradicts
		part ii) of Condition \ref{cond1}.}
	
	\new{Therefore no nonempty $S\subseteq I$ strongly blocks $x$ at $\succ$; that
		is, $x$ lies in the weak core for $\succ$.} Hence, $x$ is WC-rationalizable.
\end{proof}

	\section{Proofs of Lemmas \ref{lem-cp-profile} and \ref{lem-weakcore-dag} and Theorem \ref{Th-SCWC}}\label{AppendixLemmas}

\subsection{Proof of Lemma \ref{lem-cp-profile}}\label{app-proof-cp}

\begin{proof}
		Recall (Definition~\ref{def-sc-profile}) that a profile is single-crossing with respect to $\rhd
		_{T}$ and $\rhd _{H}$ when, for every pair $h\rhd _{H}h^{\prime }$, the set of types preferring the
		higher-quality object,
		\begin{equation*}
			U(h,h^{\prime }):=\{t\in T:h\succ _{t}h^{\prime }\}\text{,}
		\end{equation*}
		is an \emph{up-set} of $\rhd _{T}$: if $h\succ _{t}h^{\prime }$ and $t^{\prime }\rhd _{T}t$ then
		$h\succ _{t^{\prime }}h^{\prime }$. Since $\rhd _{T}$ is a linear order on the finite set $T$, its up-sets are
		exactly the \emph{top blocks} $\{t:\operatorname{rk}(t)\leq
		k\}$, one for each $k\in \{0,1,\ldots ,\left\vert T\right\vert \}$; in
		particular a top block is determined by, and determines, its size $k$. This
		correspondence between up-sets and thresholds is what links the two
		descriptions, and it is used repeatedly below.
		
		\medskip\noindent\emph{Consistent $\kappa $ yields a single-crossing profile.}
		Fix a type $t$ and write $p=\operatorname{rk}(t)$. The relation
		$\succ_{t}^{\kappa }$ is complete and antisymmetric by construction, so it
		is a linear order if and only if it is acyclic---equivalently, since $\succ_{t}^{\kappa }$ is complete, if and only if it is transitive---and any cycle must sit inside a
		triple $h_{a}\rhd _{H}h_{b}\rhd _{H}h_{c}$. There $h_{a}\succ _{t}h_{b}$ iff $p\leq \kappa (h_{a},h_{b})$, and likewise for the other two pairs; there is a cycle for some position $p$ exactly when $\kappa (h_{a},h_{c})$ falls strictly outside $[\min \{\kappa (h_{a},h_{b}),\kappa (h_{b},h_{c})\},\max \{\kappa (h_{a},h_{b}),\kappa (h_{b},h_{c})\}]$. Consistency forbids this for every
		triple, so each $\succ_{t}^{\kappa }$ is a linear order. Moreover the
		construction makes $U(h,h^{\prime })=\{t:\operatorname{rk}(t)\leq \kappa (h,h^{\prime })\}$ a
		top block for each pair, so $\succ^{\kappa }$ is single-crossing. Conversely, if
		$\kappa $ is not consistent, then the displayed equivalence produces a type whose
		ranking cycles on some triple, so $\succ^{\kappa }$ is not even a profile of linear orders; this
		proves the first assertion in both directions.
		
		\medskip\noindent\emph{Every single-crossing profile arises from a unique
			consistent $\kappa $.} Let $\succ=(\succ_{t})_{t\in T}$ be a
		single-crossing profile compatible with the two orders. Define a candidate
		crossing-point function by \emph{counting}, for each pair $h\rhd _{H}h^{\prime }$, how many types put the higher-quality object
		first:
		\begin{equation*}
			\kappa (h,h^{\prime }):=\left\vert U(h,h^{\prime })\right\vert =\#\{t:h\succ _{t}h^{\prime }\}\text{.}
		\end{equation*}
		Note $0\leq \kappa (h,h^{\prime })\leq \left\vert T\right\vert $, so $\kappa $ is a crossing-point function. We show, in turn, that
		$\kappa $ reproduces $\succ$, that it is consistent, and that it is the only
		crossing-point function reproducing $\succ$.
		
		\emph{Step 1: $\succ^{\kappa }=\succ$.} Fix a pair $h\rhd _{H}h^{\prime }$ and put $k=\kappa (h,h^{\prime })$. By
		single-crossing $U(h,h^{\prime })$ is an up-set of $\rhd _{T}$, and the only up-set of size $k$ is the top block $\{t:\operatorname{rk}(t)\leq k\}$; hence $U(h,h^{\prime
		})=\{t:\operatorname{rk}(t)\leq k\}$. Therefore, for every type
		$t$,
		\begin{equation*}
			h\succ _{t}h^{\prime }\iff t\in U(h,h^{\prime })\iff \operatorname{rk}(t)\leq k=\kappa (h,h^{\prime })\text{,}
		\end{equation*}
		and the right-hand side is exactly the rule defining $\succ_{t}^{\kappa }$
		on the pair $\{h,h^{\prime }\}$. Thus $\succ_{t}$ and $\succ_{t}^{\kappa }$ deliver the same verdict on every pair of objects;
		being complete and antisymmetric, they coincide, and as $t$ was
		arbitrary, $\succ=\succ^{\kappa }$.
		
		\emph{Step 2: $\kappa $ is consistent.} Since $\succ=\succ^{\kappa }$ is a profile of
		linear orders, each $\succ_{t}$ is transitive, and we read consistency off
		transitivity. Fix $h_{a}\rhd _{H}h_{b}\rhd _{H}h_{c}$ and abbreviate $k_{ab}=\kappa (h_{a},h_{b})$,
		$k_{bc}=\kappa (h_{b},h_{c})$, $k_{ac}=\kappa (h_{a},h_{c})$. For the upper bound, suppose toward a contradiction that
		$k_{ac}>\max \{k_{ab},k_{bc}\}$ and take the type $t$ with
		$\operatorname{rk}(t)=\max \{k_{ab},k_{bc}\}+1$. From $\operatorname{rk}(t)\leq k_{ac}$ we get $h_{a}\succ _{t}h_{c}$,
		while $\operatorname{rk}(t)>k_{ab}$ and $\operatorname{rk}(t)>k_{bc}$ give $h_{b}\succ _{t}h_{a}$ and $h_{c}\succ _{t}h_{b}$; transitivity then forces $h_{c}\succ _{t}h_{a}$, contradicting $h_{a}\succ _{t}h_{c}$. For the lower bound, suppose $k_{ac}<\min \{k_{ab},k_{bc}\}$
		and take $t$ with $\operatorname{rk}(t)=k_{ac}+1$. From
		$\operatorname{rk}(t)\leq k_{ab}$ and $\operatorname{rk}(t)\leq
		k_{bc}$ we get $h_{a}\succ _{t}h_{b}$ and $h_{b}\succ _{t}h_{c}$, so transitivity gives $h_{a}\succ _{t}h_{c}$ and hence $\operatorname{rk}(t)\leq
		k_{ac}$, contradicting $\operatorname{rk}(t)=k_{ac}+1$. Therefore $\min
		\{k_{ab},k_{bc}\}\leq k_{ac}\leq \max \{k_{ab},k_{bc}\}$, which is consistency.
		
		\emph{Step 3: uniqueness.} Suppose $\kappa ^{\prime }$ is any crossing-point
		function with $\succ^{\kappa ^{\prime }}=\succ$. For each pair $h\rhd _{H}h^{\prime }$, the definition of $\succ^{\kappa ^{\prime }}$ gives $U(h,h^{\prime })=\{t:\operatorname{rk}(t)\leq
		\kappa ^{\prime }(h,h^{\prime })\}$, a top block of size
		$\kappa ^{\prime }(h,h^{\prime })$. As a top block is
		determined by its size, $\kappa ^{\prime }(h,h^{\prime
		})=\left\vert U(h,h^{\prime })\right\vert =\kappa (h,h^{\prime })$. Hence $\kappa ^{\prime }=\kappa $, and $\kappa $ is the
		unique consistent crossing-point function with $\succ^{\kappa }=\succ$.
	\end{proof}

\subsection{Proof of Lemma \ref{lem-weakcore-dag}}\label{app-proof-dag}

\begin{proof}
		A coalition strongly blocks $x$ exactly when it can reallocate its members'
		endowments so that every member is strictly better off. Such a reallocation is
		a permutation of the coalition's endowments and decomposes into cycles; in each
		cycle every agent receives the endowment of the next and strictly improves, so
		the cycle is a directed cycle of $D(\succ)$. Conversely, any directed cycle of $D(\succ)$ visits distinct agents, and assigning each the endowment of its successor
		is a feasible blocking reallocation. A self-loop is the singleton block in
		which an agent prefers its own endowment to $x$, i.e. a violation of individual
		rationality. Hence $x$ is in the weak core at $\succ$ if and only if $D(\succ)$ has no
		directed cycle.
	\end{proof}

\subsection{Proof of Theorem \ref{Th-SCWC}}\label{app-proof-SCWC}

\begin{proof}
		By Lemma~\ref{lem-cp-profile} the single-crossing profiles compatible with the
		two orders are exactly the profiles $\succ^{\kappa }$ for consistent $\kappa $, and
		by Lemma~\ref{lem-weakcore-dag} such a profile places $x$ in the weak core if
		and only if $D(\succ^{\kappa })$ is acyclic. Single-crossing WC-rationalizability
		asks for the existence of a single-crossing profile placing $x$ in the weak
		core, which is therefore equivalent to Condition~\ref{cond2}.
	\end{proof}

\paragraph*{Data and code availability.} The CORE social housing lettings
	data are Crown copyright, accessed through the UK Data Service under a
	Special License agreement, and cannot be redistributed. The complete
	construction and testing pipeline --- the scripts that split the raw
	lettings into local-authority markets, build the reallocation datasets
	and the $100$ pairing draws, run the PI, WC, and single-crossing tests --- is available from the authors.

	\bibliographystyle{elsarticle-harv}
	\bibliography{Bibliography}

@inproceedings{abraham2004pareto,
	title={Pareto optimality in house allocation problems},
	author={Abraham, David J and Cechl{\'a}rov{\'a}, Katar{\'\i}na and Manlove, David F and Mehlhorn, Kurt},
	booktitle={International Symposium on Algorithms and Computation},
	pages={3--15},
	year={2004},
	organization={Springer}
}

@article{shapley1974cores,
	title={On cores and indivisibility},
	author={Shapley, Lloyd and Scarf, Herbert},
	journal={Journal of mathematical economics},
	volume={1},
	number={1},
	pages={23--37},
	year={1974},
	publisher={Elsevier}
}

@article{roth2004kidney,
  title={Kidney exchange},
  author={Roth, Alvin E and S{\"o}nmez, Tayfun and {\"U}nver, M Utku},
  journal={The Quarterly journal of economics},
  volume={119},
  number={2},
  pages={457--488},
  year={2004},
  publisher={MIT Press}
}

@article{Teytelboym,
Author = {Delacrétaz, David and Kominers, Scott Duke and Teytelboym, Alexander},
Title = {Matching Mechanisms for Refugee Resettlement},
Journal = {American Economic Review},
Volume = {113},
Number = {10},
Year = {2023},
Month = {October},
Pages = {2689-2717},
DOI = {10.1257/aer.20210096},
URL = {https://www.aeaweb.org/articles?id=10.1257/aer.20210096}}

@article{hylland1979efficient,
	title={The efficient allocation of individuals to positions},
	author={Hylland, Aanund and Zeckhauser, Richard},
	journal={Journal of Political economy},
	volume={87},
	number={2},
	pages={293--314},
	year={1979},
	publisher={The University of Chicago Press}
}

@article{deng1994complexity,
	title={On the complexity of cooperative solution concepts},
	author={Deng, Xiaotie and Papadimitriou, Christos H.},
	journal={Mathematics of Operations Research},
	volume={19},
	number={2},
	pages={257--266},
	year={1994}
}

@article{fekete2003complexity,
	title={The complexity of economic equilibria for house allocation markets},
	author={Fekete, S{\'a}ndor P and Skutella, Martin and Woeginger, Gerhard J},
	journal={Information processing letters},
	volume={88},
	number={5},
	pages={219--223},
	year={2003},
	publisher={Elsevier}
}

@article{roth1977weak,
	title={Weak versus strong domination in a market with indivisible goods},
	author={Roth, Alvin E and Postlewaite, Andrew},
	journal={Journal of Mathematical Economics},
	volume={4},
	number={2},
	pages={131--137},
	year={1977},
	publisher={Elsevier}
}

@article{roth1982incentive,
	title={Incentive compatibility in a market with indivisible goods},
	author={Roth, Alvin E},
	journal={Economics letters},
	volume={9},
	number={2},
	pages={127--132},
	year={1982},
	publisher={Elsevier}
}

@article{zhou1990conjecture,
	title={On a conjecture by Gale about one-sided matching problems},
	author={Zhou, Lin},
	journal={Journal of Economic Theory},
	volume={52},
	number={1},
	pages={123--135},
	year={1990},
	publisher={Elsevier}
}

@article{abdulkadirouglu1998random,
	title={Random serial dictatorship and the core from random endowments in house allocation problems},
	author={Abdulkadiro{\u{g}}lu, Atila and S{\"o}nmez, Tayfun},
	journal={Econometrica},
	volume={66},
	number={3},
	pages={689--701},
	year={1998},
	publisher={JSTOR}
}

@article{abdulkadirouglu1999house,
	title={House allocation with existing tenants},
	author={Abdulkadiro{\u{g}}lu, Atila and S{\"o}nmez, Tayfun},
	journal={Journal of Economic Theory},
	volume={88},
	number={2},
	pages={233--260},
	year={1999},
	publisher={Elsevier}
}

@article{yuan1996residence,
	title={Residence exchange wanted: a stable residence exchange problem},
	author={Yuan, Yufei},
	journal={European Journal of Operational Research},
	volume={90},
	number={3},
	pages={536--546},
	year={1996},
	publisher={Elsevier}
}

@inproceedings{saban2013house,
	title={House allocation with indifferences: a generalization and a unified view},
	author={Saban, Daniela and Sethuraman, Jay},
	booktitle={Proceedings of the fourteenth ACM conference on Electronic Commerce},
	pages={803--820},
	year={2013}
}

@inproceedings{Plaxton,
  title={A simple family of top trading cycles mechanisms for housing markets with indifferences},
  author={Plaxton, C Gregory},
  booktitle={Proceedings of the 24th international conference on game theory},
  pages={1--23},
  year={2013}
}

@article{Alcalde,
  title={Exchange of indivisible goods and indifferences: The top trading absorbing sets mechanisms},
  author={Alcalde-Unzu, Jorge and Molis, Elena},
  journal={Games and Economic Behavior},
  volume={73},
  number={1},
  pages={1--16},
  year={2011},
  publisher={Elsevier}
}

@inproceedings{Aziz,
  title={Housing markets with indifferences: A tale of two mechanisms},
  author={Aziz, Haris and De Keijzer, Bart},
  booktitle={Proceedings of the 26th AAAI Conference on Artificial Intelligence},
  volume={26},
  pages={1249--1255},
  year={2012}
}

@article{Jaramillo,
  title={The difference indifference makes in strategy-proof allocation of objects},
  author={Jaramillo, Paula and Manjunath, Vikram},
  journal={Journal of Economic Theory},
  volume={147},
  number={5},
  pages={1913--1946},
  year={2012},
  publisher={Elsevier}
}

@article{saban2015complexity,
  title={The complexity of computing the random priority allocation matrix},
  author={Saban, Daniela and Sethuraman, Jay},
  journal={Mathematics of Operations Research},
  volume={40},
  number={4},
  pages={1005--1014},
  year={2015},
  publisher={INFORMS}
}

@article{biro2021complexity,
  title={Complexity of finding Pareto-efficient allocations of highest welfare},
  author={Bir{\'o}, P{\'e}ter and Gudmundsson, Jens},
  journal={European Journal of Operational Research},
  volume={291},
  number={2},
  pages={614--628},
  year={2021},
  publisher={Elsevier}
}

@article{sonmez1999strategy,
  title={Strategy-proofness and essentially single-valued cores},
  author={S{\"o}nmez, Tayfun},
  journal={Econometrica},
  volume={67},
  number={3},
  pages={677--689},
  year={1999},
  publisher={JSTOR}
}

@book{biro2022balanced,
  title={Balanced exchange in a multi-unit Shapley-Scarf market},
  author={Bir{\'o}, P{\'e}ter and Klijn, Flip and P{\'a}pai, Szilvia},
  year={2022},
  publisher={BSE, Barcelona School of Economics}
}

@article{ma1994strategy,
  title={Strategy-proofness and the strict core in a market with indivisibilities},
  author={Ma, Jinpeng},
  journal={International Journal of Game Theory},
  volume={23},
  pages={75--83},
  year={1994},
  publisher={Springer}
}

@book{manlove2013algorithmics,
  title={Algorithmics of matching under preferences},
  author={Manlove, David},
  volume={2},
  year={2013},
  publisher={World Scientific}
}

@book{gareyjohnson1979,
  title={Computers and Intractability: A Guide to the Theory of NP-Completeness},
  author={Garey, Michael R. and Johnson, David S.},
  year={1979},
  publisher={W. H. Freeman; First Edition (January 15, 1979)},
  address={USA}
}

@article{biro2010size,
  title={Size versus stability in the marriage problem},
  author={Bir{\'o}, P{\'e}ter and Manlove, David F and Mittal, Shubham},
  journal={Theoretical Computer Science},
  volume={411},
  number={16-18},
  pages={1828--1841},
  year={2010},
  publisher={Elsevier}


}

@article{compte2008voluntary,
  title={Voluntary participation and re-assignment in two-sided matching},
  author={Compte, Olivier and Jehiel, Philippe},
  journal={Unpublished. http://www. enpc. fr/ceras/compte/matchingparticipation. pdf},
  year={2008},
  publisher={Citeseer}
}

@article{tai2022revealed,
  title={Revealed Preferences of One-Sided Matching},
  author={Tai, Andrew},
  journal={arXiv preprint arXiv:2210.14388},
  year={2022}
}

@article{gale1962college,
  title={College admissions and the stability of marriage},
  author={Gale, David and Shapley, Lloyd S},
  journal={The American Mathematical Monthly},
  volume={69},
  number={1},
  pages={9--15},
  year={1962},
  publisher={Taylor \& Francis}
}

@article{demuynck2022revealed,
  title={On the revealed preference analysis of stable aggregate matchings},
  author={Demuynck, Thomas and Salman, Umutcan},
  journal={Theoretical Economics},
  volume={17},
  number={4},
  pages={1651--1682},
  year={2022},
  publisher={Wiley Online Library}
}

@article{echenique2013revealed,
  title={The revealed preference theory of stable and extremal stable matchings},
  author={Echenique, Federico and Lee, Sangmok and Shum, Matthew and Yenmez, M Bumin},
  journal={Econometrica},
  volume={81},
  number={1},
  pages={153--171},
  year={2013},
  publisher={Wiley Online Library}
}

@article{szpilrajn1930extension,
  title={Sur l'extension de l'ordre partiel},
  author={Szpilrajn, Edward},
  journal={Fundamenta mathematicae},
  volume={16},
  number={1},
  pages={386--389},
  year={1930},
  publisher={Polska Akademia Nauk. Instytut Matematyczny PAN}
}

@article{bachmann2006testable,
  title={Testable Implications of Pareto Efficiency and Individual Rationality},
  author={Bachmann, Ruediger},
  journal={Economic Theory},
  volume={29},
  number={3},
  pages={489--504},
  year={2006},
  publisher={Springer}
}

@article{chambers2025empirical,
  title={Empirical welfare economics},
  author={Chambers, Christopher P and Echenique, Federico},
  journal={Theoretical Economics},
  volume={20},
  number={3},
  pages={911--939},
  year={2025}
}

@article{choo2006marries,
  title={Who marries whom and why},
  author={Choo, Eugene and Siow, Aloysius},
  journal={Journal of political Economy},
  volume={114},
  number={1},
  pages={175--201},
  year={2006},
  publisher={The University of Chicago Press}
}

@article{afriat1967construction,
  title={The construction of utility functions from expenditure data},
  author={Afriat, Sydney N},
  journal={International economic review},
  volume={8},
  number={1},
  pages={67--77},
  year={1967},
  publisher={JSTOR}
}

@article{duggan1999general,
  title={A general extension theorem for binary relations},
  author={Duggan, John},
  journal={Journal of Economic Theory},
  volume={86},
  number={1},
  pages={1--16},
  year={1999},
  publisher={Elsevier}
}

@InCollection{karp2009reducibility,
  author    = {Richard M. Karp},
  title     = {Reducibility among combinatorial problems},
  booktitle = {Complexity of Computer Computations},
  editor    = {Raymond E. Miller and James W. Thatcher},
  publisher = {Plenum Press},
  address   = {New York},
  year      = {1972},
  pages     = {85--103}
}

@article{zhang2026revealed,
  title={Revealed preference tests of stability in aggregate matching with single-peaked preferences},
  author={Zhang, Tina D and Carvajal, Andr{\'e}s},
  journal={Economic Theory},
  pages={1--30},
  year={2026},
  publisher={Springer}
}

@article{galichon2022cupid,
  title={Cupid's Invisible Hand: Social Surplus and Identification in Matching Models},
  author={Galichon, Alfred and Salani{\'e}, Bernard},
  journal={The Review of Economic Studies},
  volume={89},
  number={5},
  pages={2600--2629},
  year={2022},
  publisher={Oxford University Press}
}

@article{fox2018estimating,
  title={Estimating matching games with transfers},
  author={Fox, Jeremy T},
  journal={Quantitative Economics},
  volume={9},
  number={1},
  pages={1--38},
  year={2018},
  publisher={Wiley Online Library}
}

@article{agarwal2015empirical,
  title={An Empirical Model of the Medical Match},
  author={Agarwal, Nikhil},
  journal={American Economic Review},
  volume={105},
  number={7},
  pages={1939--1978},
  year={2015}
}

@article{chiappori2016econometrics,
  title={The Econometrics of Matching Models},
  author={Chiappori, Pierre-Andr{\'e} and Salani{\'e}, Bernard},
  journal={Journal of Economic Literature},
  volume={54},
  number={3},
  pages={832--866},
  year={2016}
}

@article{ekici2024pair,
  title={Pair-efficient reallocation of indivisible objects},
  author={Ekici, {\"O}zg{\"u}n},
  journal={Theoretical Economics},
  volume={19},
  number={2},
  pages={551--564},
  year={2024}
}

@article{goel2026ttc,
  title={{TTC} domains},
  author={Goel, Sumit and Tamura, Yuki},
  journal={working paper},
  year={2026}
}

@article{echenique2008matchings,
  author  = {Echenique, Federico},
  title   = {What Matchings Can Be Stable? The Testable Implications of Matching Theory},
  journal = {Mathematics of Operations Research},
  volume  = {33},
  number  = {3},
  pages   = {757--768},
  year    = {2008},
  doi     = {10.1287/moor.1080.0318}
}

@article{de2023empirical,
  title={Empirical content of classic assignment methods: jungle and market economy: G. de Clippel, K. Rozen},
  author={de Clippel, Geoffroy and Rozen, Kareen},
  journal={Economic Theory},
  volume={76},
  number={3},
  pages={813--825},
  year={2023},
  publisher={Springer}
}

@article{rodriguez2024school,
  title={School choice with transferable student characteristics},
  author={Rodr{\'\i}guez-{\'A}lvarez, Carmelo and Romero-Medina, Antonio},
  journal={Games and Economic Behavior},
  volume={143},
  pages={103--124},
  year={2024},
  publisher={Elsevier}
}

@article{hu2020revealed,
  title={The revealed preference theory of stable matchings with one-sided preferences},
  author={Hu, Gaoji and Li, Jiangtao and Tang, Rui},
  journal={Games and Economic Behavior},
  volume={124},
  pages={305--318},
  year={2020},
  publisher={Elsevier}
}
	
\end{document}


\maketitle

This supplementary appendix contains material omitted from the main text.
Sections S1--S3 concern the data: the full construction of the market-level
allocation instances, a worked example of how iterative proportional fitting,
rounding and pairing operate together, and a placebo experiment showing that
individuals cannot be tracked in the CORE data. Section \ref{S-proof-DPWC2}
proves Theorem \ref{DP-Rationalizability-WC-2} and Section \ref{S-mip} gives the
mixed-integer formulation of Condition \ref{cond2} summarized in the main text.
All other proofs are in the appendices of the paper. Section \ref{S-relations} establishes
that PI- and WC-rationalizability are logically independent and that their
conjunction is weaker than joint rationalizability (Propositions
\ref{prop-nonnested}--\ref{prop-joint}), and Section \ref{S-examples}
collects the five examples cited in the main text as Examples
\ref{example-sc}--\ref{ex-unheld}. References to theorems, conditions and tables point to the
main text. Notation follows the main text throughout.

\section{Construction of the market-level allocation instances}\label{S1}

This section gives in full the construction summarized in Section 5 of the
main text.

Our tests need, for each market, an aggregate allocation instance
$\langle I, H, q, \tau, T, x, \omega \rangle$: a population of
individuals, a set of objects with their copies, a grouping of
individuals into types, an initial allocation, and an observed final
allocation. The lettings data do not come in this form --- each row is a
single letting event --- so we build every market's instance in five
steps: define agent types and objects (Step 1), split the year into an initial
and a final period (Step 2), restrict the sample and select the
analyzable markets (Step 3), match the two periods into one population
(Step 4), and fix the granularity of the type structure (Step 5).

\textbf{Step 1 -- Agent types and objects.}
For each letting we construct an agent type and a dwelling profile.
Agent types combine seven characteristics: age band, gender, ethnic
origin, nationality, economic status, household composition, and a
vulnerability indicator equal to one if the household reports a long-term
illness or requires wheelchair access. The age of the lead tenant is
grouped into nine bands: up to 19, 20--25, 26--30, 31--35, 36--40, 41--45,
46--50, 51--64, and 65 and over. The remaining characteristics are the CORE
categories as recorded: gender ($4$ values), ethnic origin ($19$), nationality
($17$), and economic status ($10$, such as full-time work, part-time work,
jobseeker, retired, student, or unable to work through long-term sickness);
household composition takes $7$ values (single adult, two adults, single elder,
two adults including an elder, a single adult with one or more children, two or
more adults with one or more children, and other), and so distinguishes whether
a household has children but not how many. Each of these characteristics is
unlikely to change within six months, so we treat an agent type as a stable
attribute over the two half-year periods. The agent types
define the type
function $\tau$ and the set of types $T$. Dwelling profiles combine four
characteristics: number of bedrooms, building type (purpose-built or converted
from another property), unit type (house, flat or maisonette, bungalow, bedsit,
shared house, shared flat or maisonette, or other), and wheelchair
accessibility --- geography no longer enters the profile,
because a market \emph{is} a single local authority. Each distinct
dwelling profile is one object $h \in H$, and $q_{h}$ counts its lettings
in the market. A small number of records with a missing number of
bedrooms are completed by sampling from the market's own empirical
distribution of that variable.

\textbf{Step 2 -- Two allocation periods.}
We partition the 2021--22 lettings year into two six-month periods: an
initial period $H_{1}$ covering April--September 2021 and a final period
$H_{2}$ covering October 2021--March 2022. The $H_{1}$ allocation serves
as the initial endowments $\omega$ and the $H_{2}$ allocation serves as
the observed allocation $x$. 

\textbf{Step 3 -- Sample restrictions and market selection.}
Within each market we impose two restrictions. \textit{Restriction 1.} CORE records whether each letting is to a new
tenant or to an existing tenant. We use this distinction: for $H_{1}$ we keep
only new tenants, who enter social housing for the first time, and for $H_{2}$
only existing tenants, who made a voluntary internal transfer. \textit{Restriction 2.} An agent type or
an object observed in only one of the two periods cannot contribute an
initial--final comparison, so we retain only agent types and objects that
appear in both restricted samples, repeating the intersection until no
further agent type or object drops out. Finally, we require an analyzable
market to contain at least $50$ individuals after these restrictions.

The market selection is itself informative about the institutional
landscape. Of the $309$ local authorities in the data, $154$ have no
general needs lettings of their own --- these are authorities that
transferred their housing stock to housing associations --- and a further
$137$ are too small once the restrictions are applied ($96$ below the
size threshold, $41$ with no overlap between the two periods). This
leaves $\mathbf{18}$ analyzable markets, comprising $2{,}266$ individuals
(between $50$ and $273$ per market), between $6$ and $15$ dwelling
profiles per market, and $1{,}783$ recorded transfers in $H_{2}$.

\textbf{Step 4 -- Matching the two periods.}
The data record two separate cross-sections, not the same individuals
twice: the $H_{1}$ lettings show where each agent type \emph{starts},
and the $H_{2}$ lettings show where agent types \emph{end up}
after transferring. To turn the two cross-sections into one reallocation
--- the same population of individuals, each with an initial and a final
dwelling --- we proceed in four small steps, separately in each market.

\begin{itemize}
\item[(i)] \textit{Count.} For each period we build a table counting, for
every agent type $t$ and dwelling profile $h$, how many lettings of that
combination occurred.
\item[(ii)] \textit{Align the totals.} It is likely that the two periods do not contain the
same number of lettings per agent type or per dwelling profile. We therefore
rescale the $H_{2}$ table proportionally --- alternating between rows and
columns, a procedure known as Iterative Proportional Fitting (IPF) --- until
each agent type and each dwelling profile has (approximately) the same total as
in $H_{1}$. We then round the rescaled entries to integers, row by row,
so that every type has exactly as many final positions as it has
individuals.
\item[(iii)] \textit{Pair initial and final dwellings.} Within each agent type,
individuals whose dwelling profile appears in both periods are treated as
staying where they are; the remaining individuals of the agent type are matched
to the remaining final positions at random.
\item[(iv)] \textit{Balance.} The framework requires the final allocation
to be a redistribution of the initial dwellings: every dwelling profile
$h$ must house exactly $q_{h}$ individuals at the start and at the end.
Small discrepancies remain after step (iii), and we eliminate them by
reassigning the final dwellings of the affected individuals.
\end{itemize}

Section \ref{appendix:ipf} works through steps (i)--(iv) on a small
numerical example. Step (iii) involves randomness, so each constructed dataset is one draw
of a random matching. Rather than committing to a single draw, we
generate $100$ independent draws per market, with deterministic seeds for
reproducibility, and report how the verdicts and the indices vary across
draws. Two accounting measures describe how much of each dataset is
anchored in recorded moves. First, in the median market $52\%$ of final
dwellings are backed by a recorded $H_{2}$ transfer of the same type and
dwelling profile (the share ranges from $37\%$ to $82\%$ across markets);
the remainder is imputed by the proportional fitting. Second, the
balancing step (iv) reassigns a median of $10\%$ of final dwellings
(ranging from $0\%$ to $50\%$; the one market at $50\%$, Brent, is also
the smallest).

\textbf{Step 5 -- Type partitions.}
How finely should individuals be divided into types? The homogeneity
assumption --- individuals of the same type share the same preferences
--- is more credible when types are fine, but fine types leave the data
with little to say. Rather than committing to one choice, we run every
test at three levels of granularity. Two individuals share a type if and
only if they agree on every characteristic used by the partition:

\begin{itemize}
\item[] \textit{Coarsest} ($6$--$35$ types per market): age band and
household composition;
\item[] \textit{Middle} ($6$--$77$ types per market): age band, household
composition, economic status, and vulnerability;
\item[] \textit{Finest} ($6$--$99$ types per market): all seven
characteristics of Step 1.
\end{itemize}

Age band and household composition anchor the coarsest partition because
they are the demographics most directly tied to housing need and are
recorded for every letting; the middle level adds the two remaining
need-related characteristics, and the finest adds the identity
characteristics. The three partitions are nested --- each coarser type is a union of finer
ones --- and, within each market and draw, the individuals, the
dwellings, and the two allocations are identical across them. Any
difference in the results is therefore due to the type structure alone.

\section{A worked example of the dataset construction}\label{appendix:ipf}

This appendix illustrates, on a miniature market, how the two
half-years of lettings are turned into one reallocation dataset ---
that is, steps (i)--(iv) of the matching procedure in
Section~\ref{UKapplication}. The miniature market has two types,
$t_{1}$ and $t_{2}$, and three dwelling profiles, $h_{1}$, $h_{2}$, and
$h_{3}$. We follow the four steps in order, and along the way each of
the devices used in the construction --- IPF,
rounding, pairing, balancing --- appears once and does some visible
work.

\emph{Step (i): count.} The raw data consist of two separate
cross-sections. The initial period contributes eight lettings and the
final period six transfers, and all the construction ever uses from
them are the two tables counting how many lettings of each
type--profile combination occurred:
\begin{equation*}
	M_{1}=
	\begin{array}{c|ccc|c}
		& h_{1} & h_{2} & h_{3} & \\ \hline
		t_{1} & 2 & 1 & 1 & 4 \\
		t_{2} & 1 & 2 & 1 & 4 \\ \hline
		& 3 & 3 & 2 &
	\end{array}
	\qquad
	M_{2}=
	\begin{array}{c|ccc|c}
		& h_{1} & h_{2} & h_{3} & \\ \hline
		t_{1} & 1 & 1 & 1 & 3 \\
		t_{2} & 1 & 1 & 1 & 3 \\ \hline
		& 2 & 2 & 2 &
	\end{array}
\end{equation*}
Reading $M_{1}$: two individuals of type $t_{1}$ started on profile
$h_{1}$, one on $h_{2}$, one on $h_{3}$, and so on. The row totals say
that each type starts with four individuals, so the market has $N=8$
individuals in total, and the column totals give the object profile:
three dwellings of profile $h_{1}$, three of $h_{2}$, and two of
$h_{3}$. The table $M_{2}$ records where transferring tenants of each
type ended up during the second half-year. Notice that its totals do
not match those of $M_{1}$: there are only six transfer records for
eight individuals, and the transfers are spread evenly over the three
profiles while the initial stock is not. This mismatch is not a flaw
of the example but the normal situation in the data, and repairing it
is precisely the job of the next step.

\emph{Step (ii): align the totals.} The final allocation we are
building must be comparable with the initial one: every type must end
the year with exactly as many individuals as it started with, and every
profile must house at the end exactly as many individuals as it did at
the beginning. We therefore rescale $M_{2}$ until its row and column
totals agree with those of $M_{1}$, using iterative proportional
fitting, which alternates between two elementary moves. The row move
brings the row totals into line: type $t_{1}$ has three transfer
records but must account for four individuals, so its row is multiplied
by $4/3$, and the same happens to $t_{2}$; after this move every entry
of the table equals $4/3$. The column move then does the same for the
columns: the column of $h_{1}$ now sums to $8/3$ but must sum to $3$,
so it is multiplied by $3\big/\tfrac{8}{3}=\tfrac{9}{8}$; the column of
$h_{2}$ is likewise multiplied by $\tfrac{9}{8}$, and the column of
$h_{3}$, which must sum to $2$, by $2\big/\tfrac{8}{3}=\tfrac{3}{4}$.
The result is
\begin{equation*}
	M_{2}^{\mathrm{IPF}}=
	\begin{array}{c|ccc|c}
		& h_{1} & h_{2} & h_{3} & \\ \hline
		t_{1} & \tfrac{3}{2} & \tfrac{3}{2} & 1 & 4 \\[2pt]
		t_{2} & \tfrac{3}{2} & \tfrac{3}{2} & 1 & 4 \\ \hline
		& 3 & 3 & 2 &
	\end{array}
\end{equation*}
and one can check that both the row totals and the column totals now
match those of $M_{1}$, so the procedure has already converged --- in
this example after a single round; in general the two moves keep
alternating, because fixing the columns typically disturbs the rows a
little, until the discrepancy falls below a tolerance.

The fitted table, however, speaks of one and a half individuals, and
individuals come in whole numbers. We therefore round each row to
integers in a way that preserves its total, by the largest-remainder
rule. Take the row $(\tfrac{3}{2},\tfrac{3}{2},1)$ of type $t_{1}$,
whose entries must sum to four. Rounding everything down gives
$(1,1,1)$, which accounts for only three individuals, so one unit is
still missing; that unit is awarded to the entry that lost the most in
the rounding down --- the one with the largest fractional remainder.
Here the remainders are $\tfrac{1}{2}$, $\tfrac{1}{2}$, and $0$, the
tie between the first two is broken by the fixed column order, and the
missing unit goes to $h_{1}$. Both rows are rounded in the same way,
which yields
\begin{equation*}
	M_{2}^{\mathrm{int}}=
	\begin{array}{c|ccc}
		& h_{1} & h_{2} & h_{3} \\ \hline
		t_{1} & 2 & 1 & 1 \\
		t_{2} & 2 & 1 & 1
	\end{array}
\end{equation*}

\emph{Step (iii): pair.} At this point each type has an initial row
(from $M_{1}$) and a final row (from $M_{2}^{\mathrm{int}}$) with the
same total, and what remains is to decide, individual by individual,
which initial position goes together with which final position. The
rule is: stayers first. Within each type, as many individuals as
possible are declared to remain on their profile --- on each profile,
the minimum of the initial and final counts --- and only the leftover
positions are matched across profiles. For type $t_{1}$ the initial
row $(2,1,1)$ and the final row $(2,1,1)$ coincide, so all four of its
individuals are stayers and nothing is left to match. For type $t_{2}$
the rows are $(1,2,1)$ and $(2,1,1)$: one individual stays on each
profile, which exhausts the final positions on $h_{2}$ and $h_{3}$,
and leaves exactly one initial position (on $h_{2}$) and one final
position (on $h_{1}$) unmatched. These two must go together, so one
individual of type $t_{2}$ moves from $h_{2}$ to $h_{1}$. In this
miniature example the leftover matching is forced; in the actual data
several initial and several final positions are typically left over
within a type, and the final positions are then dealt out to the
initial ones by a uniformly random permutation --- this is the
randomness over which the $100$ draws of the empirical section are
taken.

\emph{Step (iv): balance.} One requirement is still violated. Adding
up the final positions across the two types, profile $h_{1}$ now
houses four individuals although only $q_{h_{1}}=3$ dwellings of that
profile exist, while profile $h_{2}$ houses two individuals in its
three dwellings. The culprit is the rounding step: awarding the
missing unit of each row to $h_{1}$ moved one individual, in effect,
from column $h_{2}$ to column $h_{1}$. The balancing step repairs
this directly: it selects, uniformly at random, one individual whose
final profile is over-supplied --- here, one of the four individuals
ending on $h_{1}$ --- and reassigns his final profile to the
under-supplied $h_{2}$. After the repair every profile houses exactly
$q_{h}$ individuals at the beginning and at the end of the year, as
the framework requires. The repair changed the final dwelling of one
of the eight individuals, and this proportion --- $12.5\%$ here --- is
what the empirical section reports as the overwrite share of a market.

\section{Why individual households cannot be tracked}\label{appendix:placebo}

The special licence data contain no tenant identifiers, but they do
contain characteristics that are fixed over time (gender, ethnic origin,
nationality) or that evolve deterministically (every age increases by
exactly the elapsed time). One may therefore ask whether an individual
household can be recognized when it appears twice: once when it enters
social housing, and once when it later transfers to another property.
This appendix explains why the answer is no.

\textbf{The matching rule.}
We pooled four years of lettings (April 2018--March 2022), giving
124{,}348 entries into social housing and 86{,}264 internal transfers. We
paired an entry record with a transfer record only if every logically
necessary condition of identity held: same gender; the transfer's
previous location equal to the local authority of the entry dwelling; the
lead tenant's age increased by exactly the time elapsed between the two
lettings; every recorded child's age increased by exactly the same
amount; and the numbers of adults and children changed by at most what
the elapsed time allows. Pairs satisfying these conditions were then
scored by the Fellegi--Sunter method: each additional characteristic in
agreement (ethnic origin, nationality, economic status, household type,
disability indicators) adds evidence in favour of a true match, with rare
characteristics counting for more, and each disagreement counts against.
A pair is accepted only if its score exceeds a threshold and it is the
unique best candidate for both records.

\textbf{The placebo test.}
To measure how often this rule accepts \emph{false} matches, we ran the
identical procedure backwards in time: We paired each entry with transfers
that occurred \emph{before} the household entered social housing. Such
pairs cannot be the same household observed twice, so every match the
rule produces in this direction is false by construction. The number and
quality of these placebo matches therefore estimate the false match rate
of the rule.

\textbf{The result.}
The placebo direction produced 12{,}471 matches; the real direction
produced 12{,}282---almost exactly the same number, with the same
distribution of match scores. In other words, the matches found in the
real direction are statistically indistinguishable from coincidences. The
reason is a lack of identifying information: the social tenant population
is homogeneous on the recorded characteristics. Agreement on ethnic
origin, for example, provides almost no evidence, because the large
majority of tenants report the same category; the same holds for
nationality and the disability indicators. Summing the evidence over all
available characteristics falls far short of what is needed to single out
one household among roughly one hundred thousand. Only two pairs in the
real direction scored above the best placebo score.

\textbf{Conclusion.}
The characteristics released in the CORE special licence data do not
carry enough information to identify individual households across
lettings, and any dataset built by such matching would consist mostly of
false pairs. We therefore construct the aggregate allocation instance at
the level of types, which requires no claim about individual identity.
Access to pseudonymous tenancy identifiers would remove this limitation
and would considerably improve the scope for research on mobility within
the social housing sector.

\section{Proof of Theorem \ref{DP-Rationalizability-WC-2}}\label{S-proof-DPWC2}

\begin{proof}
		
		\noindent Given an aggregate allocation instance, we can create its mixed graph. If an orientation $\gamma$ is given, then we can check Condition \ref{cond1} in polynomial time: part i) is cycle detection within each type, and part ii) also becomes cycle detection after splitting every vertex into two copies --- one for ``entered by an intra-arc'', one for ``entered by an inter-arc'' --- and rewiring each arc to the appropriate copies. Hence, DP-Rationalizability-WC is in NP.
		
		\medskip
		\noindent To show that the problem is NP-hard, we reduce from the maximum independent set problem, one of Karp's original NP-complete problems \citep{karp2009reducibility}.
		
		\begin{decision}[Maximum Independent Set (MIS)]
			\textit{Instance}: A graph $G=(V,E_{G})$ with $V=\{v_{1},\ldots,v_{n}\}$ and an integer $k$ with $1\leq k\leq n$.
			
			\smallskip
			\noindent\textit{Question}: Is there an \emph{independent} set $S\subseteq V$ --- no two of its vertices joined by an edge --- with $|S|\geq k$?
		\end{decision}
		
		The plan is the following. From the MIS instance, we first construct a mixed graph whose orientations satisfying Condition \ref{cond1} correspond exactly to the independent sets of size at least $k$. We then turn this mixed graph into an aggregate allocation instance. We illustrate the construction with the following running example:
		\begin{equation*}
			G^{\ast }: \quad v_{1}\;\text{---}\;v_{2}\;\text{---}\;v_{3}, \qquad k=2,
		\end{equation*}
		whose only independent set of size $2$ is $\{v_{1},v_{3}\}$. Which vertices are ``selected'' will be encoded by the orientation. For a given selection, let $\sigma _{i}$ denote the number of selected vertices among the first $i$ vertices $v_{1},\ldots,v_{i}$; note that $\sigma _{i}$ grows by at most one per step.
		
		\medskip
		\noindent\textbf{Undirected edges.} We have two types of undirected edges. The first type is the \emph{switch}. A switch $a$ consists of two vertices, $La$ and $Ra$, joined by one undirected edge. The switch is \emph{on} when the edge is oriented $La\rightarrow Ra$ and \emph{off} otherwise. We create one switch $x_{i}$ for every vertex $v_{i}$ in $V$; ``$x_{i}$ on'' means ``$v_{i}$ is selected''. The second type is the \emph{counter switch}. A counter switch $c_{i,j}$ also consists of two vertices, $Lc_{i,j}$ and $Rc_{i,j}$, joined by one undirected edge, and is \emph{on} when the edge is oriented $Lc_{i,j}\rightarrow Rc_{i,j}$. ``$c_{i,j}$ on'' makes the \emph{claim}: at least $j$ vertices are selected among $v_{1},\ldots,v_{i}$. We create one counter switch $c_{i,j}$ for every pair $(i,j)$ with $1\leq j\leq k$ and $j\leq i\leq n$.
		
		In the example $G^{\ast }$ ($n=3$, $k=2$) there are five counter switches, whose claims read as follows:
		\begin{center}
			\begin{tabular}{cl}
				\toprule
				switch & when on, it claims\\
				\midrule
				$c_{1,1}$ & at least $1$ of $\{v_{1}\}$ is selected --- that is, ``$v_{1}$ is selected''\\
				$c_{2,1}$ & at least $1$ of $\{v_{1},v_{2}\}$ is selected\\
				$c_{3,1}$ & at least $1$ of $\{v_{1},v_{2},v_{3}\}$ is selected\\
				$c_{2,2}$ & at least $2$ of $\{v_{1},v_{2}\}$ are selected --- that is, ``both $v_{1}$ and $v_{2}$''\\
				$c_{3,2}$ & at least $2$ of $\{v_{1},v_{2},v_{3}\}$ are selected\\
				\bottomrule
			\end{tabular}
		\end{center}
		
		For instance, when $c_{2,1}$ is on, it claims that among the first $2$ vertices of the list, $v_{1}$ and $v_{2}$, at least $1$ is selected. When a counter switch is off, its claim is simply not made. The last counter switch, $c_{3,2}$, carries the claim the construction cares about: at least $k=2$ vertices are selected in total. Two cells are missing from the table: $c_{1,2}$ does not exist because ``at least $2$ selected among $\{v_{1}\}$'' is impossible (hence the requirement $j\leq i$), and $c_{3,3}$ does not exist because the target $k$ is $2$ (hence the requirement $j\leq k$).
		
		\medskip
		\noindent\textbf{Black arc.} There are two further vertices, $P$ and $Q$, joined by the only black arc of the construction, $P\rightarrow Q$. We call this pair the \emph{relay}. The pair $P,Q$ carries no undirected edge, so its direction is fixed once and for all.
		
		\medskip
		\noindent\textbf{Blue arcs.} A cycle can travel through a switch (counter switch) only in the direction of its oriented edge: through an \emph{on} switch (counter switch) from the left vertex to the right one, through an \emph{off} switch (counter switch) from the right vertex to the left one. We call such a trip a \emph{passage}. This gives us a way to forbid any combination of switch (counter switch) positions: write down the passage that each prescribed position allows, and add the blue arcs that chain these passages into a loop --- one blue arc from the end of each passage to the start of the next, closing the loop at the end. If the forbidden combination is chosen, all its passages are open and the loop closes; black/red and blue arcs alternate along it, so it is an alternating cycle and Condition \ref{cond1} fails. If at least one switch (counter switch) is in the opposite position, its edge points against the loop, and this loop cannot form. We call the set of blue arcs added for one forbidden combination a \emph{device}. Below we list all the devices of the construction: first the forbidden combination, then the blue arcs, obtained by applying the rule above to the switches (counter switches) in the order listed.
		
		\begin{itemize}
			\item[(IND)] For every edge $\{v_{u},v_{w}\}\in E_{G}$, forbid ``$x_{u}$ on, $x_{w}$ on''. The two prescribed passages are $Lx_{u}\rightarrow Rx_{u}$ and $Lx_{w}\rightarrow Rx_{w}$; the blue arcs chain them:
			\begin{equation*}
				Rx_{u}\rightarrow Lx_{w}\text{,}\qquad Rx_{w}\rightarrow Lx_{u}\text{.}
			\end{equation*}
			\item[(N1)] For every $2\leq j\leq k$ and $j<i\leq n$, forbid ``$c_{i,j}$ on, $c_{i-1,j}$ off, $c_{i-1,j-1}$ off''. The passages are $Lc_{i,j}\rightarrow Rc_{i,j}$ (on), $Rc_{i-1,j}\rightarrow Lc_{i-1,j}$ (off) and $Rc_{i-1,j-1}\rightarrow Lc_{i-1,j-1}$ (off); the blue arcs run from the end of each passage to the start of the next:
			\begin{equation*}
				Rc_{i,j}\rightarrow Rc_{i-1,j}\text{,}\qquad Lc_{i-1,j}\rightarrow Rc_{i-1,j-1}\text{,}\qquad Lc_{i-1,j-1}\rightarrow Lc_{i,j}\text{.}
			\end{equation*}
			For $i=j$ (every $2\leq j\leq k$), forbid ``$c_{j,j}$ on, $c_{j-1,j-1}$ off'', with blue arcs
			\begin{equation*}
				Rc_{j,j}\rightarrow Rc_{j-1,j-1}\text{,}\qquad Lc_{j-1,j-1}\rightarrow Lc_{j,j}\text{.}
			\end{equation*}
			\item[(N2)] For every $1\leq j\leq k$ and $j<i\leq n$, forbid ``$c_{i,j}$ on, $c_{i-1,j}$ off, $x_{i}$ off'', with blue arcs
			\begin{equation*}
				Rc_{i,j}\rightarrow Rc_{i-1,j}\text{,}\qquad Lc_{i-1,j}\rightarrow Rx_{i}\text{,}\qquad Lx_{i}\rightarrow Lc_{i,j}\text{.}
			\end{equation*}
			For $i=j$ (every $1\leq i\leq k$), forbid ``$c_{i,i}$ on, $x_{i}$ off'', with blue arcs
			\begin{equation*}
				Rc_{i,i}\rightarrow Rx_{i}\text{,}\qquad Lx_{i}\rightarrow Lc_{i,i}\text{.}
			\end{equation*}
			\item[(GATE)] Forbid ``$c_{n,k}$ off''. Here the loop is closed through the relay:
			\begin{equation*}
				Lc_{n,k}\rightarrow P\text{,}\qquad Q\rightarrow Rc_{n,k}\text{,}
			\end{equation*}
			so that the cycle it threatens is $Rc_{n,k}\rightarrow Lc_{n,k}\rightarrow P\rightarrow Q\rightarrow Rc_{n,k}$, alternating the off passage of $c_{n,k}$ and the relay with the two blue arcs.
		\end{itemize}
		When two devices prescribe the same blue arc --- as (N1) and (N2) do for $Rc_{i,j}\rightarrow Rc_{i-1,j}$ --- the arc appears once in the mixed graph.
		
		Figure \ref{fig-device} shows two examples: the left panel forbids ``$x_{u}$ on and $x_{w}$ on'' and the right panel forbids ``$c_{3,1}$ on, $c_{2,1}$ off, $x_{3}$ off''.
		
		\begin{figure}[htb]
			\begin{center}
				\begin{minipage}[b]{0.44\textwidth}
					\centering
					\begin{tikzpicture}[xscale=1.5,yscale=1.5,
						vtx/.style={draw,circle,fill,inner sep=1.6pt},
						redl/.style={red,dotted,very thick},
						bluearrow/.style={blue, dashed,postaction={decorate},decoration={markings,mark=at position 0.55 with {\arrow{>}}}}]
						\node[vtx,label=above:{$Lx_{u}$}] (Lu) at (0,1) {};
						\node[vtx,label=above:{$Rx_{u}$}] (Ru) at (1.6,1) {};
						\node[vtx,label=below:{$Lx_{w}$}] (Lw) at (0,0) {};
						\node[vtx,label=below:{$Rx_{w}$}] (Rw) at (1.6,0) {};
						\draw[redl] (Lu)--(Ru); \draw[redl] (Lw)--(Rw);
						\draw[bluearrow] (Ru) to[bend left=25] (Lw);
						\draw[bluearrow] (Rw) to[bend left=25] (Lu);
					\end{tikzpicture}
				\end{minipage}
				\hfill
				\begin{minipage}[b]{0.52\textwidth}
					\centering
					\begin{tikzpicture}[xscale=1.5,yscale=1.1,
						vtx/.style={draw,circle,fill,inner sep=1.6pt},
						redl/.style={red,dotted,very thick},
						bluearrow/.style={blue, dashed,postaction={decorate},decoration={markings,mark=at position 0.55 with {\arrow{>}}}}]
						\node[vtx,label=above:{$Lc_{3,1}$}] (L3) at (0,2) {};
						\node[vtx,label=above:{$Rc_{3,1}$}] (R3) at (1.6,2) {};
						\node[vtx,label=left:{$Lc_{2,1}$}] (L2) at (0,1) {};
						\node[vtx,label=right:{$Rc_{2,1}$}] (R2) at (1.6,1) {};
						\node[vtx,label=below:{$Lx_{3}$}] (Lx) at (0,0) {};
						\node[vtx,label=below:{$Rx_{3}$}] (Rx) at (1.6,0) {};
						\draw[redl] (L3)--(R3); \draw[redl] (L2)--(R2); \draw[redl] (Lx)--(Rx);
						\draw[bluearrow] (R3) to[bend left=25] (R2);
						\draw[bluearrow] (L2) to[bend left=15] (Rx);
						\draw[bluearrow] (Lx) to[bend left=30] (L3);
					\end{tikzpicture}
				\end{minipage}
			\end{center}
			\caption{Two forbidden cases.}\label{fig-device}
		\end{figure}
		
		In words: (IND) forbids selecting two adjacent vertices. (N1) and (N2) force every claim to be \emph{justified}: if $c_{i,j}$ claims ``at least $j$ among the first $i$'', then either the shorter claim $c_{i-1,j}$ is also on, or the weaker claim $c_{i-1,j-1}$ is on \emph{and} $v_{i}$ itself is selected. (GATE) forces the final claim ``at least $k$ selected in total'' to be on.
		
		Finally, Figure \ref{fig-full} assembles the \emph{complete} mixed graph of the running example: the three switches, the five counter switches, the relay, and the twenty-three blue arcs.
		
		\begin{sidewaysfigure}[p]
			\begin{center}
				\begin{tikzpicture}[xscale=2.35,yscale=1.55,
					vtx/.style={draw,circle,fill,inner sep=1.7pt},
					redl/.style={red,dotted,very thick},
					blackarrow/.style={postaction={decorate},decoration={markings,mark=at position 0.55 with {\arrow[scale=1.15]{>}}}},
					bluearrow/.style={blue, dashed,postaction={decorate},decoration={markings,mark=at position 0.55 with {\arrow[scale=1.15]{>}}}}]
					\node[vtx,label=below:{\small $Lx_{1}$}] (Lx1) at (0,0) {};
					\node[vtx,label=below:{\small $Rx_{1}$}] (Rx1) at (1.5,0) {};
					\node[vtx,label=below:{\small $Lx_{2}$}] (Lx2) at (3.4,0) {};
					\node[vtx,label=below:{\small $Rx_{2}$}] (Rx2) at (4.9,0) {};
					\node[vtx,label=below:{\small $Lx_{3}$}] (Lx3) at (6.8,0) {};
					\node[vtx,label=below:{\small $Rx_{3}$}] (Rx3) at (8.3,0) {};
					\node[vtx,label=above left:{\small $Lc_{1,1}$}] (L11) at (0,2.1) {};
					\node[vtx,label=above:{\small $Rc_{1,1}$}] (R11) at (1.5,2.1) {};
					\node[vtx,label=above left:{\small $Lc_{2,1}$}] (L21) at (3.4,2.1) {};
					\node[vtx,label=above:{\small $Rc_{2,1}$}] (R21) at (4.9,2.1) {};
					\node[vtx,label=above:{\small $Lc_{3,1}$}] (L31) at (6.8,2.1) {};
					\node[vtx,label=above:{\small $Rc_{3,1}$}] (R31) at (8.3,2.1) {};
					\node[vtx,label=above left:{\small $Lc_{2,2}$}] (L22) at (3.4,4.2) {};
					\node[vtx,label=above:{\small $Rc_{2,2}$}] (R22) at (4.9,4.2) {};
					\node[vtx,label=above left:{\small $Lc_{3,2}$}] (L32) at (6.8,4.2) {};
					\node[vtx,label=above:{\small $Rc_{3,2}$}] (R32) at (8.3,4.2) {};
					\node[vtx,label=above:{\small $P$}] (P) at (6.8,5.7) {};
					\node[vtx,label=above:{\small $Q$}] (Q) at (8.3,5.7) {};
					\draw[redl] (Lx1)--(Rx1); \draw[redl] (Lx2)--(Rx2); \draw[redl] (Lx3)--(Rx3);
					\draw[redl] (L11)--(R11); \draw[redl] (L21)--(R21); \draw[redl] (L31)--(R31);
					\draw[redl] (L22)--(R22); \draw[redl] (L32)--(R32);
					\draw[blackarrow] (P)--(Q);
					\draw[bluearrow] (Rx1) to[bend right=20] (Lx2);
					\draw[bluearrow] (Rx2) to[bend right=45] (Lx1);
					\draw[bluearrow] (Rx2) to[bend right=20] (Lx3);
					\draw[bluearrow] (Rx3) to[bend right=45] (Lx2);
					\draw[bluearrow] (R11) to[bend left=15] (Rx1);
					\draw[bluearrow] (Lx1) to[bend left=15] (L11);
					\draw[bluearrow] (R21) to[bend right=15] (R11);
					\draw[bluearrow] (L11) to[bend right=15] (Rx2);
					\draw[bluearrow] (Lx2) to[bend left=15] (L21);
					\draw[bluearrow] (R31) to[bend right=15] (R21);
					\draw[bluearrow] (L21) to[bend right=15] (Rx3);
					\draw[bluearrow] (Lx3) to[bend left=15] (L31);
					\draw[bluearrow] (R22) to[bend right=15] (R11);
					\draw[bluearrow] (L11) to[bend left=25] (L22);
					\draw[bluearrow] (R32) to[bend right=15] (R22);
					\draw[bluearrow] (L22) to[bend left=10] (R21);
					\draw[bluearrow] (L21) to[bend left=15] (L32);
					\draw[bluearrow] (R22) to[bend left=25] (Rx2);
					\draw[bluearrow] (Lx2) to[bend left=40] (L22);
					\draw[bluearrow] (L22) to[bend right=25] (Rx3);
					\draw[bluearrow] (Lx3) to[bend right=25] (L32);
					\draw[bluearrow] (L32) to[bend left=15] (P);
					\draw[bluearrow] (Q) to[bend left=15] (R32);
				\end{tikzpicture}
			\end{center}
			\caption{The complete mixed graph of the running example ($G^{\ast}$: the path $v_{1}$---$v_{2}$---$v_{3}$, $k=2$). Bottom layer: the switches. Middle layer: the counter switches with $j=1$. Top layer: the counter switches with $j=2$, and, above them, the relay.}\label{fig-full}
		\end{sidewaysfigure}
		
		\medskip
		\noindent\textbf{From the mixed graph to an allocation instance.} So far we have drawn a mixed graph. The reduction, however, must produce an aggregate allocation instance $\langle I,H,q,\tau ,T,x,\omega \rangle$ whose mixed graph is exactly this drawing. This is done as follows. Every vertex of the drawing is an object. Every switch, every counter switch, and the relay is a type, holding its own two objects. We add one extra \emph{sink} type holding one extra object $h_{0}$. The individuals are:
		\begin{itemize}
			\item[(R1)] for every blue arc from a vertex $h$ to a vertex $h'$: one individual of the type of $h'$, endowed with $h$ and assigned $h'$;
			\item[(R2)] one relay individual, endowed with $P$ and assigned $Q$;
			\item[(R3)] for every object that nobody is assigned so far: one individual of its own type, endowed with $h_{0}$ and assigned that object;
			\item[(R4)] balancing individuals: whenever an object is assigned more often than it is endowed, individuals of the sink type endowed with it and assigned $h_{0}$; in the opposite case, individuals of its own type endowed with $h_{0}$ and assigned it.
		\end{itemize}
		After (R4) every object is endowed exactly as often as it is assigned (for $h_{0}$ this then holds automatically, because every individual is endowed with one object and assigned one object). Letting $q_{h}$ be this common number, $\omega$ and $x$ are allocations of the same object profile, so the instance is feasible. Its size is polynomial: $O(nk)$ types and objects and $O(nk+|E_{G}|)$ individuals.
		
		Four simple facts confirm that the mixed graph of this instance is the drawing. \emph{First}, every object is assigned only to individuals of its own type. So the vertices are as drawn (plus the sink vertex), and each individual of (R1) creates exactly one inter-arc --- the blue arc we wanted --- because his endowment is held by exactly one type. \emph{Second}, the only individual endowed with an object of his own type is the relay individual. So the only intra-arc is $P\rightarrow Q$, and no revealed comparison orders any switch pair: the undirected edges are as drawn. \emph{Third}, the sink vertex has no intra-arc, while every vertex of an alternating cycle needs one intra-arc and one inter-arc of the cycle; so the sink never lies on an alternating cycle, and the arcs created by (R3)--(R4) are harmless. \emph{Fourth}, every type holds at most two objects, so every type sub-digraph has at most one arc, and Condition \ref{cond1}(i) holds for every orientation. Putting these together with Theorem \ref{Th-WC}: \emph{the instance is WC-rationalizable if and only if the switches can be oriented so that no alternating cycle appears.}
		
		It remains to prove that such an orientation exists if and only if $G$ has an independent set of size at least $k$.
		
		\medskip
		\noindent\textbf{From a valid orientation to an independent set.} Let $\gamma$ be an orientation with no alternating cycle, and select the vertices whose switches are on: $S=\{v_{i}: x_{i}\text{ on}\}$. Since every forbidden combination would close its device's cycle, no forbidden combination is chosen under $\gamma$. By (IND), no two adjacent vertices are selected: $S$ is independent. By (GATE), $c_{n,k}$ is on. Finally, every claim that is on is \emph{true}. We show this by induction on $i$: if $c_{i,j}$ is on, then $\sigma _{i}\geq j$.
		\begin{itemize}
			\item If $i=j$: by (N2), $x_{i}$ is on; and if $j\geq 2$, by (N1), $c_{i-1,i-1}$ is on, so $\sigma _{i-1}\geq i-1$ by induction, and adding $v_{i}\in S$ gives $\sigma _{i}\geq i$. For $i=j=1$, $x_{1}$ on already gives $\sigma _{1}\geq 1$.
			\item If $i>j$: by (N1) and (N2), either $c_{i-1,j}$ is on --- then $\sigma _{i}\geq \sigma _{i-1}\geq j$ by induction --- or both $c_{i-1,j-1}$ and $x_{i}$ are on --- then $\sigma _{i}=\sigma _{i-1}+1\geq j$. (For $j=1$ the claim $c_{i-1,0}$ is not needed: (N2) alone gives ``$c_{i-1,1}$ on or $x_{i}$ on'', which is enough.)
		\end{itemize}
		Applying this to $c_{n,k}$, which is on: $|S|=\sigma _{n}\geq k$. So $S$ is an independent set of size at least $k$.
		
		\medskip
		\noindent\textbf{From an independent set to a valid orientation.} Let $S$ be an independent set with $|S|\geq k$. Orient the switches \emph{truthfully}: $x_{i}$ on if and only if $v_{i}\in S$, and $c_{i,j}$ on if and only if $\sigma _{i}\geq j$ --- every claim is made exactly when it is true. In the running example, for $S=\{v_{1},v_{3}\}$ we have $(\sigma _{1},\sigma _{2},\sigma _{3})=(1,1,2)$, so $x_{1},x_{3},c_{1,1},c_{2,1},c_{3,1},c_{3,2}$ are on and $x_{2},c_{2,2}$ are off. One checks directly that no forbidden combination is chosen.
		
		Checking the devices one by one is, however, not enough. Blue arcs of \emph{different} devices can meet and form cycles that no single device intended. In the example, three devices lend their arcs to the tour that passes through $x_{2}$ off, then $c_{2,1}$ on, then $c_{1,1}$ off, then $c_{2,2}$ on, and back --- a potential alternating cycle that is not the cycle of any single device. But look at what this tour would say: nothing is selected among $\{v_{1}\}$ (that is $c_{1,1}$ off), yet at least two are selected among $\{v_{1},v_{2}\}$ (that is $c_{2,2}$ on) --- impossible, since $\sigma$ grows by at most one per step. We now show that \emph{every} conceivable cycle is impossible for a reason of this kind.
		
		Under the truthful orientation, every passage states a true fact: an on passage of $x_{i}$ means $v_{i}\in S$, an off passage means $v_{i}\notin S$, an on passage of $c_{i,j}$ means $\sigma _{i}\geq j$, an off passage means $\sigma _{i}\leq j-1$. Moreover, an alternating cycle can only be a closed tour of passages joined by blue arcs: every vertex of the cycle carries exactly one intra-arc of the cycle, and the only intra-arcs available are the oriented switch edges and the relay. So it suffices to show that no such tour can close. Reading the blue arcs listed above, here is where a tour can travel:
		\begin{itemize}
			\item after the on passage of $x_{u}$: only to the on passage of a neighbouring $x_{w}$ \hfill (IND);
			\item after the off passage of $x_{i}$: only to the on passage of some $c_{i,j}$ \hfill (N2);
			\item into the on passage of $c_{i,j}$, a tour can arrive only from the off passage of $x_{i}$ (N2) or from the off passage of $c_{i-1,j-1}$ \hfill (N1);
			\item after the on passage of $c_{i,j}$: to the off passage of $c_{i-1,j}$ if $j<i$ (N1),(N2); to the off passage of $c_{i-1,i-1}$ (N1) or of $x_{i}$ (N2) if $j=i$;
			\item after the off passage of $c_{i,j}$: to the off passage of $c_{i,j-1}$ (N1), to the off passage of $x_{i+1}$ (N2), to the on passage of $c_{i+1,j+1}$ (N1), or --- for $c_{n,k}$ only --- through the relay \hfill (GATE).
		\end{itemize}
		Now follow any alternating cycle and rule everything out in turn.
		\begin{itemize}
			\item[(i)] \emph{The tour contains no on passage of a switch $x_{u}$.} Its only exit leads to the on passage of an adjacent $x_{w}$; the two passages together would put two adjacent vertices in $S$ --- impossible, $S$ is independent.
			\item[(ii)] \emph{The tour contains no on passage of a counter switch $c_{i,j}$.} Consider how the tour would enter it. From the off passage of $c_{i-1,j-1}$: then $\sigma _{i-1}\leq j-2$ and $\sigma _{i}\geq j$ --- impossible; $\sigma$ grows by at most one per step (this is exactly the tour of the example above). From the off passage of $x_{i}$: then $v_{i}\notin S$ and $\sigma _{i}\geq j$. If $j=i$, the claim $\sigma _{i}\geq i$ puts every vertex up to $v_{i}$ in $S$, contradicting $v_{i}\notin S$. If $j<i$, the only exit leads to the off passage of $c_{i-1,j}$, so $\sigma _{i-1}\leq j-1$; but $v_{i}\notin S$ gives $\sigma _{i}=\sigma _{i-1}\leq j-1$, contradicting $\sigma _{i}\geq j$.
			\item[(iii)] \emph{Nothing else can close a tour.} The off passage of a switch $x_{i}$ exits only towards on passages of counter switches, excluded by (ii). The relay is reached only from the off passage of $c_{n,k}$, and $c_{n,k}$ is truthfully on because $|S|\geq k$. What remains are off passages of counter switches, whose remaining exits lead to excluded passages or to off passages of counter switches with a strictly smaller second index --- along which no tour can come back to its starting point.
		\end{itemize}
		Hence the truthful orientation produces no alternating cycle, and the instance is WC-rationalizable.
		
		\medskip
		The construction is polynomial in the size of $(G,k)$, so DP-Rationalizability-WC is NP-complete.
	\end{proof}

\section{A mixed-integer formulation of Condition \ref{cond2}}\label{S-mip}

\clr{Condition~\ref{cond2} asks whether a certain structure exists: a consistent crossing-point function whose induced profile leaves no blocking coalition. A question of this form can be handed to a mixed-integer solver. We describe the structure by a collection of integer and binary variables, translate each requirement of the definitions into linear inequalities over those variables, and let the solver either return values satisfying every inequality (which exhibit the structure) or prove that no such values exist.} The translation
	requires one standard device. Some of our requirements are
	either--or statements, and a linear inequality alone cannot express
	``or''. The device is a binary variable that selects the branch,
	combined with a constant $M$ large enough that adding $M$ to the
	right-hand side of an inequality makes it vacuous: writing one branch
	with slack $M(1-y)$ and the other with slack $My$ activates exactly the
	branch the binary selects and switches the other off. Since every
	crossing point lies in $\{0,\ldots ,\left\vert T\right\vert \}$, the
	constant $M=\left\vert T\right\vert $ is large enough everywhere below.
	Throughout, write $r_{i}=\operatorname{rk}(\tau (i))$ for the rank of
	individual $i$'s type.
	
	\emph{Variables.} \clr{The structure we are looking for is described by four groups of variables.}
	\begin{itemize}
		\item[] One \emph{integer} variable $\kappa (h,h^{\prime })\in
		K(h,h^{\prime })$ per pair $h\rhd _{H}h^{\prime }$: the candidate
		crossing-point function itself. Note that the admissible sets of
		Definition~\ref{def-admissible} enter simply as the variable bounds, so
		the movers' revealed comparisons are built in before the solver starts.
		\item[] One \emph{binary} variable $y_{abc}$ per triple $h_{a}\rhd
		_{H}h_{b}\rhd _{H}h_{c}$, recording which of the two inner crossing
		points of the triple is the smaller one.
		\item[] One \emph{binary} variable $\delta _{i,h}$ per individual $i$
		and object $h\neq x(i)$, meant to equal $1$ exactly when the type of
		$i$ prefers $h$ to $i$'s assigned object $x(i)$ under the induced
		profile.
		\item[] One variable $u_{i}\in \lbrack 1,\left\vert I\right\vert ]$ per
		individual: a \emph{position} in a ranking of the individuals, used to
		certify that the demand digraph has no cycle.
	\end{itemize}
	
	\emph{Consistency.} The sandwich of Definition~\ref{def-consistency}
	says that $\kappa (h_{a},h_{c})$ lies between the two inner crossing
	points --- but ``between'' is an either--or statement: either
	$\kappa (h_{a},h_{b})\leq \kappa (h_{a},h_{c})\leq \kappa
	(h_{b},h_{c})$, or the same chain with the two inner values exchanged.
	The binary $y_{abc}$ selects the branch:
	\begin{align*}
		\kappa (h_{a},h_{b})-M(1-y_{abc})\ \leq\ &\kappa (h_{a},h_{c})\ \leq\
		\kappa (h_{b},h_{c})+M(1-y_{abc})\text{,}\\
		\kappa (h_{b},h_{c})-My_{abc}\ \leq\ &\kappa (h_{a},h_{c})\ \leq\
		\kappa (h_{a},h_{b})+My_{abc}\text{.}
	\end{align*}
	Setting $y_{abc}=1$ makes the terms $M(1-y_{abc})$ vanish, so the first
	line becomes exactly the first branch, while the terms $My_{abc}$ relax
	the second line into vacuity; setting $y_{abc}=0$ does the opposite.
	Whatever values the inner crossing points take, one of the two branches
	is the true sandwich, so the solver can always pick the correct
	$y_{abc}$ --- and it can never satisfy the constraints while violating
	the sandwich.
	
	\emph{Preferences.} Under the induced profile, whether the type of $i$
	prefers $h$ to $x(i)$ is read off a single crossing point. If $h\rhd
	_{H}x(i)$, Definition~\ref{def-cpf} says that $h\succ _{\tau (i)}x(i)$
	exactly when $r_{i}\leq \kappa (h,x(i))$. The pair of inequalities
	\begin{equation*}
		\kappa (h,x(i))\ \geq\ r_{i}-M(1-\delta _{i,h})\text{,}\qquad \kappa
		(h,x(i))\ \leq\ r_{i}-1+M\delta _{i,h}
	\end{equation*}
	forces $\delta _{i,h}$ to \emph{equal} this indicator, and both
	inequalities are needed. The first says: if $\delta _{i,h}=1$, then
	$\kappa (h,x(i))\geq r_{i}$ --- the variable may not claim a preference
	that the crossing point denies. The second says: if $\delta _{i,h}=0$,
	then $\kappa (h,x(i))\leq r_{i}-1$ --- the variable may not conceal a
	preference that the crossing point grants. Without the second
	inequality the solver could quietly set every $\delta $ to zero, erase
	the demand digraph, and declare any allocation unblocked. If instead
	$x(i)\rhd _{H}h$, the same two-way link ties $\delta _{i,h}$ to the
	condition $\kappa (x(i),h)\leq r_{i}-1$, with the mirrored pair of
	inequalities.
	
	\emph{No blocking.} By Lemma~\ref{lem-weakcore-dag}, $x$ is in the weak
	core for the induced profile exactly when the demand digraph is
	acyclic; and by construction of that digraph, its arc $i\rightarrow j$
	is present exactly when $\delta _{i,\omega (j)}=1$. Acyclicity is
	certified by the position variables: for every ordered pair of
	individuals $(i,j)$ with $\omega (j)\neq x(i)$,
	\begin{equation*}
		u_{j}\ \geq\ u_{i}+1-\left\vert I\right\vert \left( 1-\delta _{i,\omega
			(j)}\right) \text{.}
	\end{equation*}
	When $\delta _{i,\omega (j)}=1$ the constraint reads $u_{j}\geq
	u_{i}+1$: every arc must climb strictly in the ranking. Around a
	directed cycle the position would have to climb forever while returning
	to its starting point, which is impossible --- so if the constraints are
	satisfied, no cycle exists. Conversely, if the digraph is acyclic, then
	processing the individuals in a topological order and letting $u_{i}$
	be the position of $i$ in that order satisfies every constraint. The
	formulation therefore accepts exactly the acyclic digraphs. Two details
	come for free. Taking $j=i$ in the same constraint reads $u_{i}\geq
	u_{i}+1$ unless $\delta _{i,\omega (i)}=0$: self-loops --- violations of
	individual rationality --- are excluded without any separate constraint.
	And when $\delta _{i,\omega (j)}=0$ the constraint is vacuous, since
	positions never differ by more than $\left\vert I\right\vert $.
	
	\emph{Correctness and size.} Feasible values of the variables are
	precisely the witnesses of Condition~\ref{cond2}: the $\kappa $-block
	delivers a crossing-point function within the admissible sets, the
	$y$-block forces it to be consistent, the $\delta $-block reconstructs
	the induced profile's preferences without error, and the $u$-block
	certifies that the resulting demand digraph is acyclic. The allocation
	$x$ is therefore single-crossing WC-rationalizable with respect to the
	given orders if and only if the system is feasible. The formulation is
	small: $O(\left\vert H\right\vert ^{2})$ integer variables,
	$O(\left\vert H\right\vert ^{3}+\left\vert I\right\vert \left\vert
	H\right\vert )$ binaries and $O(\left\vert I\right\vert ^{2})$
	constraints; in implementations, individuals sharing the type and the
	assigned object are interchangeable in $D(\succ)$ and can be merged into
	one class, which reduces the acyclicity block accordingly.

\renewcommand{\theproposition}{S.\arabic{proposition}}
\setcounter{proposition}{0}
\section{Relations between the properties}\label{S-relations}

Section 5 of the main text summarizes the two results established here and states
where each is used. This section gives them in full. The third relation, between
our properties and the strict core, is Proposition \ref{prop-tai} of the paper.

\medskip\noindent\textbf{Logical independence.} Example \ref{example-wc} of the
main text is PI-rationalizable and not WC-rationalizable; Example S.2 below runs
the other way. The two notions discipline different trades: Pareto efficiency
restricts exchanges of \emph{assigned} objects, whereas the weak core restricts
exchanges of \emph{endowments}, so individuals who envy one another's assignments
need not be able to block, because the objects they covet need not be endowments
of the group. That weak core allocations may fail Pareto efficiency is known since
\citet{roth1977weak}.

Examples \ref{example-wc} and S.2 together establish the following.

	\begin{proposition}\label{prop-nonnested}
		PI-rationalizability and WC-rationalizability are logically independent:
		neither property implies the other.
	\end{proposition}

	\new{\paragraph{Joint rationalizability}
		Proposition \ref{prop-nonnested} says that neither PI- nor WC-rationalizability
		implies the other. A different question is whether the two can always be met
		\emph{together} whenever each can be met separately. Both definitions are
		existential, and nothing requires the profile that makes $x$ efficient to be
		the profile that makes $x$ weak core-stable. Say that $x$ is \emph{jointly
			rationalizable} if there is a single profile $\succ$ for which
		$x$ is Pareto efficient, individually rational \emph{and} in the weak core.
		Joint rationalizability trivially implies both PI- and WC-rationalizability.
		The converse fails.}
	
	\begin{proposition}\label{prop-joint}
		\new{There is an aggregate allocation instance that is PI-rationalizable and
			WC-rationalizable but not jointly rationalizable. Hence the condition of
			Theorem \ref{PI} and the condition of Theorem \ref{Th-WC} are each necessary
			for joint rationalizability, but their conjunction is not sufficient.}
	\end{proposition}
	
	\new{Example S.3 gives such an instance, and the
		mechanism is not an artifact of its numbers. Everything turns on a single
		comparison the data leave open, and the two requirements pin it down in
		opposite directions, because they discipline different objects. Pareto
		efficiency is about \emph{assignments}: it asks whether individuals would like
		to exchange what they were given. The weak core is about \emph{endowments}: it
		asks whether they would like to undo the reallocation and trade what they
		brought. One direction of the open comparison creates a profitable exchange of
		assignments and destroys efficiency; the other creates a profitable exchange of
		endowments and destroys the weak core. Each requirement consumes the same degree
		of freedom, and consumes it the opposite way.}
	
	\new{This sharpens Proposition \ref{prop-nonnested} in a way that matters for
		how the tests are used. Logical independence says only that the two properties
		cut the data differently. Proposition \ref{prop-joint} says something stronger:
		the set of instances rationalizable as both efficient and weak core-stable by one profile
		is a \emph{strict} subset of the intersection of the two sets our theorems
		characterize. Running both tests and finding that each accepts therefore does
		not establish that the observed reallocation is consistent with a market that
		is simultaneously efficient and weak core-stable. In particular, when the empirical
		frequencies of Section \ref{UKapplication} are read together, the fraction of
		instances that are jointly rationalizable is bounded above by the smaller of
		the two reported frequencies and may be strictly below it.}
	
	\new{We do not characterize joint rationalizability here, and we regard it as
		the natural next question. It is not the conjunction of Theorem \ref{PI} and
		Condition \ref{cond1}.}

\renewcommand{\theexample}{S.\arabic{example}}
\setcounter{example}{0}
\section{Examples omitted from the main text}\label{S-examples}

Examples S.1--S.5 are referred to in the main text and collected here.
Numbering follows the order in which the main text cites them.

	\begin{example}\label{example-sc}
		\new{Let $T=\left\{ t_{1},t_{2}\right\} $ and let there be three objects
			$h_{1},h_{2},h_{3}$, each with one copy. The three individuals are as follows.}
		\begin{center}
			\begin{tabular}{cccc}
				\toprule
				Individual & Type & Endowment $\omega(i)$ & Assignment $x(i)$ \\
				\midrule
				$1$ & $t_{1}$ & $h_{2}$ & $h_{3}$ \\
				$2$ & $t_{1}$ & $h_{3}$ & $h_{1}$ \\
				$3$ & $t_{2}$ & $h_{1}$ & $h_{2}$ \\
				\bottomrule
			\end{tabular}
		\end{center}
		\new{Individual $1$ reveals $h_{3}\succ_{t_{1}}h_{2}$ and individual $2$
			reveals $h_{1}\succ_{t_{1}}h_{3}$, so individual rationality pins down
			$\succ_{t_{1}}$ as $h_{1}\succ h_{3}\succ h_{2}$; individual $3$ reveals
			$h_{2}\succ_{t_{2}}h_{1}$. Consider the profile}
		\begin{equation*}
			\begin{tabular}{c|c}
				$\succ_{t_{1}}$ & $\succ_{t_{2}}$ \\ \hline
				$h_{1}$ & $h_{2}$ \\
				$h_{3}$ & $h_{1}$ \\
				$h_{2}$ & $h_{3}$%
			\end{tabular}%
		\end{equation*}
		\new{with objects listed from best to worst. The allocation $x$ is individually
			rational for this profile. Individual $2$ receives $h_{1}$ and individual $3$
			receives $h_{2}$, each the most preferred object of his own type, so neither
			can be made better off; only individual $1$ could gain, and only by obtaining
			$h_{1}$, which individual $2$ would then have to surrender for an object his
			type ranks strictly lower. Hence no allocation Pareto dominates $x$ and no
			coalition strongly blocks it, so $x$ is both PI-rationalizable and
			WC-rationalizable.}
		
		\new{Yet $x$ is not SC-rationalizable. Let $\succ$ be any profile for which $x$
			lies in the strict core. By part i) of Proposition \ref{prop-tai}, $x$ is
			individually rational for $\succ$, and individual $2$'s move from $h_{3}$ to
			$h_{1}$ therefore gives $h_{1}\succ_{t_{1}}h_{3}$. Take $S=\left\{ 1,3\right\}
			$ and let}
		\begin{equation*}
			\new{x^{\prime }(1)=\omega(3)=h_{1}\text{,}\qquad x^{\prime }(3)=\omega(1)=h_{2}
				\text{.}}
		\end{equation*}
		\new{This redistributes the endowments of $S$ among its members. Individual $3$
			receives $x^{\prime }(3)=h_{2}=x(3)$ and is therefore no worse off, while
			individual $1$ receives $h_{1}\succ_{t_{1}}h_{3}=x(1)$ and is strictly better
			off. So $S$ weakly blocks $x$, contradicting the assumption. As $\succ$ was
			arbitrary, no type-measurable profile places $x$ in the strict core.}
	\end{example}

	\begin{example}\label{example-wc-not-pi}
		Let $T=\{t_{1},t_{2},t_{3},t_{4}\}$ and let there be four objects:
		$h_{1}$ and $h_{2}$ with two copies each ($q_{h_{1}}=q_{h_{2}}=2$), and
		$h_{3},h_{4}$ with one copy each. The six individuals are as follows.
		\begin{center}
			\begin{tabular}{cccc}
				\toprule
				Individual & Type & Endowment $\omega(i)$ & Assignment $x(i)$ \\
				\midrule
				$1$ & $t_{1}$ & $h_{1}$ & $h_{2}$ \\
				$2$ & $t_{1}$ & $h_{3}$ & $h_{1}$ \\
				$3$ & $t_{2}$ & $h_{2}$ & $h_{1}$ \\
				$4$ & $t_{2}$ & $h_{4}$ & $h_{2}$ \\
				$5$ & $t_{3}$ & $h_{1}$ & $h_{3}$ \\
				$6$ & $t_{4}$ & $h_{2}$ & $h_{4}$ \\
				\bottomrule
			\end{tabular}
		\end{center}
		The allocation $x$ is not PI-rationalizable: type $t_{1}$ contains the
		mover $h_{1}\rightarrow h_{2}$ (individual 1) while individual 2 of the
		same type holds $h_{1}$, giving the edge $h_{1}\rightarrow h_{2}$ of the
		aggregate allocation graph; symmetrically, type $t_{2}$ gives the edge
		$h_{2}\rightarrow h_{1}$. The graph has a cycle, and Theorem \ref{PI}
		rejects. Yet $x$ is WC-rationalizable, for instance by the profile
		\begin{equation*}
			\begin{tabular}{c|c|c|c}
				$\succ_{t_{1}}$ & $\succ_{t_{2}}$ & $\succ_{t_{3}}$ & $\succ_{t_{4}}$ \\ \hline
				$h_{2}$ & $h_{1}$ & $h_{3}$ & $h_{4}$ \\
				$h_{1}$ & $h_{2}$ & $h_{1}$ & $h_{2}$ \\
				$h_{3}$ & $h_{3}$ & $h_{2}$ & $h_{1}$ \\
				$h_{4}$ & $h_{4}$ & $h_{4}$ & $h_{3}$%
			\end{tabular}%
		\end{equation*}
		Individual rationality holds, and individuals 1, 3, 5 and 6 each receive
		their type's most preferred object, so no blocking coalition can contain
		them. Only individuals 2 and 4 remain: individual 2 could improve only
		by obtaining $h_{2}$, individual 4 only by obtaining $h_{1}$, but their
		endowments are $h_{3}$ and $h_{4}$ --- redistributing these between
		themselves helps neither. Hence no coalition strongly blocks $x$, and
		$x$ is in the weak core for this profile. The example is the classic
		phenomenon of \citet{roth1977weak} in our setting: individuals 1 and 3
		would both gain by exchanging their \emph{assignments}, which is why
		efficiency fails, but neither holds an \emph{endowment} the other wants,
		so they cannot block. The weak core disciplines trades of endowments, and
		is therefore silent about this improvement.
	\end{example}
	
	Examples \ref{example-wc} and \ref{example-wc-not-pi} together establish
	the following.

	\begin{example}\label{example-joint}
		\new{Let $T=\left\{ t_{1},t_{2},t_{3}\right\} $ and let there be four objects:
			$h_{1}$, $h_{2}$ and $h_{4}$ with two copies each and $h_{3}$ with one copy.
			The seven individuals are as follows.}
		\begin{center}
			\begin{tabular}{cccc}
				\toprule
				Individual & Type & Endowment $\omega(i)$ & Assignment $x(i)$ \\
				\midrule
				$1$ & $t_{1}$ & $h_{1}$ & $h_{1}$ \\
				$2$ & $t_{1}$ & $h_{3}$ & $h_{2}$ \\
				$3$ & $t_{2}$ & $h_{2}$ & $h_{1}$ \\
				$4$ & $t_{2}$ & $h_{4}$ & $h_{2}$ \\
				$5$ & $t_{3}$ & $h_{1}$ & $h_{4}$ \\
				$6$ & $t_{3}$ & $h_{4}$ & $h_{3}$ \\
				$7$ & $t_{3}$ & $h_{2}$ & $h_{4}$ \\
				\bottomrule
			\end{tabular}
		\end{center}
		\new{Individual rationality forces $h_{2}\succ_{t_{1}}h_{3}$ from individual
			$2$; $h_{1}\succ_{t_{2}}h_{2}$ and $h_{2}\succ_{t_{2}}h_{4}$ from individuals
			$3$ and $4$; and $h_{4}\succ_{t_{3}}h_{1}$, $h_{3}\succ_{t_{3}}h_{4}$,
			$h_{4}\succ_{t_{3}}h_{2}$ from individuals $5$, $6$ and $7$. Individual $1$ is
			a stayer and reveals nothing, so how $t_{1}$ ranks $h_{1}$ against $h_{2}$ is
			left open by the data --- and it is the only comparison that matters here.}
		
		\new{\emph{The allocation is PI-rationalizable.} Take}
		\begin{equation*}
			\begin{tabular}{c|c|c}
				$\succ_{t_{1}}$ & $\succ_{t_{2}}$ & $\succ_{t_{3}}$ \\ \hline
				$h_{1}$ & $h_{1}$ & $h_{3}$ \\
				$h_{2}$ & $h_{2}$ & $h_{4}$ \\
				$h_{3}$ & $h_{3}$ & $h_{1}$ \\
				$h_{4}$ & $h_{4}$ & $h_{2}$%
			\end{tabular}%
		\end{equation*}
		\new{with objects listed from best to worst. Individual rationality is
			immediate. Individuals $1$ and $3$ receive $h_{1}$, the best object of their
			types, and individual $6$ receives $h_{3}$, the best object of $t_{3}$, so none
			of them can be made better off. The remaining individuals can improve only in
			one way each: $2$ and $4$ only by obtaining $h_{1}$, and $5$ and $7$ only by
			obtaining $h_{3}$. Any such change would take $h_{1}$ from individual $1$ or
			$3$, or $h_{3}$ from individual $6$, each of whom would then hold an object his
			type ranks strictly lower. Hence no allocation Pareto dominates $x$, and $x$ is
			PI-rationalizable. (Equivalently, the aggregate allocation graph has the two
			edges $h_{2}\rightarrow h_{1}$ and $h_{4}\rightarrow h_{3}$ and is acyclic, so
			Theorem \ref{PI} accepts.)}
		
		\new{\emph{The allocation is WC-rationalizable.} Take instead}
		\begin{equation*}
			\begin{tabular}{c|c|c}
				$\succ_{t_{1}}$ & $\succ_{t_{2}}$ & $\succ_{t_{3}}$ \\ \hline
				$h_{2}$ & $h_{1}$ & $h_{3}$ \\
				$h_{1}$ & $h_{2}$ & $h_{4}$ \\
				$h_{3}$ & $h_{4}$ & $h_{1}$ \\
				$h_{4}$ & $h_{3}$ & $h_{2}$%
			\end{tabular}%
		\end{equation*}
		\new{which differs from the first profile only in that $t_{1}$ now ranks
			$h_{2}$ above $h_{1}$ (and in an immaterial reordering of the bottom two
			objects for $t_{2}$). Individuals $2$, $3$ and $6$ receive the best object of
			their type and so cannot belong to a blocking coalition. Individuals $5$ and
			$7$ can improve only by obtaining $h_{3}$, which is the endowment of individual
			$2$ alone; since $2$ cannot belong to a blocking coalition, neither can $5$ or
			$7$. This leaves individuals $1$ and $4$, whose endowments are $h_{1}$ and
			$h_{4}$; individual $1$ can improve only by obtaining $h_{2}$, which is not
			among them, and individual $4$ alone cannot improve on $h_{2}$ with $h_{4}$.
			Hence no coalition strongly blocks $x$, and $x$ is WC-rationalizable.}
		
		\new{\emph{The allocation is not jointly rationalizable.} Let $\succ$ be any profile for which $x$ is individually rational, as it must be
			under either requirement. Exactly one of two cases arises.}
		
		\new{If $h_{2}\succ_{t_{1}}h_{1}$, let $y$ agree with $x$ except that
			$y(1)=h_{2}$ and $y(4)=h_{1}$. Then $y$ is an allocation, individual $1$
			strictly gains because $h_{2}\succ_{t_{1}}h_{1}=x(1)$, and individual $4$
			strictly gains because individual rationality at individual $3$ forces
			$h_{1}\succ_{t_{2}}h_{2}=x(4)$. So $y$ Pareto dominates $x$ and $x$ is not
			Pareto efficient.}
		
		\new{If instead $h_{1}\succ_{t_{1}}h_{2}$, let $S=\left\{ 2,5\right\} $ and let
			$x^{\prime }(2)=\omega(5)=h_{1}$ and $x^{\prime }(5)=\omega(2)=h_{3}$. This
			redistributes the endowments of $S$; individual $2$ strictly gains because
			$h_{1}\succ_{t_{1}}h_{2}=x(2)$, and individual $5$ strictly gains because
			individual rationality at individual $6$ forces $h_{3}\succ_{t_{3}}h_{4}=x(5)$.
			So $S$ strongly blocks $x$ and $x$ is not in the weak core.}
		
		\new{Since $\succ_{t_{1}}$ is a linear order, one of the two cases must hold,
			and no single profile can make $x$ both Pareto efficient and weak core-stable.}
	\end{example}
	
	\new{The example is worth reading slowly, because the mechanism is not an
		artifact of the numbers. Everything turns on a single comparison the data leave
		open --- how type $t_{1}$ ranks $h_{1}$ against $h_{2}$ --- and the two
		requirements pin it down in opposite directions.}
	
	\new{The reason they do is that efficiency and the weak core discipline
		different objects. Pareto efficiency is about \emph{assignments}: it asks
		whether individuals would like to exchange what they were given. Ranking
		$h_{2}$ above $h_{1}$ makes individual $1$, who holds $h_{1}$, want individual
		$4$'s assignment $h_{2}$, while individual $4$ already wants $h_{1}$, because
		individual $3$ of his type gave up $h_{2}$ to get it. The two can simply trade
		assignments, and efficiency fails. The weak core is about \emph{endowments}: it
		asks whether individuals would like to undo the reallocation and trade what
		they brought. Ranking $h_{1}$ above $h_{2}$ makes individual $2$, who holds
		$h_{2}$, want $h_{1}$. And $h_{1}$ is not only somebody's assignment but
		also individual $5$'s endowment. Individual $5$ is willing to give it up
		for $h_{3}$, because individual $6$ of his type gave up $h_{4}$ to get
		$h_{3}$. The
		two can trade endowments, and the weak core fails. Each requirement consumes
		the same degree of freedom, and they consume it in opposite ways.}

	\begin{example}[Choosing admissible crossing points independently is not enough]
		\label{ex-betweenness}
		Let $\rhd _{T}:t_{1}\rhd t_{2}\rhd t_{3}$
		and $\rhd _{H}:h_{1}\rhd h_{2}\rhd h_{3}$. \clr{There are three agents, $1,2,3$, each in a type of its own; their endowments and assignments are below.}
		\begin{equation*}
			\begin{tabular}{c|ccc}
				agent & $1$ & $2$ & $3$ \\ \hline
				type & $t_{1}$ & $t_{2}$ & $t_{3}$ \\
				$\omega$ & $h_{1}$ & $h_{2}$ & $h_{3}$ \\
				$x$ & $h_{2}$ & $h_{3}$ & $h_{1}$
			\end{tabular}
		\end{equation*}
		
		\clr{Because every agent is alone in its type, $x$ is trivially WC-rationalizable: for instance $\succ_{t_{1}}:h_{2}\succ h_{1}\succ h_{3}$, $\succ_{t_{2}}:h_{1}\succ h_{3}\succ h_{2}$, and $\succ_{t_{3}}:h_{1}\succ h_{2}\succ h_{3}$ leaves $x$ in the weak core.}
		
		\par\smallskip
		\begin{center}
			\begin{tikzpicture}[x=1cm,y=1cm]
				\node[vtx] (n12) at (3.4,0)    {$(t_1,h_2)$};
				\node[vtx] (n31) at (0,-2.6)   {$(t_3,h_1)$};
				\node[vtx] (n23) at (6.8,-2.6) {$(t_2,h_3)$};
				\draw[inter] (n31) -- (n12);
				\draw[inter] (n12) -- (n23);
				\draw[inter] (n23) -- (n31);
			\end{tikzpicture}\\[3pt]
		\end{center}
		\par\smallskip
		
		Each mover defines a constraint on crossing
		points. Agent $1$ of type $t_{1}$ and of the rank-$1$, receives the
		lower-quality $h_{2}$ as his final object and his endowment was $h_{1}$, so $\kappa (h_{1},h_{2})\leq 0$; agent $2$ of type $t_{2}$ and of the rank-$2$, has the lower-quality $h_{3}$ as his final object and his initial endowment was $h_{2}$, so $\kappa (h_{2},h_{3})\leq 1$; agent $3$ of type $t_{3}$ and
		of the rank-$3$ is moving from $h_{3}$ to the higher-quality $h_{1}$, so $\kappa (h_{1},h_{3})\geq 3$. Therefore, the admissible sets are all nonempty,
		\begin{equation*}
			K(h_{1},h_{2})=\{0\},\qquad K(h_{2},h_{3})=\{0,1\},\qquad K(h_{1},h_{3})=\{3\}.
		\end{equation*}
		However, no single-crossing profile is consistent, since consistency on
		the triple $(h_{1},h_{2},h_{3})$ would require
		\begin{equation*}
			\kappa (h_{1},h_{3})\ \leq \ \max \{\kappa (h_{1},h_{2}),\kappa (h_{2},h_{3})\}\ \leq \ 1,
		\end{equation*}
		contradicting $\kappa (h_{1},h_{3})=3$. Equivalently, $\kappa
		(h_{1},h_{3})=3$ forces $h_{1}\succ _{t_{2}}h_{3}$, which with $h_{3}\succ _{t_{2}}h_{2}$
		yields $h_{1}\succ _{t_{2}}h_{2}$, contradicting $\kappa (h_{1},h_{2})=0$. Hence $x$ is \emph{not} single-crossing WC-rationalizable, although the per-pair admissible sets are all
		nonempty and $x$ is WC-rationalizable.
	\end{example}

	\begin{example}[The mixed graph is still not sufficient]
		\label{ex-unheld}
		Let $\rhd _{T}:t_{1}\rhd t_{2}$ and $\rhd _{H}:h_{1}\rhd h_{2}\rhd h_{3}$. \clr{There are four agents; their types, endowments, and assignments are below, with the aggregate allocation matrix on the right.}
		\begin{equation*}
			\begin{tabular}{c|cccc}
				agent & $1$ & $2$ & $3$ & $4$ \\ \hline
				type & $t_{2}$ & $t_{1}$ & $t_{1}$ & $t_{1}$ \\
				$\omega$ & $h_{1}$ & $h_{3}$ & $h_{3}$ & $h_{2}$ \\
				$x$ & $h_{2}$ & $h_{3}$ & $h_{1}$ & $h_{3}$\end{tabular}\qquad X(x)=\begin{pmatrix}
				1 & 0 & 2 \\
				0 & 1 & 0\end{pmatrix}\text{,}
		\end{equation*}
		so $H_{t_{1}}=\{h_{1},h_{3}\}$ and $H_{t_{2}}=\{h_{2}\}$. The movers are agent $1$
		(from $h_{1}$ to $h_{2}$), agent $3$ (from $h_{3}$ to $h_{1}$) and agent $4$ (from $h_{2}$ to $h_{3}$). Hence we have
		\begin{equation*}
			h_{2}\succ _{t_{2}}h_{1}\text{,}\qquad h_{1}\succ _{t_{1}}h_{3}\text{,}\qquad h_{3}\succ _{t_{1}}h_{2}\text{.}
		\end{equation*}
		
		\emph{A forced ranking of an unheld object.} Only the middle comparison,
		$h_{1}\succ _{t_{1}}h_{3}$, joins two objects
		held by the same type; it is the unique intra-arc. The other two comparisons
		involve unheld objects---$h_{1}\notin H_{t_{2}}
		$ and $h_{2}\notin H_{t_{1}}$---so neither is
		represented in the mixed graph. The three comparisons bound the crossing
		points as before. Agent $1$, of the rank-$2$ type $t_{2}$, receives the
		lower-quality $h_{2}$, so $\kappa (h_{1},h_{2})\leq 1$. Agent $3$, of
		the rank-$1$ type $t_{1}$, receives the higher-quality $h_{1}$, so
		$\kappa (h_{1},h_{3})\geq 1$. And agent $4$, of the top type $t_{1}$,
		receives the lower-quality $h_{3}$, so $\kappa (h_{2},h_{3})\leq 0$,
		i.e.\ $\kappa (h_{2},h_{3})=0$. The last bound is the
		crucial one: by single-crossing, the top type's preference $h_{3}\succ _{t_{1}}h_{2}$ propagates downward and forces $h_{3}\succ _{t_{2}}h_{2}$ as well---a ranking, by
		type $t_{2}$, of the object $h_{3}$ it does not hold.
		Moreover, the bounds and consistency pin $\kappa $ down completely:
		consistency on the triple requires $\kappa (h_{1},h_{3})\leq \max \{\kappa (h_{1},h_{2}),\kappa (h_{2},h_{3})\}=\kappa (h_{1},h_{2})$, so $1\leq
		\kappa (h_{1},h_{3})\leq \kappa (h_{1},h_{2})\leq 1$, hence $\kappa (h_{1},h_{2})=\kappa (h_{1},h_{3})=1$. The \emph{unique} single-crossing profile
		compatible with the data is therefore
		\begin{equation*}
			\succ_{t_{1}}^{\kappa }:h_{1}\succ h_{3}\succ h_{2}\text{,}\qquad \succ_{t_{2}}^{\kappa }:h_{3}\succ h_{2}\succ h_{1}\text{.}
		\end{equation*}
		
		\emph{The mixed graph passes Condition~\ref{cond1}.} The set of
		undirected edges is again empty, and the mixed graph (intra-arc $(t_{1},h_{3})\!\to\!(t_{1},h_{1})$;
		inter-arcs $(t_{1},h_{1})\!\to\!(t_{2},h_{2})$ and $(t_{2},h_{2})\!\to\!(t_{1},h_{3})$) has no alternating cycle, so it passes Condition~\ref{cond1}.
		
		\emph{But a coalition blocks.} The allocation $x$ is not in the weak core of
		$\succ^{\kappa }$: agents $1$ and $2$ strongly block by
		exchanging endowments. Agent $1$ (type $t_{2}$, holding $h_{2} $) takes agent $2$'s endowment $h_{3}$ and improves because $h_{3}\succ _{t_{2}}h_{2}$; agent $2$ (type $t_{1}$, holding $h_{3}$) takes agent $1$'s endowment $h_{1}$
		and improves because $h_{1}\succ _{t_{1}}h_{3}$. The
		blocking step $h_{3}\succ _{t_{2}}h_{2}$ uses the
		object $h_{3}\notin H_{t_{2}}$, which has no
		vertex in the mixed graph, so the block is invisible to
		Condition~\ref{cond1}. Since $\succ^{\kappa }$ is the only single-crossing
		profile compatible with the data, $x$ is not single-crossing
		WC-rationalizable even though Condition~\ref{cond1} holds for a consistent
		$\kappa $. Dropping the single-crossing restriction, $x$ \emph{is}
		WC-rationalizable: e.g.\ $\succ_{t_{1}}:h_{1}\succ h_{3}\succ h_{2}$ together with $\succ_{t_{2}}:h_{2}\succ
		h_{1}\succ h_{3}$ leaves $x$ in the weak core. This profile
		ranks $h_{2}\succ _{t_{2}}h_{3}$, precisely the
		comparison single-crossing rules out (the top type has $h_{3}\succ _{t_{1}}h_{2}$), which is why no consistent $\kappa $ can
		reproduce it.
		
		\par\smallskip
		\begin{center}
			\begin{tikzpicture}[x=1cm,y=1cm]
				\node[vtx]   (m11) at (0,0)      {$(t_1,h_1)$};
				\node[vtx]   (m13) at (6.8,0)    {$(t_1,h_3)$};
				\node[vtx]   (m22) at (3.4,-2.0) {$(t_2,h_2)$};
				\node[vtxph] (m23) at (6.8,-2.0) {$(t_2,h_3)$};
				\draw[intra] (m13) -- (m11);
				\draw[inter] (m11) -- (m22);
				\draw[inter] (m22) -- (m13);
				\draw[miss]  (m22) -- (m23);
			\end{tikzpicture}\\[3pt]
			{\footnotesize Held-object mixed graph for Example~\ref{ex-unheld} (solid arcs): no alternating
				cycle, so Condition~\ref{cond1} holds. Single-crossing forces $h_3\succ_{t_2}h_2$ (dashed), yet $(t_2,h_3)$ is \emph{not} a vertex of $\mathcal{G}$; the
				block by agents $1$ and $2$ runs through this missing vertex and so is invisible to the graph.}
		\end{center}
		\par\smallskip
	\end{example}

\bibliographystyle{elsarticle-harv}
\bibliography{Bibliography}